\documentclass[11pt,a4paper]{article}

\usepackage[T1]{fontenc}
\usepackage{lmodern}
\usepackage[margin=2.5cm]{geometry}
\usepackage{microtype}
\UseMicrotypeSet[protrusion]{basictext}
\usepackage{graphicx}
\usepackage{booktabs}
\usepackage{tabularx}
\usepackage{longtable}
\usepackage{placeins}
\usepackage{array}
\usepackage{enumitem}
\usepackage{xcolor}
\usepackage[numbers,sort&compress]{natbib}
\usepackage{xurl}
\usepackage{caption}
\usepackage{listings}
\usepackage[colorlinks=true,linkcolor={blue!50!black},citecolor={blue!50!black},urlcolor={blue!50!black}]{hyperref}
\makeatletter\providecommand*{\toclevel@lstlisting}{0}\makeatother

\hypersetup{pdftitle={Everything but the Software: Stable Retention, Missing Features and the Cost of Consolidation},pdfauthor={Juan Manuel Servera Bondroit; Isabel Cabezas Martin}}
\setlist{itemsep=2pt,topsep=4pt}
\newcolumntype{Y}{>{\raggedright\arraybackslash}X}
\newcolumntype{L}[1]{>{\raggedright\arraybackslash}p{#1}}
\newcommand{\sourceref}[1]{\hyperref[prov:#1]{\textsf{#1}}}
\newcommand{\armlegend}{A\@: feature-only; B\@: consolidation; C\@: read-only
review; D\@: longer feature-only sessions; E\@: visible-specification repair.}
\title{Everything but the Software:\\[4pt]
\large Stable Retention, Missing Features and the Cost of Consolidation.\\
An Exploratory Case Study and Experiment}

\author{%
Juan Manuel Servera Bondroit%
\thanks{The views expressed are the authors' own and do not
necessarily reflect those of Microsoft.}\\[4pt]
\small Senior Partner Solution Architect\\
\small Microsoft EMEA, Zurich
\and
Isabel Cabezas Martin\\[4pt]
\small Software Engineer\\
\small Frontier Company at Microsoft, Zurich
}

\begin{document}
\maketitle

\begin{abstract}
I spent weeks refining requirements with agents before asking a coordinator
agent I had created to manage a ``herd'' of agents researching, building and
verifying the product's code. After 47 completed issues in five days, checks
were passing, but I could not sign in.
What does each feature leave for the next one?
We tested three models across five arms with 21 runs per cell, each building
ten Python features before a withheld change request, with hidden tests
after each step.
Earlier behaviour was largely preserved: one of 315 runs ended with an
open regression, and workflow replay found no witnessed losses in 626
evaluable packets, each representing one run at one endpoint.
Consolidation removed complexity hotspots, but its coverage gain over longer
sessions was 1.19 percentage points (conditional 95\% interval [0.10, 2.22]);
the controls left only a 3.04-point ceiling against the pre-specified
ten-point target.
The main failures concerned missing ranking features and permission
boundaries; read-only review completed the extension in 16/21 Gemini runs
versus 8/21 with consolidation, which used 1.84--2.25 times the delivery
credits of feature-only work in fully billed subsets.
Rollback and immutable accepted tests plausibly helped prevent decay while
also obstructing a later feature, but no arm tested their absence.
Unregistered amendments, uncertain run independence, incomplete billing and
judging, and two unknown extension outcomes limit this single-task study.
Structural cleanup improved the measured code properties without
demonstrating cheaper adaptation.
\end{abstract}

\noindent\textbf{Keywords:} agentic software engineering; coding agents;
spec-driven development; software evolution; technical debt; refactoring;
feedback loops; human-AI interaction; exploratory case study; experiment protocol.

\section{Introduction}\label{sec:intro}

For my capstone in Microsoft's internal Azure Master Architect (AMA)
programme~\cite{microsoftamabadge}, I was developing a product from an assigned synthetic
housing-retrofit use case: helping staff coordinate energy-efficiency
work on residents' homes.\footnote{``I'' identifies Servera Bondroit's participant account; ``we''
identifies the authors' experiment, analysis and paper.}
I used the HyperVelocity methodology~\cite{hvecore} to move an early
prototype towards
production readiness while leaving out some non-essential production
requirements, expecting to keep the behaviour that already worked.
Short exchanges with agents had produced a prototype and a Minimum Viable
Experiment (MVE) in a few hours, including working sign-in. I used those
results to refine the business and product requirements documents (BRD and
PRD), then gave this coordinator agent and its parallel workers a formal backlog to
implement autonomously. I call these managed worker agents a ``herd'', using
the metaphor developed in my practitioner article
\emph{The Agent Shepherd}~\cite{servera2026shepherd}.
My account records 47 completed issues out of more
than 180 in the first 5 days. I could inspect completed issues and passing
checks, but when I opened the replacement to provide feedback, I could not
sign in to the new application.

The immediate problem was a blocked deployment and sign-in workflow,
but it raised a question about the accumulated system:
when a change passes its local checks, what evidence shows that the next
change inherits working software?

Inherited faults predate coding agents: the Knight Capital incident
combined old router code with an incomplete deployment and inadequate
controls~\cite{sec2013knight}, illustrating why local feature checks can
miss the state a deployment inherits.

Established engineering practice has a long answer to inherited faults:
define which behaviour a change must preserve, test it, and keep the
running increment available for feedback, allowing refactoring to change
the implementation while those contracts continue to hold
~\cite{parnas1994aging,beck2002tdd,beck2004xp,fowler2007stamina}.
Anthropic's long-running-agent example makes this concrete: before adding
a feature to a chat application, the next session starts the server,
opens a new chat, sends a message and checks for a response
~\cite[Getting up to speed]{anthropic2025harness}. That check asks whether the previous session left a usable application for the next change, a concern revisited by Beck and Fowler~\cite{beck2026prodacity,fowler2025thoughts}. Beck's distinction between producing code and learning through working
software sharpens the research question~\cite{beck2026prodacity}:
what does a feature leave behind for the next one?

Before asking the coordinator to implement the backlog, I had spent weeks
working through the requirements with the agents and recording our decisions.
Research agents maintained the evidence base, and in one what-if exercise, I asked whether phone photos with
GPS metadata would make an assessment agent more useful; the resulting
proposal still required checked provenance and a professional decision.
Persona and journey work examined who would take over a case when information or approval was missing.
Disposable pilots and prototypes exposed workflow problems before a usable minimum viable product (MVP)
emerged, and a backlog agent
translated the resulting requirements and architecture decisions into epics
and user stories.
I expected the agents to build the product faster and make fewer mistakes
because we had already questioned and revised the requirements.

My attempts to use the new application changed what I asked the coordinator
to do next: provide a running website, fix the missing workflow and refactor.
The case analysis follows those requests to see where preparation and local
completion checks failed to preserve the workflow I needed.
The public repository, pull requests, append-only log and versioned
coordinator instructions document that evolving record~\cite{servera2026ama};
the source package pins the evidence used here.
My earlier experience on another book-search project (Aithena, with PDF
ingestion and a browser interface, discussed in the supplement) and the
external PhotoCraft inspection selected after the motivating-case analysis
are documented as contextual comparisons in the Supplement.

The initial conjecture was that feature-driven delivery carries
unresolved errors and avoidable change burden forward, and that cumulative
verification and consolidation can reduce that burden. We distinguish
the software being changed from the instructions used to change it; in this case, my
instructions grew to require the full research, plan, implement and review
(RPI) sequence, committed evidence
and recurring checks (Section~\ref{sec:coordinator}). Section~\ref{sec:background} separates these
pathways; the case then examines how they interact with completion proxies,
review capacity and product-use feedback.

Stable behaviour would count against the conjecture under the tested conditions, while a single regression would provide little evidence of progressive decay. Here
\emph{feature-driven decay} means a measurable loss of previously passing
behaviour or a rising cost and failure burden across sequential changes.
The experiment asks agents to build the same small Python
package, called S1, through ten features and then complete a withheld
change request (CR1). The arms compare feature-only work (A), consolidation (B), read-only review (C), longer feature sessions (D) and visible-specification repair (E). A \emph{contract} is a required behaviour checked by a fixed hidden test, and the analysis distinguishes the \emph{decay premise} from
the \emph{consolidation remedy} through three predictions and a check of
whether consolidation actually changed the code.

\paragraph{P1 (Short-horizon decay premise).}
Without a dedicated
post-feature pause (Arms~A and~D), sequential delivery across ten features
induces progressive behavioural loss by the final feature checkpoint
($P_{10}$): previously passing contracts still fail, failures accumulate
across cycles, or replayed workflows lose earlier capabilities.

\paragraph{P2 (Primary coverage gain threshold).}
Scheduled consolidation (Arm~B) exceeds the matched-allowance longer-session control
(Arm~D) in equal-model-weight final hidden-test coverage at $P_{10}$ by at
least the pre-specified minimum useful effect of $\Delta_*=10$ percentage
points (Section~\ref{sec:protocollimits}).

\paragraph{P3 (Change-request adaptability and short-horizon payback).}
Consolidation (Arm~B) raises held-out CR1 success and reduces CR1 or net
credit expenditure relative to feature-only delivery (Arm~A), read-only
review (Arm~C) and longer sessions (Arm~D).

\paragraph{Manipulation check (Structural treatment fidelity).}
Consolidation reduces functions above cyclomatic complexity 10, duplication
and lint density towards its stopping targets, checking whether
the intervention changed structure before the behavioural
results are interpreted.

The results changed our interpretation: most acquired behaviour survived,
while structural cleanup produced a coverage gain below the practical
benchmark (Table~\ref{tab:predictionverdicts}), directing attention towards
missing features and failures exposed by a new requirement.

Only $\Delta_*=10$ percentage points was fixed in the pre-collection
planning calculation. The public execution manifest left its formal hypothesis
array empty, so the operational criteria in P1 and P3 structure the
post-collection evaluation and do not amount to a preregistered
confirmatory test.

Three questions guide the investigation:

\begin{itemize}[leftmargin=*]
\item \textbf{RQ1.} How can the code and instructions left by one feature
carry errors or make the next feature harder to implement, and what does
existing evidence show about that problem?
\item \textbf{RQ2.} What could explain the gap between completed issues and
an application I could use in the motivating case, and which competing
explanations does the delivery record support?
\item \textbf{RQ3.} Does scheduled agent-led consolidation after each feature
improve retained behaviour and future changeability, and at what cost?
\end{itemize}

The primary comparison measures consolidation, added tests and a handover
note against a control with matching time and nominal-credit ceilings;
the secondary feature-only comparison adds resources as well as
consolidation. The experiment neither isolates refactoring nor tests the
five case mechanisms individually; Section~\ref{sec:threats} consolidates
the limits of both studies.
Table~\ref{tab:source-register} and Data availability identify the source
register, public repositories and supporting records.

\section{Background}\label{sec:background}

\subsection{What later changes inherit}\label{sec:decay}

Maintenance alongside growth is a recurring concern in software evolution
~\cite{lehman1980laws,lehman1997metrics,parnas1994aging}, and Lehman's 1997
formulation states that quality declines without maintenance, but supplies
no data for or against that law. Eick et~al.\ found mixed evidence in more
than fifteen years of change history~\cite{eick2001decay}; a systematic
review likewise found uneven support across systems~\cite{herraiz2013laws}.
Counting features or changed lines cannot establish decay: a regression
requires previously working behaviour that is still required to fail.
In our study, GPT trajectories that never implemented priority ranking
had an acquisition failure, whereas the Gemini consent contract that first failed
at $P_7$ and remained failing through $P_{10}$ instead records one
persistent regression (Section~\ref{app:reportingdisposition}).
Showing that it caused a distinct later defect would require tracing
that dependency.

Cunningham describes shipping first-time code and rewriting it as
understanding improves~\cite{cunningham1992wycash}. Fowler's design
stamina is explicitly a hypothesis~\cite{fowler2007stamina}, and reported debt costs depend on setting and measurement method: a
vendor-associated 39-codebase study reports 124\% longer issue resolution in
low-quality files~\cite{tornhill2022codered}.

\begin{figure}[!htbp]
\centering
\includegraphics[width=\linewidth]{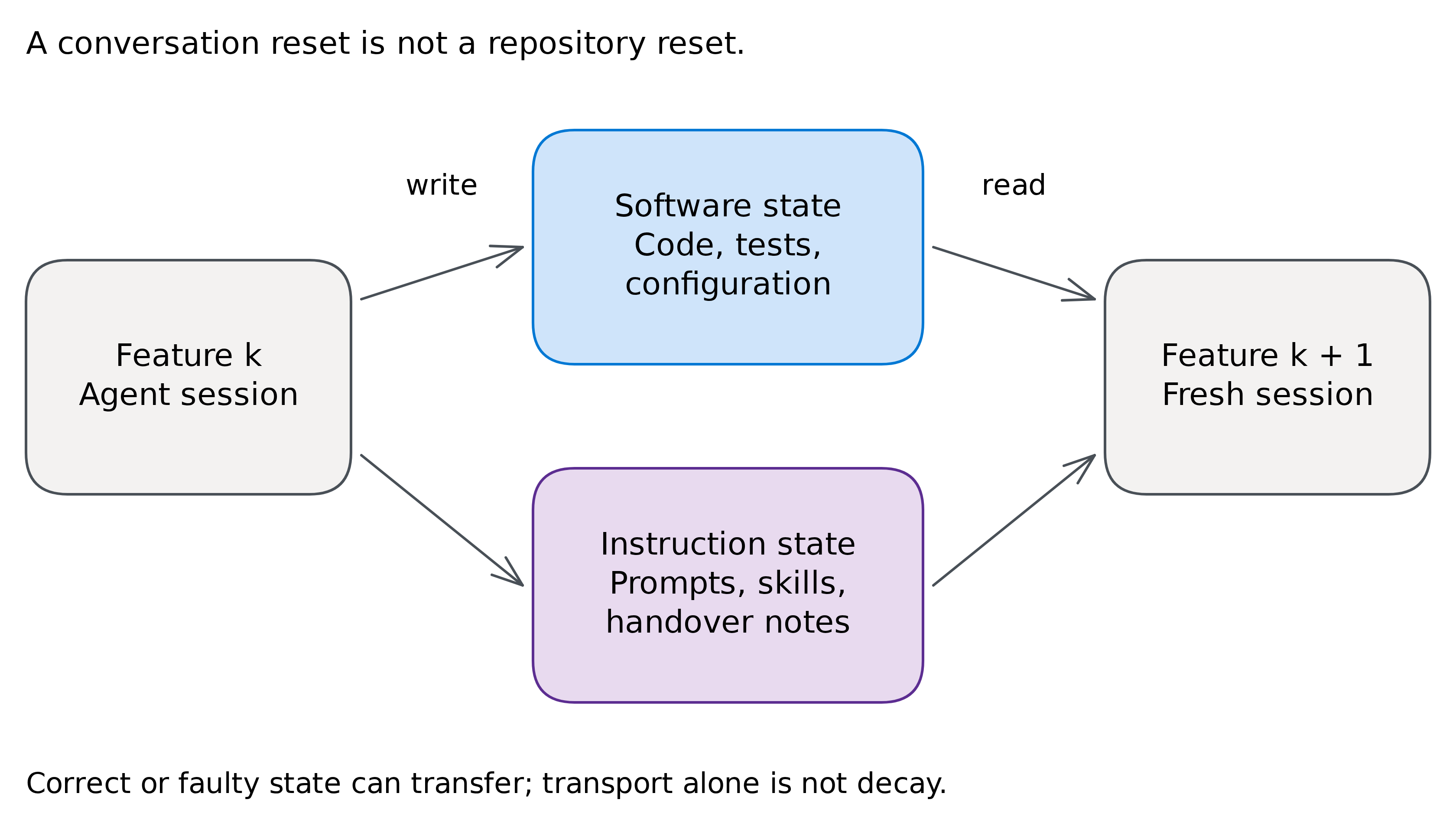}
\caption{Conceptual read/write paths available between sessions. Arrows identify
state-carrying paths.}\label{fig:stateinheritance}
\end{figure}

\paragraph{What survives the next session?}\label{sec:inheritance}
In this experiment, fresh feature sessions received the retained repository;
B and C also permitted design and review handover files
(Section~\ref{sec:protocolsummary}). The study's
\texttt{READ\_METHODS} test, for example, continued to constrain the
ninth feature after the session that wrote it had ended
(Section~\ref{sec:resultscorrectness}).
Figure~\ref{fig:stateinheritance} distinguishes the software and instruction
paths; ``software entropy'' is used here only
as an analogy for maintenance burden; the measured outcomes are retained
behaviour, adaptation and resource use.

\subsection{Evidence within and between features}\label{sec:empirical}

Conversational benchmarks identify dependence on erroneous histories.
Zhang et~al.\ study false justifications after
wrong answers~\cite{zhang2023snowball}; Sinha et~al.\ vary erroneous prior
state updates in a controlled execution task
~\cite[Secs.~2.2--3, Figs.~5--6]{sinha2025illusion}.

Pan et~al.'s SWE-Maintain more directly tests repository inheritance:
100 sequences of three maintenance tasks from 50 Python repositories.
Later tasks are matched across developer-written and agent-written histories,
with a fresh session for each task; across 1,996 paired later events, agent history reduces current-task
resolution by 5.6 points; requiring both current and earlier tests to pass gives
a 14.0-point gap. Unweighted mean current-task gaps grow from 4.0 to
7.2 points between the second and third events. In the selected
619-event clean-history subset, the current-task difference is
$-1.1$ points, interval $[-3.8,+1.6]$, and cumulative event resolution
differs by $+2.7$, interval $[-0.5,+6.1]$
~\cite[Secs.~4.1--4.2, Tables~2--3]{pan2026swemaintain}.
The smaller difference in the clean-history subset is consistent with
inherited behavioural failures contributing more than complexity alone.
Because the authors selected that subset rather than randomly repairing
the repositories, the comparison cannot estimate the benefit of
a repair intervention or establish how widely the pattern applies.

\begin{figure}[!htbp]
\centering
\includegraphics[width=\linewidth]{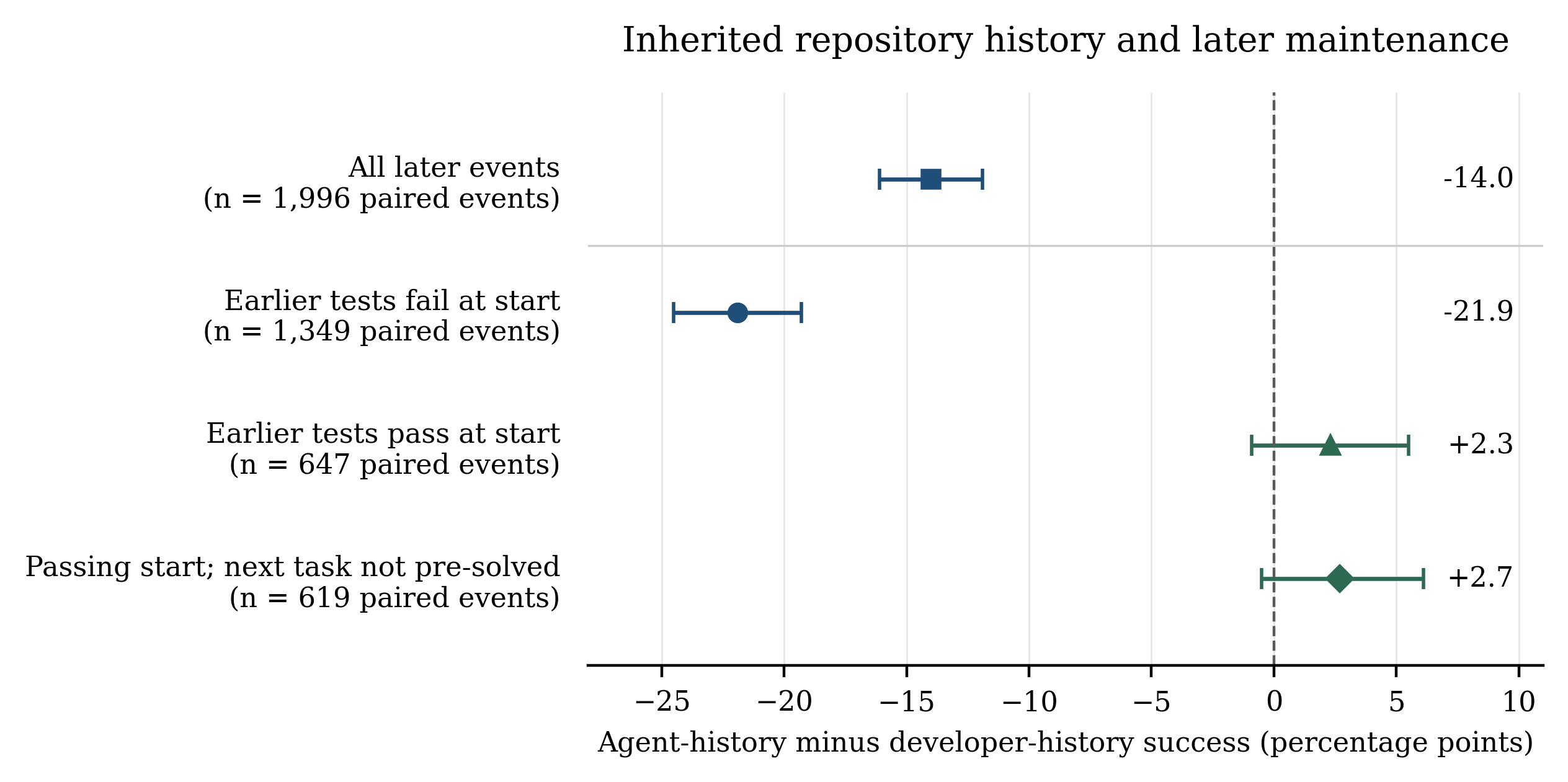}
\caption{Published event-resolution differences and repository-clustered
95\% intervals from Pan et~al., Table~3. Current and prior
non-superseded tests must pass. The overlapping selected cohorts apply
different restrictions to the same benchmark. Points and whiskers reproduce
the published estimates and intervals.}\label{fig:inheritedhistory}

\end{figure}

Their \texttt{python\_unidiff} example reuses line-splitting logic that
loses terminators in a later constructor; Pan et~al.'s Figure~1 caption introduces
it, and their Section~5.1 and Figure~5 supply the detail
~\cite{pan2026swemaintain}. SlopCodeBench and SWE-CI also evaluate agents over successive changes
~\cite{orlanski2026slopcodebench,chen2026sweci}, but evidence of downstream harm
is not uniform: Borg et~al.\ found no significant maintainability difference
for late-2024 assistants in a two-phase study where the second phase was
randomised~\cite[Secs.~1, 4]{borg2025echoes}. Differences in tool generation and collaboration
setting limit direct comparison.

\subsection{Instruction maintenance and evaluation}\label{sec:instructiondebt}

Nigh recommends runnable commands such as \texttt{npm test} and explicit
write boundaries for agent
instruction files~\cite{nigh2025agentsmd}; Anthropic's context-engineering
guide recommends compacting history and retaining structured notes
~\cite{rajasekaran2025context}.
Anthropic's harness asks each coding session to update
a progress note and leaves the next one a startup script,
feature list and Git history to inspect~\cite{anthropic2025harness}.
Mohsenimofidi et~al.\ assign 242 labels across nine categories to 169 commits,
with several labels possible per commit: 78 include instruction additions,
59 modifications and 23 removals
~\cite[Sec.~4.3, Table~2]{mohsenimofidi2026context}.
Chatlatanagulchai et~al.'s sampled maintenance labels likewise include
addition and pruning~\cite{chatlatanagulchai2026readmes};
we use \emph{instruction debt} for obsolete, duplicated, conflicting or poorly
routed directives, naming a possible maintenance burden; the commit
counts alone do not measure its effect on development.

\begin{figure}[!htbp]
\centering
\includegraphics[width=0.94\linewidth]{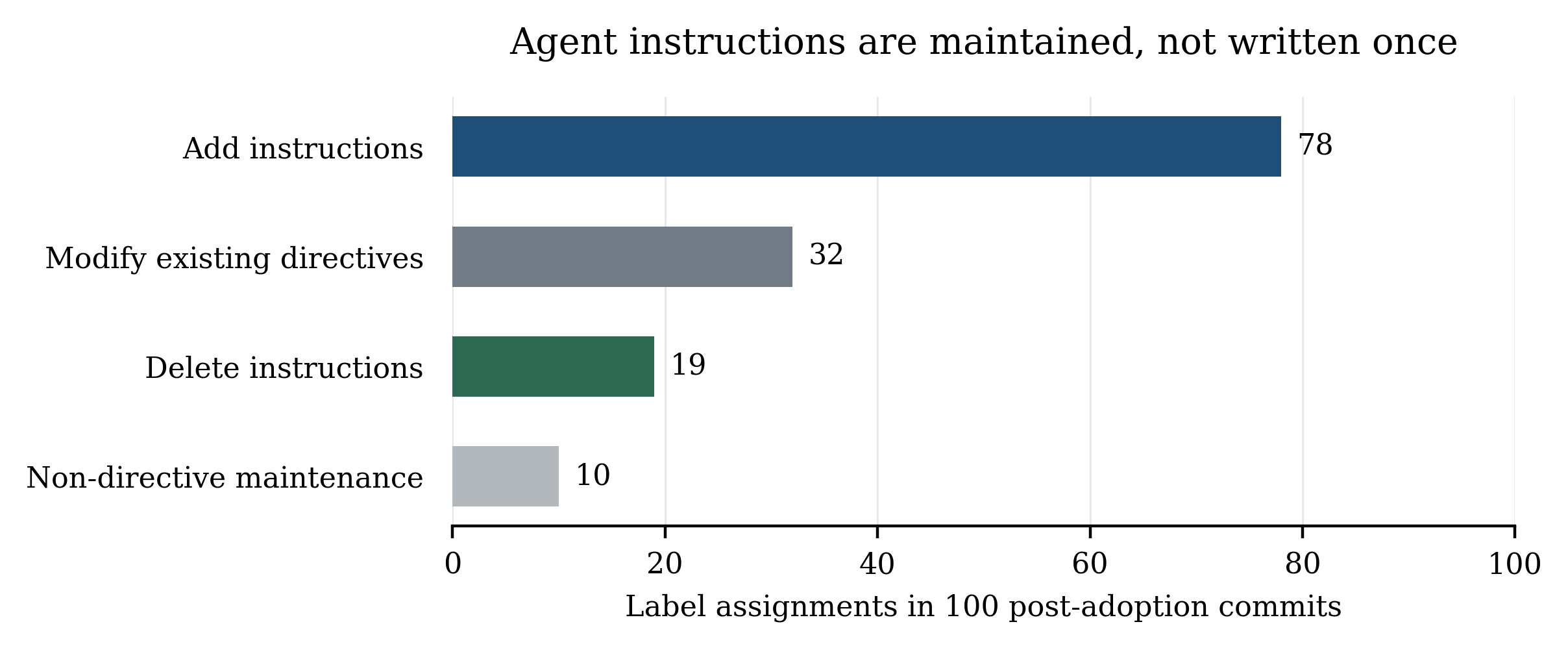}
\caption{Labels in 100 sampled post-adoption commits, from
Chatlatanagulchai et~al., Sections~4.2.2--4.2.3. Labels overlap and
count edit types, not net token growth or effects on agent success.}\label{fig:instructionmaintenance}

\end{figure}

In their two benchmark comparisons, Gloaguen et~al.\ report 20\% and 23\%
higher inference cost with LLM-written context files and success lower by
0.5 and 2 percentage points, respectively; the success differences from
the no-file baselines were not statistically significant. Developer-written files outperform
LLM-written ones on CTXbench ($p=0.038$)
~\cite[Sec.~4.2, Tables~2--3]{gloaguen2026agentsmd}.
SkillOpt uses a more selective update rule: a proposed skill edit survives
only if it improves the held-out validation score, while rejected edits
remain feedback for later proposals~\cite[How SkillOpt works]{microsoft2026skillopt};
its target model's weights stay fixed.
Jiang et~al.\ distinguish skill retrieval from downstream execution
~\cite{jiang2026skills}; loading the intended file does not by itself
show that the agent applied it successfully.

\paragraph{Tests and review.}\label{sec:evaluationwork}
In our study, a view could refuse \texttt{open\_case} in the workflow panel while permitting the later \texttt{approve\_handoff} write, illustrating how the tests covered different parts of the requested read-only rule.
RefactorBench tests refactoring tasks ~\cite{gautam2025refactorbench} and SWE-Sharp-Bench supplies reproducible repository evaluation~\cite{mhatre2025swesharp}; CRBugBench examines the
limits of automated review~\cite{yang2026crbugbench}; observations of
developer-agent collaboration describe practice without isolating its
causal effects~\cite{kumar2025agents}. In Building to the Test, seven of
twelve runs using test-oracle feedback left a required library unused or absent
~\cite{ma2026building}, and ImpossibleBench examines test exploitation more
broadly~\cite{zhong2026impossiblebench}. Protecting tests against edits is
therefore separate from checking that they cover the intended behaviour.
AgentRx provides a precedent for traceable failure diagnosis
~\cite{microsoft2026agentrx}: it derives guarded constraints from tool
schemas and domain policies, checks them at applicable steps and supplies
the resulting violations to a diagnostic judge. For example, a tool
response required to be valid JSON can be checked against that requirement
before the judge attributes the failure.

\paragraph{Bounded correction.}\label{sec:factory}
Pavlenko et~al.'s factory optimises human-architected systems against
measurable objectives; productivity estimates cover application work,
excluding infrastructure and onboarding
~\cite{pavlenko2026factory}, but that evaluation does not isolate the present
between-feature intervention and later change probe.

\paragraph{Reflection and self-improvement.}
Ng's reflection design-pattern letter explicitly recommends Self-Refine,
Reflexion and CRITIC~\cite{ng2024reflection}, attributing methods developed
by their respective authors. Self-Refine iterates feedback and revision with the same model
without training~\cite[Sec.~2]{madaan2023selfrefine}; Reflexion retains
written feedback in memory for subsequent trials rather than updating
model weights~\cite[Sec.~3]{shinn2023reflexion}. CRITIC uses
external tools to check and revise an output
~\cite[Sec.~3]{gou2024critic}; these methods revise work using feedback, but their
repeated-trial budgets and feedback differ from this experiment: hidden
evaluator outputs never reach the development agents. Whether a retained
revision helps with a different, later feature is an empirical question;
here that request added regional access and cross-region handoff
approval to an already developed read-only view.

Ng coauthored MedAgentBench, which evaluates task completion in an
interactive medical-record environment rather than chatbot answers alone
~\cite[arXiv v2, Secs.~2.3--2.4]{jiang2025medagentbench};
its evaluation of observable action outcomes informs the measurement focus
of this protocol.

\paragraph{Engineering rationale.}\label{sec:practitioners}\label{sec:theory}
Beck's alternation of features and tidying aims to restore future
options~\cite{beck2023tidy,beck2025devslow,beck2026prodacity},
and we extend that metaphor to acceptance criteria as an interpretation.
Beck distinguishes producing code from learning
through working software~\cite{beck2026prodacity}.
For the case, the proposed checkpoint is concrete: let a staff member
sign in and complete an assigned-case task before the coordinator advances
the next delivery increment. Tests, short releases and explicit design
responsibilities motivate that checklist
~\cite{beck2002tdd,beck2004xp,
naur1985theory,brooks1987silver,royce1970,gall1975,cockburn2004crystal,
hunt1999pragmatic,agile2001manifesto}.
Monitoring burden and completion proxies qualify the rationale
~\cite{bainbridge1983ironies,simkute2024ironies,tankelevitch2024metacognitive,
lee2025critical,goodhart1975,strathern1997improving,cohen2006review}.
B\"ockeler's Kiro trial illustrates the burden of applying a fixed
specification workflow to a small change: one bug fix became four user
stories with sixteen acceptance criteria
~\cite[One workflow to fit all sizes?]{boeckeler2025sdd}.
Her account is a practitioner trial, not an estimate of wasted effort.
Spec-driven handoffs still need checks of the resulting behaviour
~\cite{delimarsky2025speckit,agarwal2026sdd,waseem2025vibe}.

\paragraph{HVE Core.}\label{sec:hve}
Microsoft's Hypervelocity Engineering Core supplies file-based
Research--Plan--Implement--Review handovers, to which my coordinator added autonomous
dispatch. The reported case snapshot is
\texttt{plugins-v3.3.106}; its installation date remains unverified
~\cite{hvecore}. Additional source qualifications and reproducibility detail
are in the supplement.

\section{Case and method}\label{sec:method}

\subsection{Setting}

For this architectural training exercise, I acted as architect, developer and product owner on a supplied housing-association retrofit use case; the application used synthetic cases to connect documentary evidence
and approvals with staff and resident actions. A staff member
needed to sign in, open an assigned case and record a resident decision
without replacing the staff decision to release work. An assistant could prepare a draft, but nothing was sent automatically; the exercise
required research, requirements, architecture, a delivery plan and a working
version of the product. No operational
service or real resident data is evaluated here.

Agents performed most research synthesis, drafting and coding while I steered scope and decisions, through several weeks of discovery and requirements work followed by roughly a week of interactive MVP development and a ten-day autonomous delivery window. The project continued afterward, with a versioned repository chronology~\cite{servera2026ama} and an automated scorer reporting implementation completeness
(Section~\ref{sec:audittrail}).

\subsection{Workflow and tools}

GitHub Copilot and HVE Core first helped me learn what to build.
The agents worked across the software development life cycle (SDLC),
including BRDs, PRDs and architecture decisions, application code,
infrastructure and deployment scripts, and continuous integration and
delivery (CI/CD) workflows.
Research agents assembled source snapshots and maintained syntheses
in a knowledge base with an index, owner-curated memory and an append-only
log; the Supplement's case evidence index gives their locators.
The Design Thinking coach and stakeholder-role agents then challenged
assumptions by asking what would happen under different conditions, examining
users' responsibilities and walking through their tasks.
For example, a photo's GPS metadata could support the location of a survey,
but did not establish whether the evidence was sufficient for a retrofit
assessment. The proposed assessor would check the evidence bundle and retain professional approval.
Role-agent answers stayed hypotheses for the
exercise; direct stakeholder research was still needed.

As understanding changed, BRD and PRD builders, quality review and the
architecture decision record (ADR)
assistant helped turn conversations and owner decisions into revised
requirements and architecture records. Disposable pilots and prototypes made candidate workflows tangible, and critiques together with my product-use feedback informed a deployed MVP that I used intermittently and refined
through short feedback loops. Its landing view was
\textquotedblleft{}My cases today\textquotedblright{};
case cards showed status and the next action, and a call log and manual
paths let work continue without a model response. These features are
recorded in the historical log and prototype review identified in the
case evidence index; my intermittent MVP use is a participant account.

The handoff to autonomous delivery changed how that understanding reached implementation: a backlog agent converted the accumulated context into
epics, features and user stories, including acceptance criteria, dependencies
and indications of existing MVP coverage. Some stories exposed scope that still needed requirements reconciliation.
A custom coordinator dispatched
implementation and review against this evolving backlog
(Section~\ref{sec:coordinator}).

I expected this preparation to shorten planning because each feature's
purpose and constraints had already been discussed. The documents and
conversations show what workers were asked to build and reviewers were asked to accept, without measuring a speed or quality gain: I did not
compare the work with a human-only team or another planning approach.

\subsection{The coordinator}\label{sec:coordinator}

The coordinator ran through Copilot CLI on a VM, selecting dependency-ready
issues and launching parallel worker sessions in isolated git worktrees,
so workers could change their own copies without editing the same files.
Each worker received an issue and HVE Core's research, plan, implement and review (RPI) workflow, and the coordinator checked the resulting pull request,
updated the project board and, under owner delegation, merged changes.
Interactive and CLI model configurations changed during the project;
recorded selectors and recalled interface labels are distinguished in the
source package. The delivery workflow kept changing, and the recovered history records my steering decisions and prescribed model
selectors; the Supplement distinguishes those instructions from verified
historical execution.

Early worker sessions implemented directly, prompting me to make
the full RPI sequence and committed tracking artefacts mandatory; further
instructions required evidence-backed acceptance ticks, scheduled monitoring
and increasingly extensive local gates. I instructed workers to use the new
source layout and migrate the prerequisites first
(the supplementary coordinator audit, turn 66).
Human decisions, approvals, credentials and unresolved review findings could still block work.
An accepted architecture decision prioritised migration to a new source layout and froze feature work in the MVP\@. Coordination changed alongside migration and procedural requirements,
and the recovered history also contains repeated scheduled prompts, so a
message stored in the user field cannot automatically count as a human
intervention. Supplement Section~\ref{app:deferredmeasures} records the
selection rule and explains why the human-intervention and
intervention-demand rates (HIR and IDR) were not estimated.

\subsection{Data and analysis}

This exploratory single-case study follows the transition from interactive
discovery and MVP use to delivery through issues dispatched by a coordinator. I selected the case after encountering the unusable replacement and acted as participant-observer, using case-study guidance and the ACM SIGSOFT Empirical Standards~\cite{runeson2009guidelines,sigsoft2021standards}.

We examined versioned requirements, architecture decisions, coordinator
instructions, issues, pull requests and selected research/plan/review
records, using distinct snapshots for the original case and later follow-up.
My supplied issue-state export and recollections help reconstruct the
handoff, while committed sources show what was recorded at the time.

We traced descriptive measures and selected episodes with AI assistance,
using a focused narrative literature synthesis and purposively selected
external repositories for context rather than rerunning their applications.
Section~\ref{sec:threats} explains the selection and corroboration limits;
the Supplement preserves extraction rules and source-specific qualifications.

\subsection{What the study measures}\label{sec:constructs}

Here an \emph{agent} combines a model, tools and instructions in a loop that
selects and executes actions. In the case, I kept scope decisions,
approval and delegated merge authority (Section~\ref{sec:coordinator});
the human authors retain responsibility for this study.
A \emph{skill} is a reusable set of procedures describing when and how to
act and check the outcome, distinct from a session history and without updating model weights, a distinction that follows
Jiang et~al.~\cite[Sec.~1]{jiang2026skills}. The pilot uses a handover
note as context; it does not vary skill use as its own treatment.

\emph{Working behaviour} means completing an intended task through the
running system, including the identity, data and service dependencies needed
for that task. In this case, rejecting an unauthenticated request did not
show that an authorised staff member could sign in and open a case.
\emph{Decay} denotes loss of previously working behaviour or
increasing difficulty of understanding and changing a system. The case
records lost usability during replacement; any decline in intrinsic code
quality was not measured. Deployment drift is another explanation (Section~\ref{sec:threats}).
\emph{Adaptability} was defined in the prospective design as success and cost
on the unforeseen change request under a fixed budget, measuring one aspect of maintainability.
Here, \emph{consolidation} means scheduled structural work intended
to preserve observable behaviour, with tests checking that intention.
The later CR1 view-write failures exposed behaviour that the earlier
workflow panel did not check (Section~\ref{sec:resultsjudging}).

We distinguish five measurement categories: PRs, lines and words describe
\emph{production volume}; checks, reviews and receipts describe
\emph{validation activity}; observed user tasks describe \emph{system
behaviour}; code metrics and the change request probe \emph{internal quality
and changeability}; and corrective help requested or supplied describes
\emph{intervention}. The original observations cover the first three categories, and a later post-hoc comparison adds code metrics as indirect measures of structure,
but no controlled changeability task or systematically coded intervention
dataset.

\subsection{Evidence and my role in the analysis}\label{sec:audittrail}

Supplement Table~\ref{tab:audittrail} distinguishes observations, interpretations and proposed checks concerning how discovery was carried
into implementation, how completion was judged and how product feedback
returned. Repository sources refer to the original case pin unless a later
record or participant account is identified, and the supplementary case evidence index provides detailed locators and unresolved execution links.

I selected the episodes and designed the coordinator, so the case reflects
my experience as well as repository evidence; the independent checks still
needed are stated in Section~\ref{sec:threats}. The Supplement preserves
source-specific missing links rather than treating an unrecorded execution
as a successful one.

\FloatBarrier
\section{What happened during product development}\label{sec:what}

After using the MVP, I could no longer complete sign-in and the service
workflow in its replacement. The sequence in Table~\ref{tab:timeline}
starts with discovery and the backlog handoff, then follows the interruption
and my renewed feedback, and the source package retains the dated chronology
and changing totals.

\begin{table}[!htbp]
\centering\small
\caption{Phases of the motivating case and the state carried forward.}\label{tab:timeline}

\begin{tabularx}{\textwidth}{@{}L{2.8cm}YY@{}}
\toprule
Phase & Activity & State inherited by later work \\
\midrule
Discovery, several weeks & Research, role-agent and Design Thinking challenge,
what-if scenarios, persona and journey analysis; iterative BRD, PRD and ADR
work. & A rich but evolving requirements and decision context. \\
Interactive probes, about a week & Disposable prototypes, my critique
and short implementation/feedback cycles produce a used MVP\@. &
Working journeys, simplicity lessons and deployment prerequisites. \\
Backlog handoff & A backlog agent derives epics and user stories, criteria,
dependencies and existing-coverage labels. & A formal work hierarchy that
selects implementation and review tasks. \\
Autonomous delivery, about ten days & Parallel RPI workers deliver issues
under a coordinator while migration and gate requirements evolve. & Merged
changes and validation records; the replacement lacks a complete usable journey. \\
Renewed steering and follow-up & I prioritise usable increments,
functional checks, refinement and refactoring; stabilisation is formalised. &
Product-use feedback again shapes what subsequent work must establish. \\
\bottomrule
\end{tabularx}
\end{table}

\FloatBarrier
\subsection{Building the epics and user stories}

Preparation took sustained steering rather than a single specification prompt: I first asked whether better inputs would change the proposed
agents' responsibilities, then asked for distinct scenarios rather than
one photo/GPS example. The documented study considers structured invoices,
contractor-capacity feeds and sensor-validity metadata as well.
The desk-research journey atlas subsequently informed BRD and PRD revisions
and ADR~0003, leaving quality and approval questions open.
The case evidence index records these exchanges and the requirements
handoff under its author-review context, including PR~\#71;
none was direct stakeholder validation.

Disposable prototypes then made some assumptions tangible, and my critique
and an agent-assisted simplicity review produced requirements for
clear ownership, a visible next action, understandable blockers and manual
paths that remained usable without AI\@. The archived prototype review
also records a re-acceptance defect after handover, outside the existing
tests: a material change could leave an earlier approval valid. The resulting proposals required invalidation and fresh acceptance as short feedback cycles reshaped an
MVP around those lessons. The three disposable prototype PRs stayed unmerged as evidence, with the simplicity review proposing SA-001--SA-010.
I used the resulting MVP intermittently, including sign-in, synthetic
data and model-backed assistant paths.

The backlog manager drew on this accumulated context to generate the work hierarchy, acceptance criteria and dependencies, with existing-coverage labels distinguishing proposed work from functionality already present in the MVP\@.
The supplied project export records a batch expansion around this handoff followed by movement into completed states, tracing work organisation; accepted user journeys require their own evidence.
Requirements reconciliation continued afterward, including for scope first surfaced in backlog stories, so preparation and delivery overlapped and the issue hierarchy continued to evolve as an interpretation of the design.

\subsection{Local completion and the usable workflow}

Autonomous delivery changed the feedback path as the coordinator selected issues and workers satisfied local implementation and review obligations.
At the same time, a source-layout migration replaced the used MVP\@. The
replacement passed bounded engineering checks, including health, restart
and unauthenticated-request rejection, but lacked identity, staff and model
prerequisites needed for my complete workflow. Restoring a
public entry point left sign-in dependencies unresolved.

The earlier setup log distinguishes an application-role assignment from
the internal staff binding: a user could pass tenant and role validation
and still receive \textquotedblleft{}no staff record with this oid\textquotedblright{}.
That recorded
dependency helps explain why opening a browser entry point alone could
not restore the staff workflow.

I continued supplying approvals and resolving technical blockers, but what diminished was the ability to exercise realistic work in the
replacement and use that experience to steer the next change. Missing
prerequisites became further issues, each with research, planning, review
and correction work. Scope additions and validation costs also changed
delivery conditions, and the supplementary case evidence index documents
my account of the interruption alongside the unresolved execution links.
The automated score and failed journey still need release-and-scope linkage
before their disagreement can be assessed as a matched comparison.

\subsection{Comparing episodes within the case}\label{sec:contrasts}

Supplement Table~\ref{tab:contrasts} compares documented episodes and negative cases.
Migration, recorded model choices, workload and gate rules changed alongside coordination, so the contrasts motivate explanations for later testing.

\subsection{Renewed feedback and follow-up}\label{sec:currentstatus}

I asked the coordinator to exercise the accumulated functionality and
refactor, then still had to request ``provide a running website as soon
as possible'' (coordinator turn 1147). I needed a site where I could supply human acceptance and product feedback.
That request was the turning point. Once I could use the site again, my feedback started steering the work, and the history from then on shows refinement and regular stabilisation checkpoints. Those reviews still covered local engineering checks; live acceptance was another step.

My interaction with the project continued to change what needed building, although measuring time to restored usability still needs a dated task execution linked to a release; later repository snapshots cannot fill that gap in the original observation window. The ancillary records retain issue
totals, document volume, lexical counts and static source metrics for
reconstruction; the separate experiment tests retention and changeability.

\FloatBarrier
\section{Five possible explanations}\label{sec:mechanisms}

Why did extensive preparation and passing local checks leave me with an
application I could not use? I identified five possible explanations, called
candidate mechanisms here because the case does not identify their causes.
Table~\ref{tab:mechanisms} summarises the indicators, and the
\hyperref[app:mechanismaudit]{supplementary operational audits}
give definitions, counter-evidence, rivals and discriminating checks.
Each mechanism concerns one part of the delivery loop: how it selected work,
how it passed lessons to later sessions, or how it checked outcomes.

\begin{table}[!htbp]
\centering\footnotesize
\caption{Five candidate mechanisms, case indicators, relevant prior work
and proposed early-warning signals. Indicators are not causal tests.}\label{tab:mechanisms}
\begin{tabularx}{\textwidth}{@{}L{2.3cm}YL{3.1cm}L{3.2cm}@{}}
\toprule
Candidate & Case indicators & Related prior work & Early-warning signal \\
\midrule
Specification inflation & Expanding requirements and acceptance obligations
during discovery; the journey atlas adds requirements and ADR work &
Royce; Gall; Brooks; B\"ockeler; Beck & Specification growth versus
human-validated behaviour \\
Proxy capture & Artefacts became a merge condition; completion assessments
coexisted with blocked staff sign-in, with release/scope linkage still pending &
Goodhart; Strathern; Beck's genie; ImpossibleBench; METR &
Artefacts per behaviour change; scores that disagree with a live demo \\
Learning evaporation & Recorded simplicity and deployment lessons were
followed by missing staff bindings; actual session access is unknown & Naur; context studies;
multi-agent failure taxonomy & Lessons that exist only as prose; problems
solved twice \\
Process-gate inflation & Broad per-push gates, constrained validation
resources and PR \#445's correction exception & Bainbridge;
ironies of generative AI\@; Murphy; Goldratt & Gate time per change; review
minutes per merge \\
Feedback starvation & A used MVP was frozen; I could not exercise the
replacement and requested a running website & XP\@; incremental delivery;
tracer bullets; Agile Manifesto; DORA & Time to first human use; age of last
usable increment \\
\bottomrule
\end{tabularx}
\end{table}

The proposed mechanisms concern different parts of delivery: specifications added approval and handoff requirements through the journey atlas.
Later sessions also needed the staff-binding lesson from the working MVP\@.
The coordinator could advance a PR after its engineering checks without
my completing sign-in, and the move to local gates responded to Actions
budget pressure. These are different candidate sources of delay or failure;
the record does not estimate their individual effects, and
Table~\ref{tab:theory} in the Supplement relates this account to
accumulation, transmission and correction.

Before I requested a checkpoint, the loop had no scheduled cumulative
consolidation step. The one large restructuring, the source-layout migration,
was planned as backlog items, although local refactoring could still occur
within individual issues. The experiment measures previously passing
contracts that later fail, checks whether consolidation changed code
structure, and tests adaptation to a new request. Establishing whether an
earlier fault causes a distinct later failure would require tracing
dependencies between faults, and that analysis is deferred.

We developed the five categories in Table~\ref{tab:mechanisms} after the case events; the Supplement defines each one and lists checks that would distinguish its effects from competing explanations.

\subsection{A proposed reinforcing relationship}

\begin{figure}[t]
\centering
\includegraphics[width=0.92\linewidth]{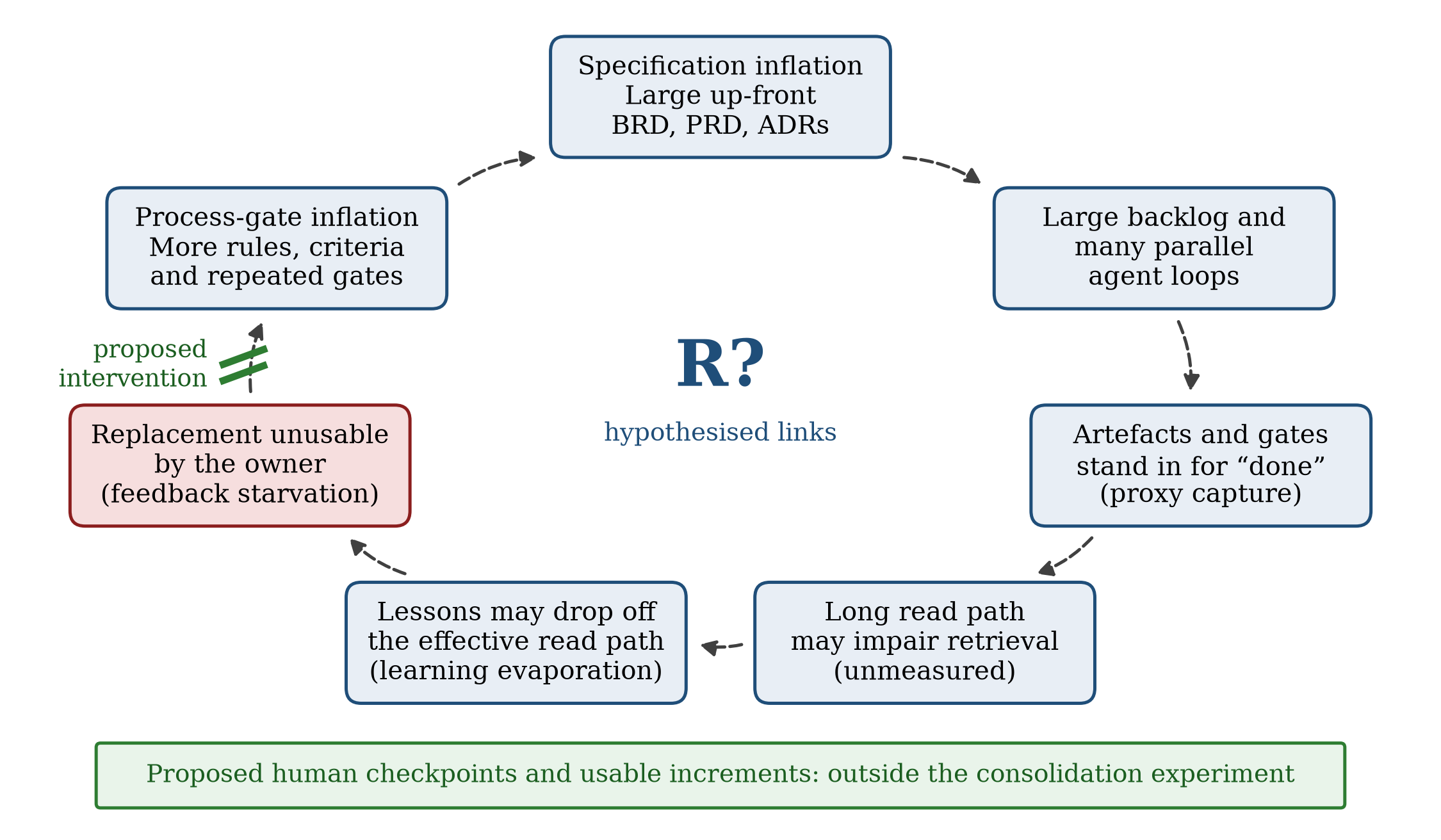}
\caption{Proposed reinforcing relationships drawn from the case interpretation.
Their causal effects were not measured.
Context overload is unmeasured. The proposed human-checkpoint intervention
differs from the consolidation package tested in Section~\ref{sec:experiment}.}\label{fig:loop}
\end{figure}

The case suggests a reinforcing loop
(Figure~\ref{fig:loop}), in which an extensive specification led into a large backlog and
parallel sessions. Committed RPI evidence became a merge condition (Section~\ref{sec:coordinator}).
The replacement lacked the previously working staff bindings, and I
could not use it to expose workflow gaps. I responded with further rules, criteria and gates, a sequence consistent with reinforcement without identifying causal links. A running increment and a human-use checkpoint
in each cycle are proposed interventions; they were not tested here.

\FloatBarrier

\section{Consolidation hypothesis and procedure}\label{sec:harness}

Restoring the case's missing identity and staff bindings required
configuration and end-to-end verification. Consolidation addresses a
different concern from the same experience: even with a correct
deployment, repeated feature work may leave a repository that is harder to
extend. The experiment narrowed the question to that inherited state by testing preserved behaviour and a withheld change request, and it
did not treat code cleanup as a remedy for the case's configuration failure.

Practitioners already structure multi-session agent handoffs around clean
repository states: Anthropic prompts each coding session to leave merge-ready
code and update a progress file~\cite{anthropic2025harness}, an account of
practice without a comparison of feature-only delivery against
consolidation between feature steps.
Darwin's continuous optimisation is a further precedent for automated
improvement, but its performance evaluations do not isolate that intervention
or test an unforeseen change request~\cite{pavlenko2026factory}
(Section~\ref{sec:factory}). RefactorBench already evaluates refactoring
agents~\cite{gautam2025refactorbench}; the narrower hypothesis concerns preserved
behaviour and later adaptation during repeated feature additions.

The consolidation procedure tests one bounded form of reflective
correction, and it changes what persists in the repository. Self-Refine revises
an output and Reflexion retains lessons for later trials
~\cite{madaan2023selfrefine,shinn2023reflexion}; the present intervention
consolidates the software and its handover between different
feature requests. CRITIC's tool feedback is another precedent
~\cite{gou2024critic}. The held-out evaluation remained separate from that
feedback, and the study tests whether the accepted changes help with an
unforeseen request. The schedule sets each pause, and the agent cannot reject
features or redirect the product; after each feature, delivery pauses for
a bounded consolidation session.
The agent runs the tests and examines structural diagnostics, then reviews
the latest diff for duplication, complexity, unclear names, dead code,
misplaced responsibilities and tests coupled to implementation details.
It attempts small refactorings through
\texttt{test \&\& commit || revert}, so a change survives
only while the tests remain green. It may add tests of current behaviour,
but the defining example tests stay protected.

To prepare the next session, the agent replaces a design note of at most 300 words instead of appending to it, recording where rules reside and what debt is outstanding.
Once the quality targets are reached or the time budget expires, the pause
ends and the next feature begins. Supplement Section~\ref{app:activeprompts}
gives the inspected B/C templates and source bindings alongside the historical
prompt, keeping their different scopes explicit. The intervention combines
refactoring, additional tests and a design-oriented note, so the study
cannot isolate refactoring alone.

The hypothesis draws on Beck's alternation of features and
futures~\cite{beck2026prodacity,beck2025devslow} and \emph{Tidy
First?}~\cite{beck2023tidy}. The procedure costs time, and its benefit remains
an empirical question. Agent
refactorings can be incomplete or incorrect~\cite{gautam2025refactorbench},
and human studies caution against equating a refactoring label with a
quality gain. Kim et~al.\ report developers' refactoring challenges and
benefits~\cite{kim2014refactoring}; Bavota et~al.\ find that observed
refactorings do not consistently target or remove code
smells~\cite{bavota2015quality}. These findings argue for checking outcomes instead of assuming improvement, because test-commit-or-revert
protects behaviour only to the extent that the tests cover it
~\cite{ma2026building}. Section~\ref{sec:experiment} specifies how
the procedure was evaluated and reports the results.

\FloatBarrier
\section{Experiment: design and results}\label{sec:experiment}

To address RQ3, we compared scheduled correction with feature-only delivery
and longer sessions on the same task. We completed the five-arm exploratory study and post-collection evaluation in November 2026, following planning informed by
agent-configuration protocols and empirical-study guidance
~\cite{lulla2026protocol,baltes2026guidelines,wohlin2012experimentation}.

\subsection{Execution status and disclosed amendments}\label{sec:studyexecution}

We completed the five-arm exploratory study under an OSF-registered base
and disclosed unregistered amendments. Data availability identifies the public
repositories and OSF registrations; the Supplement preserves the amendment chronology,
sixteen execution versions and recovery decisions without treating public
version control as retrospective registration.

Each of three models, GPT~6.1 Sol, Claude Sonnet~5.5 and Gemini~4 Argon,
ran all five arms at medium effort, with 21 trajectories per model and arm:
$3\times5\times21=315$. The scheduler used seed 2027; we divided execution and collection from design and analysis as stated in Author contributions. Section~\ref{sec:threats} explains the resulting
limits on inference.

\paragraph{Deviations from the registration.}
The OSF update dated 6 November 2026 and the registered base of
7 November 2026 specified before collection five arms; allocation options
of 12, 21 or 33 trajectories per model and arm; primary outcome final
$P_{10}$ coverage; whole-run bootstrap intervals with 10,000 draws and
seed 2027; and, for CR1, 97.5\% Clopper--Pearson component intervals
with opposite-endpoint bounds. No significance testing or $p$-value
thresholds were registered.
Some registration fields, including sample size, inference and inclusion,
retain the earlier eighteen-trajectory, three-arm wording; the update's
stated reason and sampling section describe the executed design.
B--D is explicitly registered as the primary coverage contrast.

Disclosed in Git, not registered, were the selection of $n=21$ before
collection, model-stratified scheduling, parallel execution, short calibration
and recovery handling.
Newcombe intervals, Fisher $p$ values in
Supplement Table~\ref{tab:resultscrintervals} and the verdicts in
Table~\ref{tab:predictionverdicts} are post hoc.

\subsection{Task and comparison groups}\label{sec:protocolsummary}

Each agent starts with a fresh repository, develops it through ten ordered features and then receives the withheld change request, CR1;
that complete development run is the experimental unit, called a
\emph{trajectory}. A \emph{cell} contains runs using the same model and
comparison group, or arm, and the analysis treats trajectories as independent
within cells; shared execution conditions limit that assumption
(Section~\ref{sec:protocollimits}).

The task, S1, is a standard-library Python API package called \texttt{workbench}, covering synthetic case lifecycle, handoffs, consent, permissions, ranking and audit, with consent withdrawal pausing installation unless superseded by a safety hold. Contractor views must hide contact and vulnerability fields and refuse every write, requirements that CR1 extends with postcode-prefix regions, region-scoped users and manager-approved cross-region handoffs.
The notation $F_t$ denotes the state after feature $t$, and in B, C and E,
a consolidation, review or repair session follows; $P_t$ denotes the
state after that follow-up, called a \emph{post-step}. These states
coincide in feature-only arms A and D, which have no follow-up session, so $P_{10}$ is the final ten-feature repository before the withheld extension.
Every selected model, GPT~6.1 Sol, Sonnet~5.5 and Gemini~4 Argon,
runs every arm; using seed 2027, the scheduler randomises model order within each replicate and arm order within each model. Its parallel dispatcher can admit later
eligible trajectories as workers become free; a five-arm scheduling group
is therefore not a synchronised execution batch. Feature steps, tests, checkpoints and agent sessions provide repeated
observations within a trajectory. Counting them as independent replicates would overstate the sample size.
The selected number $n$ applies
\emph{per model per arm}, equally across all fifteen cells.

\begin{table}[!htbp]
\centering\small
\caption{Executed five-arm comparison. D is the matched-allowance A+
control; E tests visible regression repair, not hidden-test feedback.}\label{tab:protocolsummary}

\begin{tabularx}{\textwidth}{@{}L{1cm}YL{3.6cm}@{}}
\toprule
Arm & Intervention and information & Per-cycle ceilings \\
\midrule
A & Feature delivery only & 30 min; 300 nominal credits \\
B & Feature, then behaviour-preserving consolidation, additional tests
and replacement design note & 30 + 15 min; 300 + 300 credits \\
C & Feature, then read-only review and replacement review note
& 30 + 15 min; 300 + 300 credits \\
D (A+) & Longer feature delivery, without a scheduled consolidation pause
& 45 min; 600 credits \\
E & Feature, then repair of previously specified behaviour using visible
specifications and examples & 30 + 15 min; 300 + 300 credits \\
\bottomrule
\end{tabularx}
\end{table}

B--D compares a scheduled consolidation session with giving the builder the same total time and nominal credits in one session: both permit 45 minutes and 600 nominal credits per feature cycle.
B divides that allowance between feature work and a separate consolidation
session; D keeps one longer feature session, with fewer fresh starts; the ceilings do not specify spending.
B--A adds the consolidation package and its resources, whereas B--C changes editing rights, tests and handover content; E--B compares repair against
visible requirements with consolidation intended to preserve behaviour.
Because these comparisons change several conditions together, they cannot
isolate refactoring or any one of the case mechanisms.
The Supplement preserves historical inspections, freezes and the design diagram
(Figure~\ref{fig:pilotdesign}), which does not depict the executed interventions
in Table~\ref{tab:protocolsummary}.

C cannot edit code during review, but its review note can guide later feature sessions, although attributing a repair to the note would require linking the
specific review item to the later change; a shared file path is insufficient. B's added tests must agree with visible
specifications; a test that accepts a defect does not satisfy this rule. Subsequent feature sessions must keep accepted tests green, although E can correct an agent-authored expectation only with a
specific contradictory specification and a recorded explanation.
Accepted Python test files are immutable in full, including imports, fixtures and module/class constants, with additions confined to new files; E has only its explicit specification-bound correction exception, with immutable example tests protected in every arm.

\subsection{Development and evaluation execution boundaries}\label{sec:runtimeboundaries}

\begin{figure}[!htbp]
\centering
\includegraphics[width=\linewidth,height=0.68\textheight,keepaspectratio]{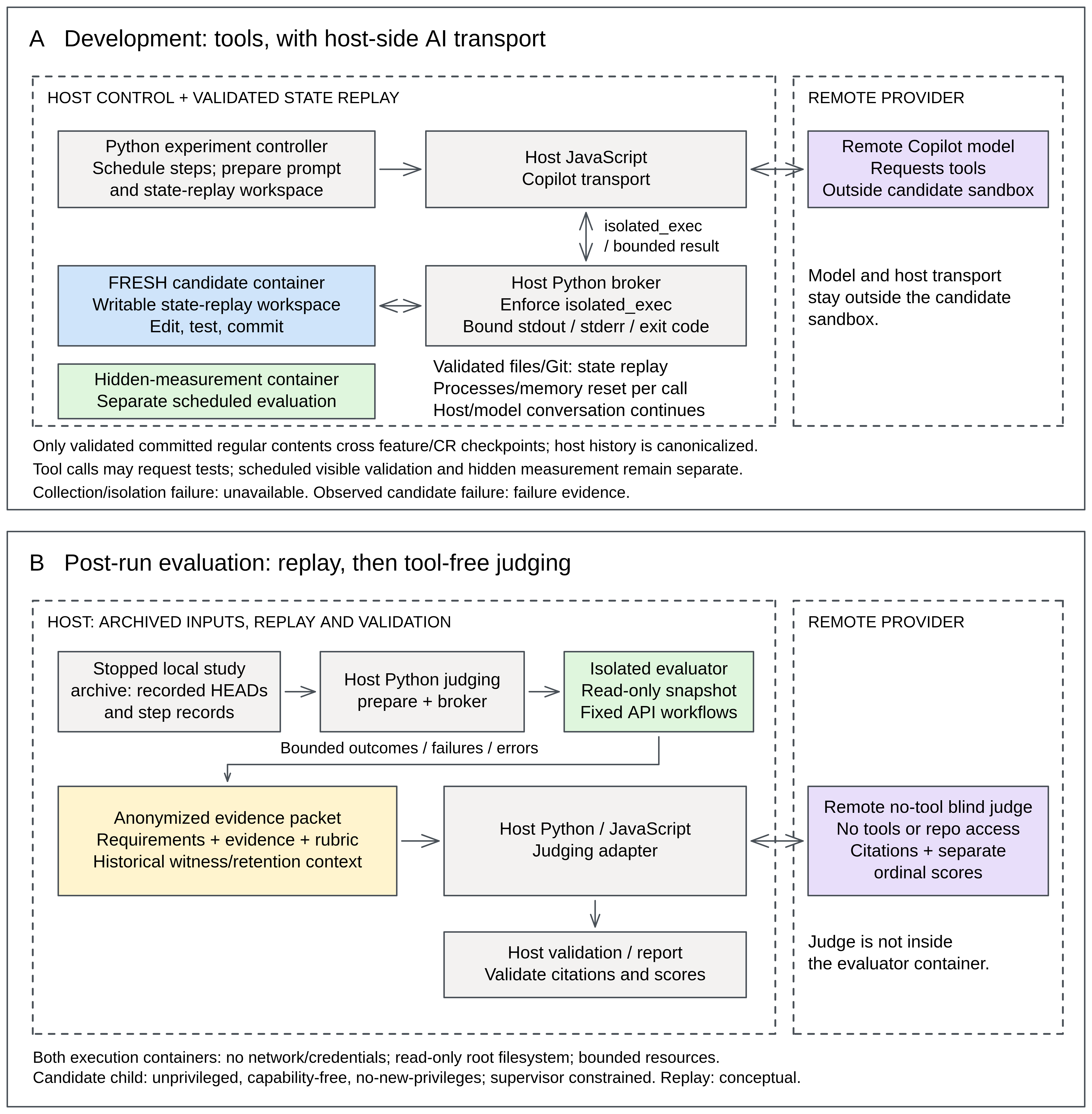}
\caption{Source-described development and post-run evaluation paths.
Development agents request isolated commands against a writable repository;
post-run judging follows read-only workflow replay and receives evidence
without tools. Python orchestration and JavaScript model transport operate
on the trusted host in both paths. Arrows show information/control
flow; the Supplement documents the historical source and implementation
scope.}\label{fig:runtimeboundaries}

\end{figure}

The trusted host runs the controller and authenticated Copilot SDK/CLI
transport; a broker executes each requested command in a fresh
container and returns bounded output and exit status
(Figure~\ref{fig:runtimeboundaries}). Validated filesystem and Git state
carry into the next writable workspace, but candidate processes and process
memory do not; the host/model conversation continues across calls.
Only validated committed regular content crosses feature/CR checkpoints,
and visible-test verification and hidden measurement run in other
evaluator containers after development steps.

For supplementary assessment of a stopped local study archive, host Python selects recorded checkpoint HEADs and asks the broker to replay
fixed API workflows.
With candidate snapshots and evaluator code mounted read-only, the replay collects bounded observations, failures and errors in anonymised
evidence packets, each a record for one trajectory at one endpoint.
Host Python then invokes a tool-free judge session, which returns ordinal
ratings, rationales and evidence citations without command tools or
candidate-repository access, with its model connection and the host orchestration
operate outside the replay container.

Candidate commands used network-disabled containers with bounded resources;
the trusted host held authentication and model transport, and the Supplement
gives the capability and mount configuration, including the distinction
between supervisor and candidate privileges. Isolation failures made a
checkpoint unavailable, whereas observed candidate failures stayed in the
assessment. This was source inspection rather than adversarial security evaluation.

\subsection{Outcomes and retained-contract loss}\label{sec:decaymeasurement}

The primary outcome is the proportion of all 127 feature contracts passed at $P_{10}$, before CR1. Each contract uses a fixed hidden test to check required behaviour, such as refusing a write through a contractor view or reproducing a ranking from the same state.
We call this \emph{behavioural coverage}, distinct from instrumented line or
branch coverage, and distinguish acquisition, retention and repair when
interpreting an increased score.
Retained-contract loss and CR1 adaptation are distinct secondary outcomes,
with legacy identifiers H2 and H6 used for data traceability.
The supplementary appendices define HIR/IDR; comprehension was not measured.

\paragraph{Who measures, and what ``quality'' means.}
During development, models implement features, consolidate code, write
review notes or repair defects as assigned treatments.
The host evaluator runs fixed hidden contracts and code-analysis tools,
then computes numerical summaries without calling an LLM-as-judge.
Supplement Table~\ref{tab:measurementprovenance} distinguishes outcomes, structural proxies and resource observations.
Complexity, duplication and lint describe selected code properties without verifying CR1 read-only enforcement in Gemini's cleaned repositories
(Section~\ref{sec:resultscorrectness}). Overall architecture quality, readability and comprehension require other evidence, and the active metric dictionary contains no maintainability-index score.

The evaluator records errors, skips and timeouts rather than treating a
missing or non-binary endpoint as a pass; Supplement
Section~\ref{app:measurementprovenance} specifies raw fields and tool settings.

\paragraph{How AI review enters the experiment.}
Arm C's model reads the quality report and feature diff, then
replaces its review note in at most 300 words using a checklist of
duplication, complex functions, names, dead code, responsibilities and
behaviour-focused tests. The review can vary between model sessions, and the kit archives the note without converting it into a numerical quality score. Arm B uses a similar checklist to guide changes, with complexity $\leq10$, duplication $<3\%$ and zero configured-lint-warning
targets serving as requested stopping conditions and treatment-fidelity checks.
\texttt{make check} runs tests and generates the metrics report without
asserting that all three structural targets have been met.
The runner enforces the time budget, protected tests and retained green
code, rather than accepting an AI statement as an independent quality verdict.
Supplement Section~\ref{app:activeprompts} gives the checked-in B/C examples, parameter substitution and saved-prompt limitations.

For trajectory $r$, $F_t$ is the accepted state after feature step $t$ and
$P_t$ the state after its assigned post-step; these source states coincide
in A and D. Let $\mathcal{R}_{r,t}$ contain eligible, unchanged contracts
that passed at least one earlier end-of-cycle checkpoint $P_u$, $u<t$.
This \emph{risk set} excludes superseded contracts and includes only behaviour that could now be lost because it worked before and is still required. At $F_t$ and $P_t$, let
$\mathcal{O}^{p}_{r,t}\subseteq\mathcal{R}_{r,t}$ contain contracts with
observed binary results $S^p_{r,t}(q)$, $p\in\{F,P\}$. Observed loss is
\begin{equation}
L^p_{r,t}=\sum_{q\in\mathcal{O}^{p}_{r,t}}
\mathbf{1}\{S^p_{r,t}(q)=0\},\qquad
d^p_{r,t}=\frac{L^p_{r,t}}{|\mathcal{O}^{p}_{r,t}|}.
\label{eq:decayloss}
\end{equation}
Here $L$ counts observed failures among previously acquired contracts, and
$d$ divides that count by the observed risk-set size. The fraction is undefined when its denominator is empty, and missing earlier checkpoints can leave acquisition unknown. The Supplement specifies richer reporting of those histories; missing observations cannot count as acquisition failures.

With $M^p_{r,t}$ missing evaluations, full-risk loss lies between
$L^p_{r,t}$ and $L^p_{r,t}+M^p_{r,t}$; never-acquired contracts are
acquisition failures rather than losses.

The final export marks all 6,300 F/P rows as having complete observation and risk histories, and the analyser withholds loss when either history is
incomplete; the Supplement distinguishes this implemented rule from the
richer reporting requirements proposed for future incomplete histories.

\begin{figure}[!htbp]
\centering
\includegraphics[width=\linewidth]{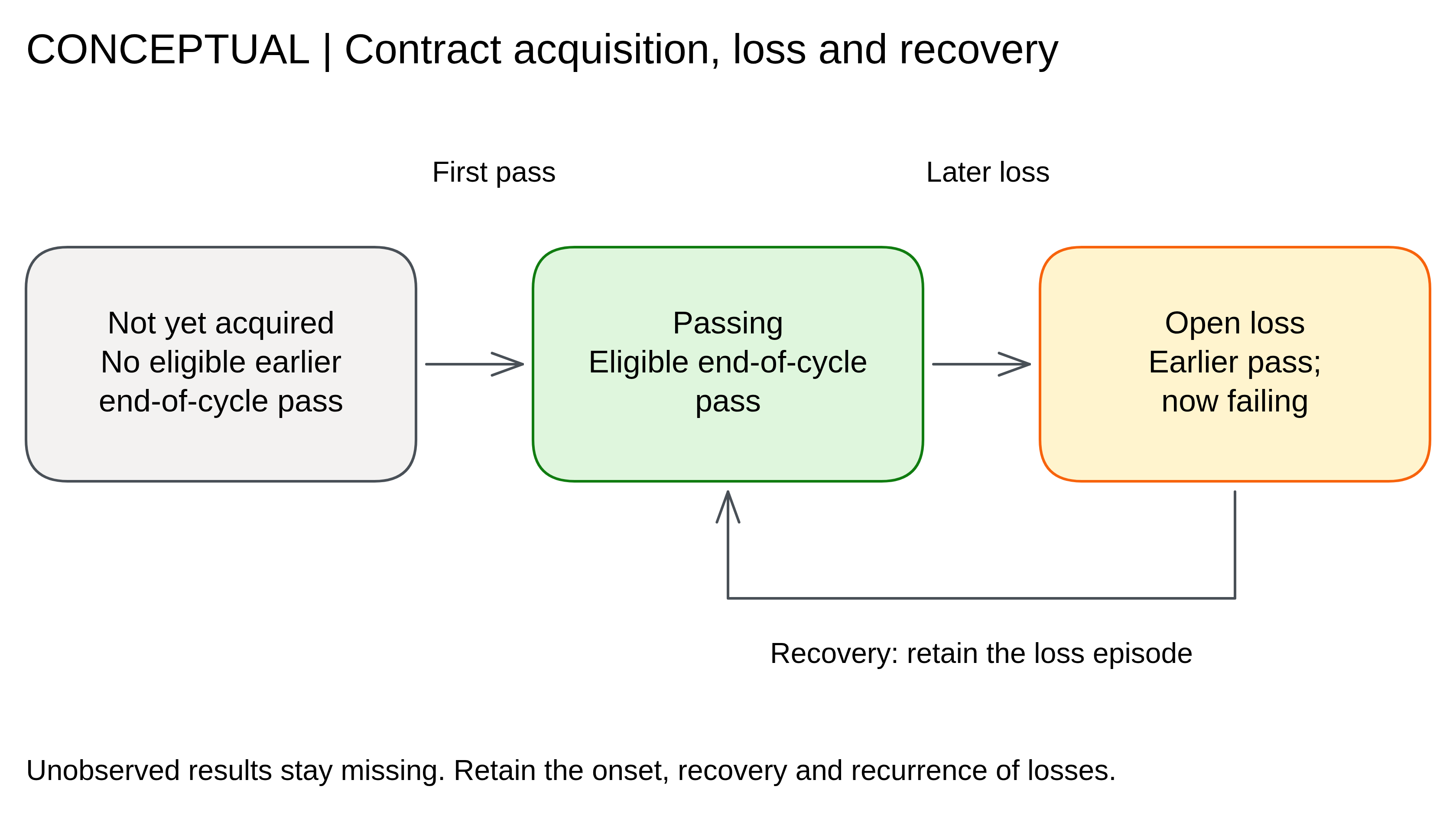}
\caption{Conceptual contract states. Acquisition, loss, recovery and
unobserved evaluation are distinct; the diagram contains no measured
frequencies.}\label{fig:contractstates}
\end{figure}

CR1 success requires all 15 CR1 contracts and preservation of every feature contract passing before CR1, and we cross-tabulate that outcome by prior F9/F10 acquisition in
Supplement Table~\ref{tab:crprerequisites}. These post-treatment groups have
different preservation burdens, so their differences cannot estimate effects
among otherwise equivalent repositories. Graded CR1 and prior-pass loss decomposition await a later study; Supplement
Figure~\ref{fig:experimentmeasures} preserves the earlier measurement plan.

We report delivery and CR1 credits alongside delivered behaviour, distinguishing credits from tokens, elapsed time and currency; the Supplement documents all-attempt accounting and the Lizard, jscpd and Ruff settings. Inline lint suppressions were not independently counted
(Section~\ref{sec:threats}).

\subsection{Supplementary post-run workflow assessment}\label{sec:postrunjudging}

Designed after development, a fixed panel replayed API workflows against archived S1 checkpoints; it left the treatment budgets and original coverage, retained-contract-loss
and CR1 endpoints unchanged. Development agents were neither rerun nor allowed to repair their repositories, and Supplement Section~\ref{app:postrunjudging} records the instrument and
assessment lineage.

The panel exercised ten feature workflows, a delivery/reopening journey and a CR1 workflow, looking in each scenario for an
earlier successful execution at an eligible end-of-cycle checkpoint in the
same development run. We call that passing execution a \emph{witness} and count a failing final scenario as witnessed loss only if it passed earlier. A scenario that never passed counted as an acquisition failure, while post-CR1 loss could include a
failure already present at $P_{10}$, so it did not isolate damage caused
by CR1. The scenarios complemented hidden contracts through longer workflows with a different, narrower set of checks.

A tool-disabled judge model, \texttt{gpt-6-luna}, rated each
evaluable endpoint twice in fresh sessions. It received anonymised
execution evidence and returned four ordinal 0--4 ratings:
functional completeness, workflow coherence, error recovery and API
usability. Ratings required final-observation citations and null values for insufficient evidence, addressing the exercised API tasks rather than constituting a human interface usability study. The packets also exposed
earlier passing witnesses and retention labels, allowing historical
anchoring despite anonymisation: a judge could see that a scenario had
worked earlier before rating its final state. The judge shares a provider/model family
with the development model \texttt{gpt-6.1-sol}.

The assessment supplies supplementary exploratory evidence, distinct from development-time review as an intervention. A
\emph{complete-frame contrast} requires observations for all allocated
trajectories in both compared arms for every development model; each model
contributes equally. Repeated ratings, scenarios
and checkpoints remained nested within trajectories. Results, missingness
and the human spot-check are reported in Section~\ref{sec:resultsjudging};
the Supplement accounts for judging costs.

\subsection{Replication and uncertainty}\label{sec:protocollimits}

We chose the ten-percentage-point minimum useful gain before collection, with conditional sample-size calculations and alternative allocations preserved in the Supplement; the executed selection was 21 per model and arm.

Coverage scores $Y$ use the 0--100 scale, with $\overline{Y}_{ma}$ denoting mean final coverage for model $m$ and arm $a$.
The primary estimand, the quantity the study aims to estimate, is the
difference in mean coverage averaged equally over the $M=3$ selected models,
each contributing one-third of the result regardless of its coverage level.
We estimate this difference from the observed trajectories as
\begin{equation}
\widehat{\Delta}_{BD}=\frac{1}{M}
\sum_{m=1}^{M}(\overline{Y}_{mB}-\overline{Y}_{mD}).
\label{eq:primarycontrast}
\end{equation}
The reported pooled within-cell SD was 3.1278 percentage points with
300 degrees of freedom (Supplement Section~\ref{app:resultsdetail}); this
describes the completed data without supplying a retrospective power claim.

The reported independent-cell analysis uses a bootstrap that samples whole trajectories with replacement within each model/arm cell, with the 2.5th and 97.5th percentiles of 10,000 resulting estimates giving each
95\% interval. Resampling treats trajectories within cells as independent and gives each
selected model equal weight; it does not jointly resample scheduling blocks.
Model-specific and other contrasts are secondary, exploratory and
unadjusted for multiple comparisons. For fully observed CR1 cells we report
success counts and binomial uncertainty; unknown outcomes receive
identification bounds instead. The coverage planning calculation supplies no power guarantee for CR1, and Section~\ref{sec:threats} consolidates the
dependence, selection and generalisation limits.

\subsection{Resource and information boundaries}\label{sec:prerun}

Beyond the session ceilings in Table~\ref{tab:protocolsummary}, the controller enforced aggregate admission against spent credits and
reserved liabilities; unknown charges never became free headroom.
The Supplement records the aggregate envelope, retry rules, historical
launch criteria and limits of the information-access evidence.

Candidate tools received only current and earlier visible specifications and examples, while hidden tests, scores, reference code and the future change request stayed external; discovery of skills and other tools was disabled. The trusted host's authenticated model connection did not expose
the evaluator through the candidate command tool.

Unsuccessful feature increments are reset; later features and scheduled
B/C/E post-steps continue on accepted green state. Hidden denominators
still include undelivered features. One infrastructure-only retry shares
the original budget; poor performance never justifies replay.
Contradictory valid oracle results halt collection.
Assisted and incomplete trajectories and all their costs are recorded,
without replacements or success-conditioned primary analyses.
Consequently, a zero loss count does not exclude broken edits, because rejected or transient failures belong to an attempt outcome; the retention endpoint measures the state allowed to survive.
Immutable accepted tests and rollback are part of every tested policy.

\FloatBarrier
\subsection{Collection completeness, resilience and credit ledger}\label{sec:resultsledger}

All 315 allocated trajectories completed, including the audited recovery;
the Supplement preserves the step, attempt and liability counts.
The study recorded 62,299.76 settled AI credits (US\$623.00), about
197.8 credits per run (about US\$1.98), and 19,200 reserved credits on
55 holds (US\$192.00).
Coverage and regression counts are complete, whereas CR1 sampling
intervals and unknown-outcome bounds answer different questions
(Supplement Section~\ref{app:resultsdetail}).
Billing holds prevent a complete all-allocation cost comparison
(Section~\ref{sec:threats}); the detailed credit ledger is in Supplement
Section~\ref{app:executiondetail}.

Isabel Cabezas Martin supplied the attestation that all 315 trajectories were
unassisted; she executed and collected the study rather than analysing it,
but is a coauthor and was not an independent auditor. The SDK journal audit corroborates the absence of in-band interactive input, without checking out-of-band hints or edits.

\subsection{End-of-delivery coverage, retention and CR1 adaptation}\label{sec:resultscorrectness}

\begin{figure}[!htbp]
\centering
\includegraphics[width=\linewidth]{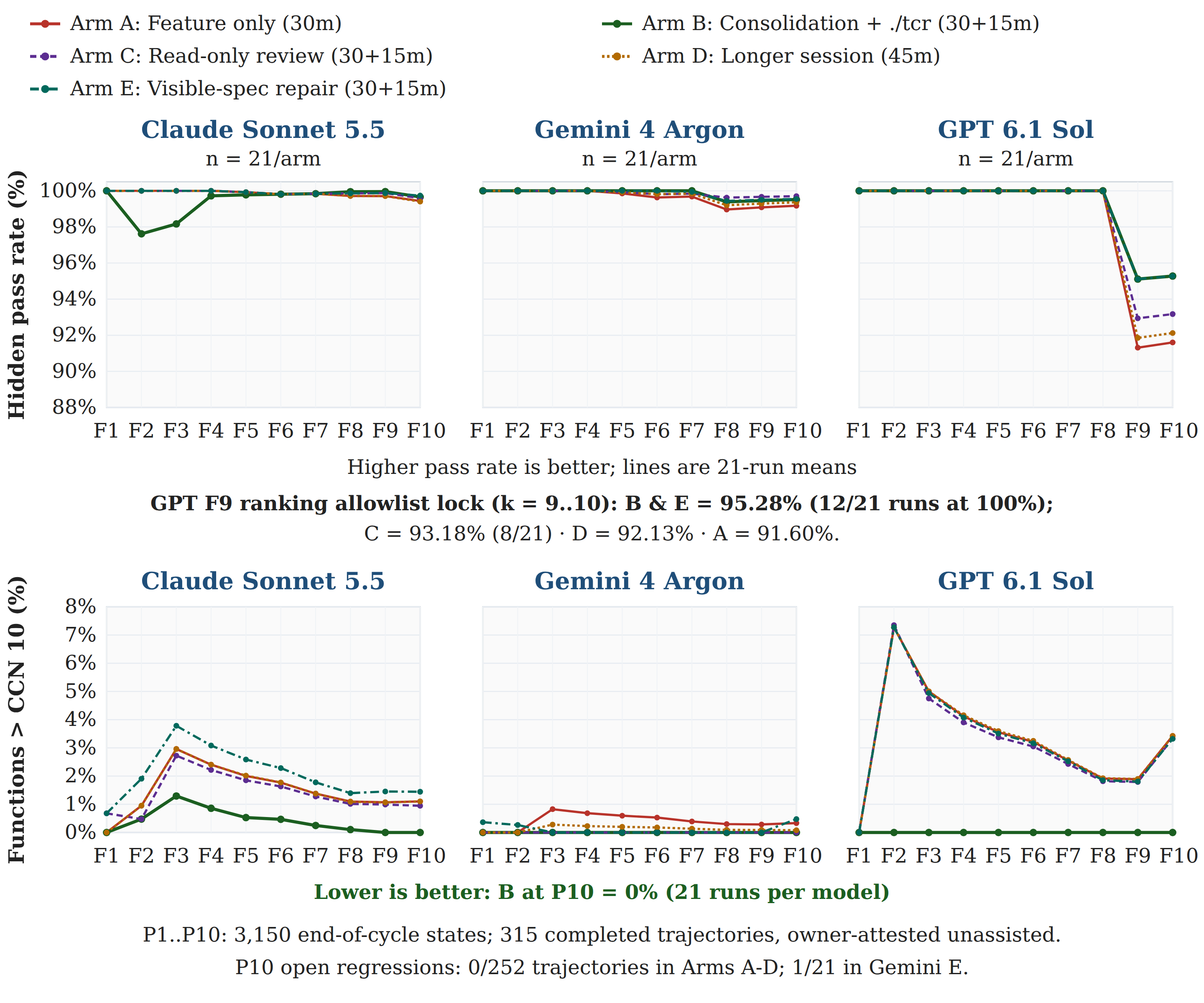}
\caption{Cycle-by-cycle trajectories across all 315/315 completed
runs ($k=1..10$, 3,150 post-step checkpoints $P_k$). Top: mean hidden-test
pass rate (\%). Bottom: mean percentage of functions exceeding
cyclomatic complexity 10. Lines are 21-run cell means without uncertainty bands. The
upper axis is zoomed to 88--100.5\%; contract sets expand across cycles.}\label{fig:studytrajectories}
\end{figure}

The agents largely preserved earlier behaviour but never acquired some required capabilities, distinguishing acquisition of a new feature from retention of one that previously passed.
Table~\ref{tab:resultsbenchmark} and
Figure~\ref{fig:studytrajectories} separate final coverage and
cycle-by-cycle progression from the number of trajectories with a
retained-contract loss. At the end of feature delivery ($P_{10}$),
Sonnet~5.5 and Gemini~4 Argon averaged 99.18\%--99.74\% of the 127 hidden
feature tests across all 21/21 trajectories per cell. The retained-contract
panel recorded one open regression among all 315 completed endpoints, in
Gemini~4 Argon E, with
zero open regressions and zero loss cycles across all 252 trajectories in
Arms~A, B, C and D.

\begin{table}[!htbp]
\centering\small
\caption{Final feature coverage, before CR1, across all 21/21 completed
trajectories per cell (315/315 total). ``All 127'' counts complete
test-suite passes; ``With loss'' counts trajectories with a previously
passing unchanged contract failing at $P_{10}$. Additional distribution and
loss statistics are in the Supplement\@. \armlegend}\label{tab:resultsbenchmark}

\begin{tabular*}{\textwidth}{@{\extracolsep{\fill}}lcrrrr@{}}
\toprule
Model & Arm & Observed & Mean (\%) & All 127 & With loss \\
\midrule
Sonnet 5.5 & A & 21/21 & 99.44 & 12 & 0 \\
Sonnet 5.5 & B & 21/21 & 99.66 & 13 & 0 \\
Sonnet 5.5 & C & 21/21 & 99.59 & 12 & 0 \\
Sonnet 5.5 & D & 21/21 & 99.40 & 11 & 0 \\
Sonnet 5.5 & E & 21/21 & 99.74 & 16 & 0 \\
\midrule
Gemini 4 Argon & A & 21/21 & 99.18 & 6 & 0 \\
Gemini 4 Argon & B & 21/21 & 99.51 & 8 & 0 \\
Gemini 4 Argon & C & 21/21 & 99.70 & 16 & 0 \\
Gemini 4 Argon & D & 21/21 & 99.36 & 7 & 0 \\
Gemini 4 Argon & E & 21/21 & 99.55 & 10 & 1 \\
\midrule
GPT 6.1 Sol & A & 21/21 & 91.60 & 5 & 0 \\
GPT 6.1 Sol & B & 21/21 & 95.28 & 12 & 0 \\
GPT 6.1 Sol & C & 21/21 & 93.18 & 8 & 0 \\
GPT 6.1 Sol & D & 21/21 & 92.13 & 6 & 0 \\
GPT 6.1 Sol & E & 21/21 & 95.28 & 12 & 0 \\
\bottomrule
\end{tabular*}
\end{table}

A falling pass rate alone does not identify loss of acquired behaviour.
GPT~6.1 Sol differed: many trajectories preserved earlier
functionality through the eighth feature (100.00\% mean pass rate in every arm) but
failed to deliver priority ranking, the ninth feature.
The handoff's exploratory inspection traced the common 88.98\% plateau to
all fourteen ranking tests failing: earlier tests had asserted that view
methods outside a fixed list must be rejected; the new ranking method
conflicted with that assertion while accepted test files were immutable. Complete feature delivery occurred in 5/21 baseline trajectories, 6/21
longer-session trajectories, 8/21 review-only trajectories and 12/21 with
either consolidation or repair. The report's inspection identifies the
over-specified earlier test and the permission requests it provoked, but
does not trace whether B and E avoided the restrictive test initially,
changed the implementation around it or used E's specification-bound
correction exception in each successful run. We therefore cannot attribute
the acquisition difference to a demonstrated repair mechanism.

Across 63 trajectories per arm, the primary equal-model-weight coverage
gain for consolidation over longer sessions (B--D) was 1.19 percentage
points (95\% bootstrap
interval [0.10, 2.22]), and 1.41 points over feature-only delivery (B--A,
[0.33, 2.45]). With C and E also complete (63 versus 63), B--C is $+0.66$
points ($[-0.45, 1.75]$) and E--B is $+0.04$ points ($[-1.07, 1.14]$). The
intervals use 10,000 stratified whole-trajectory draws with seed 2027.

All estimated coverage gains are small relative to the earlier ten-point
practical benchmark ($\Delta_*=10$ percentage points).
The realised equal-model D mean was 96.9629\%, so even if every B endpoint
had scored 100\%, its gain against these observed controls could not exceed
$100-96.9629=3.0371$ percentage points. This post-collection arithmetic
check diagnoses ceiling-limited measurement; it neither changes the
original target nor estimates an alternative treatment effect. P2 is
unmet, but this task could not have demonstrated its ten-point target
against the realised controls.
Model-specific
contrasts in Supplement Table~\ref{tab:resultsh1modelintervals} show that
the pooled B--D gain comes almost entirely from GPT~6.1 Sol
acquiring the ninth feature ($+3.15$ points, 95\% interval $[0.00,+6.30]$;
B--A $+3.67$ points, $[+0.52,+6.82]$; B--C $+2.10$ points,
$[-1.05,+5.25]$).
The B--D intervals for Sonnet~5.5 ($+0.26$
points, $[-0.15,+0.75]$) and Gemini~4 Argon ($+0.15$ points,
$[-0.15,+0.52]$) span zero.

\paragraph{The final extension exposed a different weakness.}
Table~\ref{tab:resultsadaptation} reports CR1 success: all fifteen new
contracts had to pass while preserving the feature contracts that passed
before the extension. Sonnet succeeded in nearly every trajectory, but on
Gemini, read-only review achieved 16/21 successes (76.2\%) versus 8/21
(38.1\%) with consolidation and 11/21 (52.4\%) with visible-specification
repair. Exploratory inspection in the source report attributed the recurring failure
to views that blocked known write methods but forwarded new CR1 methods.
The structural improvement therefore coexisted with a permission-boundary
failure. The Gemini B--C difference is $-38.10$ percentage points, with
registered bounds of $[-77.05,+14.69]$, which span zero, followed by the
post hoc unadjusted Newcombe hybrid-score 95\% interval of $[-59.88,-8.18]$
and two-sided Fisher exact $p=0.02778$
(Supplement Table~\ref{tab:resultscrintervals}).
Holm-adjusted across the eight Sonnet and Gemini CR1 contrasts, $p=0.22$.
This is an exploratory
comparison among many, conditional on independent binomial outcomes;
we did not trace which review actions produced the difference.

\begin{table}[!htbp]
\centering\small
\caption{CR1 outcomes across all 21 completed trajectories per cell.
Bounds assign the two unadjudicated GPT~6.1 Sol test errors (\texttt{B-08}
and \texttt{C-14}) first to failure and then to success. They describe
missing-outcome uncertainty, not sampling confidence intervals\@. \armlegend}\label{tab:resultsadaptation}

\begin{tabular*}{\textwidth}{@{\extracolsep{\fill}}lcrrrl@{}}
\toprule
Model & Arm & Pass & Fail & Unknown & Success bounds (\%) \\
\midrule
Sonnet 5.5 & A & 21 & 0 & 0 & [100.0, 100.0] \\
Sonnet 5.5 & B & 21 & 0 & 0 & [100.0, 100.0] \\
Sonnet 5.5 & C & 21 & 0 & 0 & [100.0, 100.0] \\
Sonnet 5.5 & D & 20 & 1 & 0 & [95.2, 95.2] \\
Sonnet 5.5 & E & 20 & 1 & 0 & [95.2, 95.2] \\
\midrule
Gemini 4 Argon & A & 6 & 15 & 0 & [28.6, 28.6] \\
Gemini 4 Argon & B & 8 & 13 & 0 & [38.1, 38.1] \\
Gemini 4 Argon & C & 16 & 5 & 0 & [76.2, 76.2] \\
Gemini 4 Argon & D & 7 & 14 & 0 & [33.3, 33.3] \\
Gemini 4 Argon & E & 11 & 10 & 0 & [52.4, 52.4] \\
\midrule
GPT 6.1 Sol & A & 19 & 2 & 0 & [90.5, 90.5] \\
GPT 6.1 Sol & B & 17 & 3 & 1 & [81.0, 85.7] \\
GPT 6.1 Sol & C & 17 & 3 & 1 & [81.0, 85.7] \\
GPT 6.1 Sol & D & 19 & 2 & 0 & [90.5, 90.5] \\
GPT 6.1 Sol & E & 20 & 1 & 0 & [95.2, 95.2] \\
\bottomrule
\end{tabular*}
\end{table}

CR1 success must be read with feature completeness. A trajectory that never
implemented ranking had no passing ranking contracts to preserve during
CR1. GPT~6.1 Sol A and D entered the extension without ranking in 16/21
and 15/21 cases, respectively; B carried it into 12/21. A higher CR1
success rate can therefore coexist with less complete software. The
same pattern is visible in the prerequisite table: GPT A's fourteen CR1
successes among sixteen incomplete repositories are separate from its
five successes among five repositories with both F9 and F10 complete.
The
Supplement retains the exact model-specific contrast intervals and bounds,
including the two unadjudicated errors rather than imputing them.
Supplement Table~\ref{tab:crprerequisites} cross-tabulates the existing
outcomes with F9/F10 completeness. These post-treatment groups describe
different preservation burdens and cannot identify effects among otherwise
comparable repositories.

\FloatBarrier
\subsection{Structural quality, code volume and resource consumption}\label{sec:resultsstructural}

Consolidation's clearest advantage was structural cleanup
(Table~\ref{tab:resultsstructural}). Cyclomatic complexity counts independent
control-flow paths through a function; the threshold applies to each function.
No functions exceeded cyclomatic
complexity 10 at any of the 63 B endpoints. Relative to A, lint density
fell by about 92\% on Sonnet and reached zero on GPT\@. The targets were
not uniformly achieved: Sonnet B still averaged 2.26 lint warnings per
1,000 source lines despite the zero-warning target. These metrics
measure the intervention's structural targets. Broader architecture and
correctness require distinct assessments.

Gemini's baseline metrics were already low, but some functions exceeded
complexity 10 and lint was nonzero. Its baseline therefore did not satisfy
every B target. The observed comparison supports a narrower conclusion:
structural measures can improve while a semantic permission defect persists.

\begin{table}[!htbp]
\centering\small
\caption{Mean structural metrics at $P_{10}$, over all 21 observed endpoints
per cell (315/315 total). The complexity threshold applies to individual
functions, not the mean complexity of a repository\@. \armlegend}\label{tab:resultsstructural}

\begin{tabular*}{\textwidth}{@{\extracolsep{\fill}}lcrrr@{}}
\toprule
Model & Arm & \shortstack{Functions above\\complexity 10 (\%)} & \shortstack{Duplicated\\lines (\%)} & \shortstack{Lint warnings per\\1,000 source lines} \\
\midrule
Sonnet 5.5 & A & 1.10 & 0.33 & 28.15 \\
Sonnet 5.5 & B & 0.00 & 0.05 & 2.26 \\
Sonnet 5.5 & C & 0.95 & 0.25 & 21.14 \\
Sonnet 5.5 & D & 1.10 & 0.31 & 28.84 \\
Sonnet 5.5 & E & 1.44 & 0.26 & 28.23 \\
\midrule
Gemini 4 Argon & A & 0.33 & 0.00 & 0.08 \\
Gemini 4 Argon & B & 0.00 & 0.00 & 0.00 \\
Gemini 4 Argon & C & 0.00 & 0.00 & 0.06 \\
Gemini 4 Argon & D & 0.08 & 0.05 & 0.09 \\
Gemini 4 Argon & E & 0.48 & 0.00 & 0.00 \\
\midrule
GPT 6.1 Sol & A & 3.42 & 0.00 & 16.04 \\
GPT 6.1 Sol & B & 0.00 & 0.00 & 0.00 \\
GPT 6.1 Sol & C & 3.33 & 0.11 & 12.72 \\
GPT 6.1 Sol & D & 3.42 & 0.11 & 13.01 \\
GPT 6.1 Sol & E & 3.32 & 0.05 & 20.73 \\
\bottomrule
\end{tabular*}
\end{table}

\begin{figure}[!htbp]
\centering
\includegraphics[width=\linewidth]{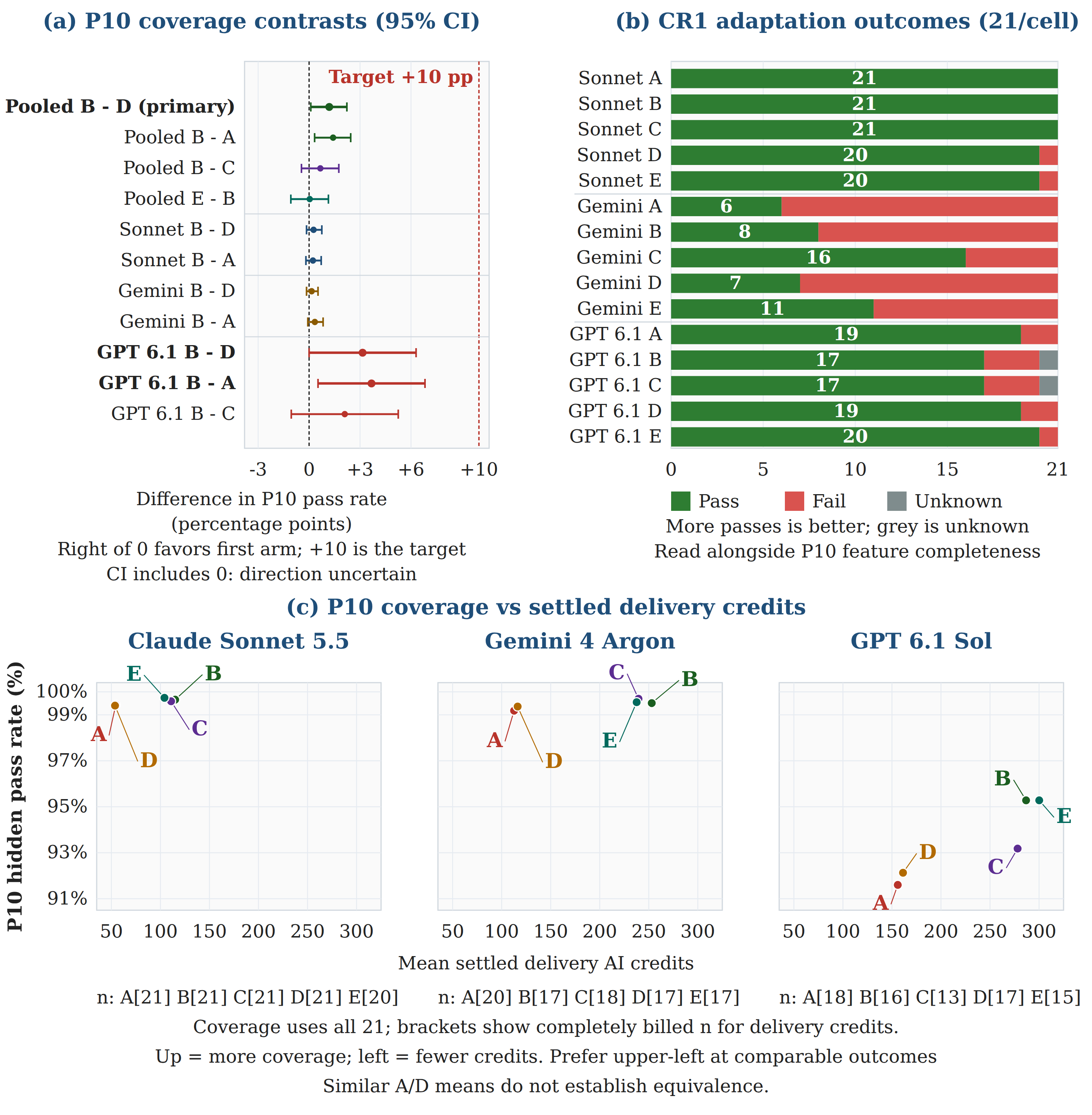}
\caption{Completed 315/315 study endpoints\@. (a) Pooled
63-versus-63 and within-model 21-versus-21 $P_{10}$ coverage contrasts
(\texttt{H1}) with 95\% bootstrap intervals relative to zero and the
$+10$~pp target\@. (b) CR1 adaptation counts (\texttt{H6}: pass, fail and
unadjudicated test error) across all 15 model-arm cells\@. (c) Mean $P_{10}$
coverage against mean settled delivery AI credits (\texttt{H5}, completely
billed sample size $n$ in brackets). The reference lines mark zero and
the primary comparison's practical benchmark; grey CR1 outcomes are
unknown. Coverage uses all 21 trajectories, whereas credit means use
the fully billed subset.}\label{fig:studyoutcomescosts}
\end{figure}

The additional sessions used more resources
(Table~\ref{tab:resultscosts} and Figure~\ref{fig:studyoutcomescosts}).
Settled delivery credits in B were 1.84 to
2.25 times those in A, with no consistent CR1 cost reduction. The longer
single feature session in D used similar resources to A despite its larger
allowance. B--D had matched ceilings, but unequal observed resource use.
Completely billed trajectory counts vary by cell, with unresolved attempt
holds excluded from these settled means (Section~\ref{sec:threats}).
Source size, mean function complexity and documentation volume are
reported in Supplement Section~\ref{app:resultsdetail}.

\begin{table}[!htbp]
\centering\small
\setlength{\tabcolsep}{4pt}
\caption{Mean credits and recorded session minutes for delivery and CR1.
Credit means use only trajectories with complete billing for that phase;
counts are in parentheses. Duration means describe all 21 completed
trajectories per cell (315/315 total). Human effort and total calendar
time were outside that measure. Outstanding charges remain outside settled
means\@. \armlegend}\label{tab:resultscosts}

\begin{tabular*}{\textwidth}{@{\extracolsep{\fill}}lcrrrr@{}}
\toprule
Model & Arm & \shortstack{Delivery\\credits ($n$)} & \shortstack{CR1\\credits ($n$)} & \shortstack{Delivery\\minutes} & \shortstack{CR1\\minutes} \\
\midrule
Sonnet 5.5 & A & 54.52 (21) & 11.07 (21) & 7.36 & 1.09 \\
Sonnet 5.5 & B & 115.02 (21) & 12.07 (21) & 16.94 & 1.15 \\
Sonnet 5.5 & C & 110.89 (21) & 11.85 (21) & 15.92 & 1.12 \\
Sonnet 5.5 & D & 53.70 (21) & 11.14 (21) & 7.32 & 1.12 \\
Sonnet 5.5 & E & 104.10 (20) & 11.05 (21) & 14.89 & 1.07 \\
\midrule
Gemini 4 Argon & A & 112.66 (20) & 27.10 (21) & 24.56 & 5.01 \\
Gemini 4 Argon & B & 253.03 (17) & 30.79 (21) & 55.64 & 5.54 \\
Gemini 4 Argon & C & 239.55 (18) & 40.53 (20) & 49.70 & 7.31 \\
Gemini 4 Argon & D & 116.29 (17) & 28.02 (21) & 26.32 & 5.38 \\
Gemini 4 Argon & E & 237.66 (17) & 30.06 (20) & 51.86 & 5.39 \\
\midrule
GPT 6.1 Sol & A & 155.93 (18) & 39.32 (20) & 25.33 & 6.45 \\
GPT 6.1 Sol & B & 286.78 (16) & 38.46 (21) & 51.54 & 6.36 \\
GPT 6.1 Sol & C & 278.14 (13) & 39.88 (21) & 46.76 & 6.52 \\
GPT 6.1 Sol & D & 161.28 (17) & 38.51 (21) & 26.27 & 6.18 \\
GPT 6.1 Sol & E & 300.18 (15) & 39.88 (20) & 54.33 & 6.68 \\
\bottomrule
\end{tabular*}
\end{table}

\FloatBarrier
\subsection{Post-collection workflow replay and anonymised AI judging}\label{sec:resultsjudging}

The workflow replay found no witnessed endpoint losses among 626 evaluable packets, with failures mainly concerning ranking and the delivery journey
where ranking had never been completed.
The 630 allocated packets covered both $P_{10}$ and CR1, with four
unavailable because judging preceded completion of the interrupted runs
(Section~\ref{sec:threats}). Replay used 3,457 recorded checkpoints and the twelve-scenario
panel described in Section~\ref{sec:postrunjudging}.

Zero workflow losses and the hidden-contract regression count answer
different questions, because the workflow panel exercised selected journeys whereas the hidden suite checked additional invariants. In particular, the workflow's
\texttt{privacy:3} scenario checked refusal of \texttt{view.open\_case},
but did not exercise \texttt{view.define\_region} or
\texttt{view.approve\_handoff}. Those new methods could still write through a view that passed the first check, a coverage gap that explains why near-perfect workflow ratings can coexist with
poor hidden-suite CR1 success.

The \texttt{gpt-6-luna} judge completed two fresh ratings per
evaluable packet: 1,252 calls costing 418.64 settled AI credits (US\$4.19) in
the completed assessment. Its complete-frame differences in functional completeness
and API usability favoured B by about 0.06 on the 0--4 scale at $P_{10}$
(Supplement Table~\ref{tab:resultsjudgecontrasts}), largely reflecting GPT's greater
ranking completion. These differences are small descriptive summaries on an ordinal scale, for which the companion supplies exploratory bootstrap intervals. At CR1,
all four intervals for these two dimensions include zero; no practical-importance threshold was established, and the intervals are
not adjusted for multiplicity: their stated coverage applies to each
comparison individually. The Supplement retains the detailed
contrasts, all cell means, judge-version accounting, incomplete-frame bounds
and scenario abstentions.

\FloatBarrier
Our human-calibration spot-check used one human rater to score 90 rows, of which 79 were scorable and represented sixteen distinct
requirement/evidence contexts. Guidance was clarified after initial AI
scoring, the review remained in progress and AI-assisted triage focused
inspection on failures or mismatches. There were no numeric disagreements of two or more points, although one disagreement concerned a
\texttt{ValidationError} rejection at the last step: did that demonstrate
error recovery when no subsequent action was observed?
The high agreement and strong ceiling
concentration do not provide independent inter-rater reliability; Supplement Table~\ref{tab:resultscalibration} gives the raw and context-weighted
agreement figures and abstentions.

The SDK audit found no interactive user-input RPCs, although some sessions
ended by requesting permission and left their proposed change uncommitted.
The Supplement gives invocation counts and model-specific breakdowns.

Table~\ref{tab:predictionverdicts} maps the original predictions to the completed results.

\begin{table}[!htbp]
\centering\small
\caption{Predictions, criteria, observed ten-feature outcomes and exploratory
verdicts across all 315/315 completed trajectories (\texttt{H1}--\texttt{H7}).
Only the $\Delta_*=10$ percentage-point B--D threshold was pre-specified in
the planning calculation; P1 and P3 structure the post-collection evaluation.
The combined portfolio decision is \textbf{adjust}.}\label{tab:predictionverdicts}

\begin{tabularx}{\textwidth}{@{}L{0.22\textwidth}L{0.24\textwidth}YL{0.20\textwidth}@{}}
\toprule
Prediction & Criterion & Observed result (315/315 complete) & Exploratory verdict \\
\midrule
P1: Decay (\texttt{H2}) & A, D: open loss ($L^P_{r,10}>0$) or witnessed workflow loss at $P_{10}$ & 0/126 A, D runs with open loss at $P_{10}$ (1/315 overall, Gemini~E); Sonnet/Gemini coverage 99.18\%--99.74\%; workflow losses 0/626 & Not supported at $N=10$ \\
P2: Coverage gain (\texttt{H1}, \texttt{H7}) & B--D equal-model $P_{10}$ gain $\Delta_*\geq10$ pp & $+1.19$ pp (95\% CI $[+0.10,+2.22]$); B--A $+1.41$ pp, $[+0.33,+2.45]$; B--C $+0.66$ pp, $[-0.45,+1.75]$; GPT $+3.15$ pp $[0.00,+6.30]$ (F9: 12/21 vs 6/21); Sonnet $+0.26$ pp and Gemini $+0.15$ pp CIs span zero & Unmet and ceiling-limited (single-model F9 recovery) \\
P3: Adaptation and payback (\texttt{H5}, \texttt{H6}) & B raises CR1 success and reduces CR1 or net credits versus A, C, D & Equal-model CR1 B--D, B--A bounds $[0.00,+1.59]$ pp; B--C $[-14.29,-11.11]$ pp; Gemini C 16/21 vs B 8/21; delivery credits $1.84\times$--$2.25\times$ A, no consistent CR1 saving & Not supported (\texttt{H5} unmet; \texttt{H6} inconclusive for B; higher observed Gemini CR1 success under review) \\
Manipulation check: Structure (\texttt{H3}, \texttt{H4}) & B lowers functions above complexity 10, duplication and lint towards stopping targets & 0.00\% functions $>10$ complexity at all 63 B endpoints; Sonnet lint $-92\%$ (2.26/kloc); GPT $-100\%$ (0.00/kloc); \texttt{src} NLOC $+3.2\%$ to $+7.2\%$ & Structural change observed, not all targets met. Hybrid B+C+E untested. \\
\bottomrule
\end{tabularx}
\end{table}

\FloatBarrier
\section{Discussion}\label{sec:discussion}

\subsection{Observed costs and possible payback}\label{sec:costprojection}

Consolidation improved structure at 1.84--2.25 times feature-only delivery
credits in fully billed subsets; its 1.19-point coverage advantage
over longer sessions faced a 3.04-point ceiling against the original
practical target. Matching resource ceilings did not match spending,
and the extra delivery cost brought no consistent reduction in CR1 cost.
Some consolidation runs delivered more complete software, so cost
comparisons must consider that difference rather than treating cheap
failure as efficiency.

The assumed payback model in Supplement Section~\ref{app:economics} asks whether lower costs on enough later features could recover the cleanup investment, but we did not fit its cost-growth parameter $\delta$ to these
data: capped budgets allow harder work to produce less complete features
without increasing spending. Testing long-term payback needs longer
sequences, resolved billing and comparable delivered behaviour.

\subsection{What worked}

For the product described in Section~\ref{sec:method}, the agents were useful
before coding began: they helped me question what to build. Research agents
gathered and analysed source material, and role-based agents challenged
the proposed design. An evidence-assessment task still
needed surveyor judgement; a resident-contact task still needed consent
and a manual route. The prototypes made those boundaries visible in a
case card and resident-decision record. I built and reshaped the MVP
through targeted exchanges and feedback.
These observations come from my use of the tools; we did not measure a
human-only duration or run a controlled comparison of feedback cadences.
The Supplement documents the discovery, prototype and backlog handoffs.

\subsection{Human judgement and reflective work}

I supplied several kinds of reflective work: choosing what not to build,
recognising the unusable replacement and deciding which earlier lessons
mattered. Naur's theory-building account
describes the understanding I carried forward~\cite{naur1985theory}, and
Beck similarly separates producing code from learning through working
software~\cite{beck2026prodacity}. These case-specific judgements
raise the question of whether agents could make comparable ones.
The experiment asks something narrower: whether bounded post-feature
correction preserves behaviour and later changeability.

An approval record alone cannot establish that a person evaluated the
alternatives independently. In Anthropic's
randomised study, 61 analysts were recruited and 50 used the agent. Those
50 analysts matched their agent's
randomly assigned preference on 86\% of four analytical choices
~\cite[Study design; Results]{herdagdelen2026judgment}. The preferences were
deliberately strengthened in one farming-analysis task with one model; the
rate concerns a different task and setting from coding
~\cite[Our study's limitations]{herdagdelen2026judgment}.
The study did not assess the decisions reported in this case. Its authors suggest
asking experts for their view first or making the best case for each option
~\cite[What this means]{herdagdelen2026judgment}. I propose a separate step:
record the alternatives and a person's reasons before accepting an agent's
recommendation. This study did not test that procedure.

\subsection{Protection can also constrain the next feature}

Every arm rolled back unsuccessful increments and protected immutable
accepted tests, a harness that plausibly explains why the expected decay
did not appear. The same protection obstructed GPT~6.1 Sol's ninth feature:
an earlier test fixed the allowed view methods, so adding ranking broke a
test the agent could not edit. This is a hypothesis about the harness,
not an estimated treatment effect, because no arm ran without those rules.

The case and experiment also suggest a shared problem at interfaces that
local checks did not exercise. In my application, rejecting unauthorised
requests did not verify identity and internal-staff binding for an
authorised user; in CR1, refusing an established write method did not
verify refusal of newly introduced view methods. We interpret this
gap between local completion and usable behaviour as ``everything but the
software'', without treating the coding experiment as a reproduction of
the deployment failure. Gemini's 16/21 review successes versus 8/21 under
consolidation illustrate the distinction between structural hygiene and
extension-safe permissions, although we did not trace the responsible review
actions.

\subsection{What this study teaches a future experiment}\label{sec:futureexperiment}

We expected decay, but the agents largely preserved what they had
implemented, changing the next study's question from whether cleanup prevents
inevitable loss to when it improves acquisition or adaptation enough to
justify its cost. A longer, coupled brownfield task could test that
trade-off, distinguishing rejected edits from accepted failures and
allowing stable retention as an outcome.

Several change requests on isolated branches would probe adaptation without
altering the continuing development history, and component comparisons could
distinguish refactoring from review, new tests and handover notes.
Discovery was disabled here; comparing full RPI with feature-only delivery
would answer another question. Supplement Section~\ref{app:futuredesign}
preserves the numerical candidates and reporting requirements for a
prospectively registered study, deferred until its design and resources
are approved. We plan no additional experimental or judge batch for this paper.

\subsection{Implications for practice}\label{sec:feedbackfirst}

The report recommends \emph{adjust} before adopting unconditional consolidation and proposes a triggered hybrid of cleanup, review and specification-bound repair (Supplement Section~\ref{app:practicecandidate}), but we have not tested that combination,
so its proposed efficacy cannot justify replacing an existing review
workflow.

For my application, the useful checkpoint would have been signing in and
opening an assigned case with both identity bindings checked together.
Carrying that prerequisite into later briefs or executable checks is a
proposed response to the case, subject to the existing safety and approval
boundaries; any context compaction must preserve adverse evidence and use
archive-backed, restore-verified selection.

\subsection{Threats to validity}\label{sec:threats}

The case is a retrospective participant-observer account of a project whose
success interested me: I designed the coordinator and selected the episodes.
We analysed selected records without exhaustive coding or independent
corroboration of every observation. The failed deployed task still lacks a
matched release-linked trace. Migration, scope, model/tool changes,
configuration and owner priorities changed with coordination; document
counts cannot identify their effects or measure context overload,
per-change gate burden or lost knowledge~\cite{devanbu2016belief}.

The experiment covers one synthetic Python suite, ten features and one
fixed change request under rollback and protected-test rules, so it
does not estimate the five case mechanisms or unrestricted development.
Consolidation combines refactoring, added tests and a handover note;
matching resource ceilings leaves spending and session structure
unequal. Near-ceiling controls prevented the planned useful coverage gain;
post-treatment prerequisite groups and fully billed subsets describe
different delivered functionality rather than quality-matched effects.

The executed design includes unregistered amendments and sixteen source
versions, without a complete chronology of outcomes inspected when
amendments were chosen. Shared scheduling and provider conditions could
violate the within-cell independence assumed by the reported intervals,
which are exploratory and unadjusted for multiplicity; exposed model names
cannot guarantee fixed provider builds~\cite{chen2023drift}. Operationally
distinct execution and analysis do not provide independent scientific
review, and the unassisted-run attestation cannot exclude unrecorded
out-of-band intervention.

Two GPT CR1 outcomes are unknown, and 55 billing holds leave all-allocation
cost and payback unresolved. Judging used the earlier 313-run snapshot,
leaving four packets unavailable; scenario abstentions and an incomplete,
AI-assisted human spot-check limit rating interpretation. Packet
anonymisation did not remove historical anchoring or judge/development
family overlap (Section~\ref{sec:postrunjudging}), and the workflow panel
missed permission checks that the hidden suite exercised.
Inline suppressions could also affect lint counts because the evaluator
did not independently count them. Human effort, opportunity cost,
provider-active time and currency prices were not measured.

\FloatBarrier
\section{Conclusion}\label{sec:conclusion}

Across the ten-feature S1 task, previously acquired behaviour was largely
preserved. None of the 126 feature-only or longer-session runs had an open
regression at the final checkpoint ($P_{10}$); only one trajectory across
all arms did, and workflow replay found no witnessed losses across 626
evaluable packets. Rollback and immutable accepted tests plausibly helped
protect earlier behaviour while also constraining ranking, but their effect
was not tested against an unprotected arm.

Consolidation (B) removed complexity hotspots at all 63 endpoints, but its
primary coverage advantage over longer feature-only sessions (D) was
1.19 percentage points (conditional 95\% interval [0.10, 2.22]), below the
ten-point minimum useful threshold. Even perfect consolidation endpoints
could exceed the observed longer-session results by only 3.04 points,
exposing a ceiling in the task and measurement design.
GPT~6.1 Sol supplied most of the observed gain by
completing priority ranking. Sonnet~5.5 and Gemini~4 Argon showed small B--D gains whose
intervals spanned zero. Read-only review (C) achieved 16/21 Gemini CR1
successes versus 8/21 with consolidation.
Consolidation used 1.84--2.25 times feature-only delivery credits in
fully billed subsets without consistently lowering CR1 cost.

The motivating case concerns a different scale and failure: a migration
interrupted a usable workflow after extensive agent-assisted preparation.
Its candidate explanations informed a narrower experiment whose missing features
and permission failures redirect attention towards acquisition and adaptation as
well as retention; subject to the consolidated threats
in Section~\ref{sec:threats}, the evidence supports structural improvement
at additional delivery cost, without demonstrated short-horizon payback.
A later study needs longer, coupled work and several adaptation probes
under its own approved protocol and resources.

\section*{Author contributions}
Following the Contributor Roles Taxonomy (CRediT, ANSI/NISO~\nolinkurl{Z39.104-2022}),
Juan Manuel Servera Bondroit contributed Conceptualization, Methodology
(study design and evaluation criteria), Investigation (historical exploratory
case), Formal analysis, Validation (post-collection analysis and
human-calibration spot-check), Writing (original draft) and Writing (review
and editing). Isabel Cabezas Martin contributed Investigation (independent
experimental execution and data collection across the 315 allocated
trajectories under the predefined exploratory protocol and disclosed
amendments) and Writing (review and editing, including critical review for
clarity and methodological recommendations for a subsequent study).
Separating execution and data collection from design and analysis was intended
to reduce experimenter bias. Both authors are Microsoft employees; this
division of work does not establish institutionally independent analysis
or scientific review.

\section*{Use of generative AI}
GitHub Copilot with HVE Core supported development; Microsoft 365
Copilot assisted summarisation and reference discovery. The deployed
assistant used \texttt{gpt-4.1-mini-2025-04-14}, distinct from coding models.
Copilot in VS Code assisted log analysis, mechanism naming, drafting,
bibliography and source checks, figures and local validation, including
this editorial pass and the interval correction. Copilot CLI
assisted source integration; the AMA Presentation Communications Strategist
supplied bounded narrative suggestions.

Scientific Agent Skills supported writing and method checks, including
SymPy, statistical-analysis, scientific-critical-thinking and
experimental-design~\cite{kassis2026skills}. Coach-guided assessments used
the Data Science and Engineering Coach and, earlier, a general-purpose
agent. Gemini~4 Argon proposed concepts that we checked and adapted
into editable diagrams; empirical charts use committed data, while the
economics graphic uses assumptions. The package records tools,
access limits, superseded findings and revision-specific checks.

I supplied case observations and judgements, checked repository figures and
Azure expenditure, and confirmed the Beck paraphrase from the essay and talk.
The incomplete 90-row human spot-check used AI-assisted mismatch triage
without inspecting every passing row; neither it nor assisted case
interpretation provides independent reliability or replication. Editorial
work collected no new outcomes and changed neither scoring nor registration.
We are responsible for exact-version approval, content and venue-policy compliance.

\section*{Competing interests}
Both authors are Microsoft employees. The study concerns Microsoft and
GitHub tools, and its historical case arose within a Microsoft training
programme. I designed the coordinator and analysed a case in which I
participated. We divided experimental execution and data collection from
design and analysis. These professional and
participant-observer interests remain relevant despite the operational
separation. The authors' views do not imply company endorsement.

\section*{Data availability}
Both repositories are public: the motivating case and paper sources at
\url{https://github.com/jmservera/ama}, and the experiment code and sanitised
results at \url{https://github.com/jmservera/ama-thought-mve}.
The experiment's
\href{https://github.com/jmservera/ama-thought-mve/releases/tag/results-exploratory-s1-v1}
{results-exploratory-s1-v1 release}
provides the selected public deposit with checksums.
The Supplement's source register identifies the exact snapshots and
analysis exports; historical freezes and later amendments retain their
own scope.

The OSF registrations comprise the earlier three-arm protocol at
\url{https://osf.io/ywsrv/overview} and the registered base of
7 November 2026 at \url{https://osf.io/xcpt4}.
Both registrations are public.
The five-arm design and allocation options were registered before collection;
later scheduling, execution and recovery amendments are disclosed in Git, not retroactively registered.

No post-collection OSF deposit was created; the registrations link to the
experiment repository. Raw packets, specimen mappings and runtime
workspaces stay outside Git to preserve future reviewer anonymisation,
while the public exports include actual assignments, failures and unknown
outcomes. The Supplement preserves deposit exclusions and historical
access records.

\appendix
\clearpage
\section{Supplement}\label{sec:supplement}

The Supplement records source qualifications, measurement definitions,
detailed results and execution history. It also retains proposed checks of
the case explanations, an assumed cost model and follow-up design candidates.
Those proposals are not results; their execution and approval status is
specified below. For example, the cost model assumes a growing implementation
premium, while the experiment measured only ten-feature spending.
Section~\ref{sec:experiment} explains the registered base and the disclosed,
unregistered amendments. Historical freezes below are version-specific.

For the result tables, $P_{10}$ means the final ten-feature checkpoint and
CR1 the withheld regional-access extension\@. \armlegend\ Full source and
digest identifiers appear in the end register
(Table~\ref{tab:source-register}); historical records retain their own scope.

\subsection{Additional source qualifications}\label{app:additionalevidence}

Retained literature claims reflect specific scope and formatting details in
the inspected primary texts. Gloaguen et~al.'s Table~3 records the
developer-versus-LLM comparison on CTXbench as $p=0.038$, alongside
percentage-formatted p-values in the surrounding prose; the manuscript uses
the table value. Borg et~al.'s two-phase study (151 participants, 95\%
professional developers) was conducted in late 2024 with IDE and chat
assistants predating autonomous coding agents: Phase~1 was observational,
while Phase~2 randomised human evolution of Phase~1 solutions without AI
assistance and found no significant differences in completion time or
CodeHealth~\cite[Secs.~1, 2.2, 4]{borg2025echoes}. In the SEC's Knight
Capital order (Exchange Act Release No.~\nolinkurl{34-70694}, Paragraphs~1, 12--17) and
companion press release \nolinkurl{2013-222}, a 2005 change moved cumulative-quantity
tracking in discontinued Power Peg code without retesting it, and a July
2012 deployment omitted one of eight SMARS servers, triggering millions of
child orders, more than four million executions from 212 parent orders, and
a loss exceeding USD~460 million~\cite{sec2013knight}. Non-essential
citations that had only abstract, excerpt or unverified third-party caption
access were removed rather than retained with unresolved gaps.

\begin{table}[!htbp]
\centering\footnotesize
\caption{Source-specific qualifications relevant to the proposed measurements.}\label{tab:sourcequalifications}

\begin{tabularx}{\textwidth}{@{}L{2.2cm}YY@{}}
\toprule
Source & Additional detail & Consequence for interpretation \\
\midrule
Sinha et al.~\cite[Sec.~3.2, Figs.~5--6]{sinha2025illusion} &
The artificial-history experiment varies erroneous prior updates.
Qwen3 with thinking enabled avoids the reported effect in that setup,
whereas scaling non-thinking models does not remove it. &
Earlier-error sensitivity depends on configuration; it is not a universal
property of all models or a measured repository-decay rate. \\
Pan et al.~\cite[Sec.~5.1, Fig.~4]{pan2026swemaintain} &
In the \texttt{parse\_crontab} episode, a local Python 3 change satisfies
current tests but omits module-wide compatibility aliases later code uses. &
An initially successful patch can leave a latent dependency failure.
This supplies a second mechanism alongside the main text's line-terminator example. \\
Gloaguen et al.~\cite[Secs.~4.2--4.4]{gloaguen2026agentsmd} &
File length does not explain success in their analysis, and generated
context helps when other documentation is removed. &
Instruction length alone is not an adequate exposure or burden measure;
available alternative documentation matters. \\
Jiang et al.~\cite[Secs.~5.2--5.5]{jiang2026skills} &
Skill misapplication and missing runtime verification remain failure modes.
The retrieval diagnostic is separate from downstream execution. &
Lower exact-match access alongside sustained task success does not isolate
an effect of retrieval; accessed, applied and sufficient guidance differ. \\
\bottomrule
\end{tabularx}
\end{table}

Merge-outcome studies measure different populations and outcomes.
Peralta et al.\ started with 9,799 human-reviewed agent PRs and manually
inspected 717. The 35.7\% clear-failure share among rejections and 15.4\%
explicit reviewer-involvement share among merges describe that manual sample
~\cite{peralta2026merged}. Murphy-Hill et al.\ estimate roughly 24\% more merges among adopters in an
observational rollout study~\cite{murphyhill2026agents}. Delivered value
was outside that output measure. AIDev contains
932,791 agent-authored PRs from five agents across 116,211 repositories
~\cite{li2026aidev}. Its size does not establish how many user outcomes were accepted.

Pavlenko et al.'s World Model offers an additional context-management
design: preserve raw evidence, derive rebuildable views and retrieve compact,
task-specific context with provenance and access controls
~\cite{pavlenko2026factory}. Applied to this case, that would mean giving
a sign-in worker the relevant role/staff-binding decision and its source,
while preserving the larger discovery record for other tasks.
Its effect on the AMA case was not tested.

\subsubsection{Case detail moved from the main article}

\paragraph{Historical deployment example}
Software development had this problem long before coding agents existed, as
public post-mortems and classic maintenance accounts show. On the 1st of
August 2012, for instance, Knight Capital's automated router sent millions
of orders and obtained more than four million executions while processing
212 customer orders, costing the firm more than USD~460 million. The SEC
traced that public incident to a defective function left in the router after
a 2005 change, an incomplete deployment for a new trading programme across
one of eight servers, and inadequate controls; the new deployment activated
old code that was no longer meant to run~\cite{sec2013knight}. The failure
crossed a feature boundary and a seven-year interval, so understanding it
required examining the code and deployment state that the new programme
inherited.

\paragraph{Original article roadmap}
\noindent The investigation moves from earlier studies to my project and
then to a separate coding experiment:

\begin{enumerate}[leftmargin=*]
\item We review earlier studies of errors carried through conversations,
inherited code and maintained instructions, including published comparisons
of agent-written and developer-written histories (Section~\ref{sec:background}).
\item We trace the move from questioning requirements with agents to
coordinated feature delivery, then my return to trying the product and
requesting fixes. Five possible explanations are examined against episodes
in the delivery record
(Sections~\ref{sec:method}--\ref{sec:mechanisms}).
\item We test scheduled consolidation in a 315-trajectory, five-arm
exploratory study, then replay earlier workflows and have a separate AI
judge rate the execution evidence. The results distinguish implementing a
feature from preserving it and adapting it later. The study also exposed
a task too easy to demonstrate the planned coverage gain and incomplete
billing. Its earlier planning drew on the registered-report example of
Lulla et~al.~\cite{lulla2026protocol} and HVE Core's Minimum Viable Experiment method
(Sections~\ref{sec:harness} and~\ref{sec:experiment}).
\end{enumerate}

\paragraph{Coordinator-history recovery and model selectors}
Read-only recovery of the original coordinator history corroborates my
requests for RPI tracking, tmux workers and recurring checks, followed by
migration, budget and functional-checkpoint decisions. Later committed
instructions selected \texttt{claude-sonnet-5.5} for worker launches while
leaving the coordinator on Automatic. Actual historical CLI versions,
reviewer models and backend builds remain unavailable in the inspected
records. The evidence index distinguishes prescribed selectors, stored
instructions and assistant-reported outcomes.

\paragraph{Case extraction and analysis scope}
The original analysis and later follow-up use separate repository snapshots
(source records \sourceref{R01} and \sourceref{R02} in the end register).
Sources include the knowledge-base
log and index, requirements, architecture decisions, coordinator definition,
issues, pull requests and retained RPI records. My review adds
process clarification and an export of project issue states, retained as
\nolinkurl{anc/project-insights-2026-11-01.csv}. Source dates, historical
counts and extraction rules remain in the ancillary records. The main
account uses them to reconstruct the sequence and handoffs.

I analysed descriptive measures, traced references to recorded lessons and
used AI assistance to interpret selected episodes. I did not systematically
code every log entry or review every specification. Early entries were
reconstructed from history; later entries were written during the project.
The evidence record distinguishes my recollection from committed source and
execution observations. It also records missing links between assessment
scores, releases and deployed tasks.

External repositories and literature were selected for relevance to
inherited faults, instruction maintenance and outcome verification.
They supply contextual comparisons and a focused narrative synthesis.
The published-evidence figures reproduce reported values with source
locators; the external applications and original studies were not rerun.
The ancillary observation record preserves its purposive issue sample,
exclusions and retrieval boundaries. These sources are analysed separately
from the motivating project.

\paragraph{Participant-observer analysis and independent checking}
I selected the episodes, designed the coordinator and had an interest in
successful delivery. Those roles favour a coherent retrospective
explanation. Experience-based beliefs can differ from project
evidence~\cite{devanbu2016belief}; the same caution applies to AI-written
records and AI-assisted synthesis. Independent checking of O1--O5 against
retained source and execution records remains outstanding.
Jiang et~al.\ make mechanism analysis inspectable through matched trajectories,
evidence-linked labels and a separate human check of taxonomy construction,
while distinguishing that check from exhaustive human coding~\cite[Sec.~4, App.~A.2]{jiang2026skills}.
Here the five mechanisms organise selected episodes.
Their prevalence and causal effects require a different
design. The evidence index retains null release or execution identifiers
where a link still needs checking.

\paragraph{Original case validity and rival-explanation account}

The five categories overlap in this case. Missing identity and staff
bindings could reflect deferred migration, incomplete deployment acceptance
or a lesson that workers did not apply. The deployment comparison and
session-input trace in Table~\ref{tab:audittrail} could help distinguish
these explanations, while document and citation counts cannot. The available
evidence also cannot establish context overflow from a growing read path, or
disproportionate cost from an expensive gate alone. I did not measure
per-change gate time, avoided defects or human effort.

The source-layout migration, owner-selected priorities and delegated merging
changed alongside coordination. Scope additions, budget constraints, model/tool
changes and deployment conditions also differ between periods. Any of these
could help explain the outcome attributed to autonomy. The
\hyperref[sec:supplement]{supplementary rival-explanation register} retains the full
register, separating recorded circumstances from plausible but untested
alternatives.

Construct validity is limited by whitespace counts, lexical indicators and
selectively reviewed records. Appropriate caution can raise boundary-phrase
density, and the records cannot show whether a missing reference reflects
absent knowledge or missing documentation. I was coordinator designer,
participant and analyst, so I selected the episodes and may favour a coherent
retrospective explanation. AI-assisted synthesis shares these source limits
and needs independent checking.
Public records support recomputation but not independent corroboration of
every statement. The single exercise I designed and the purposively selected
external examples do not establish prevalence or general causal effects.
The separate experiment has its own identification limits in
Section~\ref{sec:protocollimits}.

\subsubsection{Case observation and contrast tables}

\begin{table}[!htbp]
\centering\footnotesize
\caption{Evidence relevant to the delivery transition and checks that would
distinguish the candidate explanations.}\label{tab:audittrail}
\begin{tabularx}{\textwidth}{@{}L{0.7cm}L{4.7cm}YY@{}}
\toprule
ID & Observation and provenance & Interpretation & Discriminating evidence \\
\midrule
O1 & The log and participant account describe a used MVP followed by a
replacement lacking identity, staff and model prerequisites. & Delivery
failed to preserve a complete working journey. & Execute the same journey
against versioned deployments and configurations. \\
O2 & Pull requests and automated assessments recorded completion while
my sign-in remained blocked. & The completion signals covered a
different or insufficiently linked outcome. & Link the assessed capability
and release to the failed deployed task. \\
O3 & Requirements and process records expanded through discovery and delivery;
PR \#445's small source correction carried a separate evidence record. &
Documented context may outgrow what a worker or reviewer can use effectively.
& Compare actual context access and review effort with validated behaviour. \\
O4 & Simplicity and deployment lessons were recorded, but later replacement
work lacked registration and staff bindings. & Lesson availability and
application may diverge. & Trace the relevant session inputs and whether
the lesson was actionable or deliberately deferred. \\
O5 & Coordinator rules formalised gates and delegated merging; I continued
resolving blockers and later requested stabilization. & Approval
activity and product-use feedback served different functions. & Compare
outcome checks, steering decisions and subsequent implementation changes. \\
\bottomrule
\end{tabularx}
\end{table}

\begin{table}[!htbp]
\centering\footnotesize
\caption{Contrasts relevant to context, completion and feedback.}\label{tab:contrasts}

\begin{tabularx}{\textwidth}{@{}L{3.1cm}YY@{}}
\toprule
Contrast & Recorded difference & Analytical consequence \\
\midrule
Interactive MVP versus autonomous replacement & Product use supported short
feedback cycles; the replacement could not support the same task. & Feedback
is a candidate explanation alongside the changed migration and environment. \\
Requirements before versus during autonomy & Expansion and review obligations
already existed during discovery. & The burden has a longer history than the
coordinator, and some detail preserves important controls. \\
Engineering checks versus participant task & Bounded checks passed while
real-user sign-in remained blocked. & Acceptance needs coverage of the whole
journey and its deployment dependencies. \\
Exploratory challenge versus backlog execution & Role agents questioned the
initial proposal; delivery workers implemented specified stories. & Task
framing changed with the tools and environment; its contribution was not isolated. \\
\bottomrule
\end{tabularx}
\end{table}

\paragraph{Contextual comparisons from the opening.}

The migration instruction at coordinator turn 66 was ``build all new in
src, but migrate what's needed first''. The article paraphrases this
instruction; the case evidence index supplies the original locator.

Cunningham's 1992 account of WyCash described a financial product
continually adapted to market demands, where mature code was repeatedly
revised and rewritten; his technical-debt metaphor explained why shipping a
first-time implementation required later rewriting as understanding
improved~\cite{cunningham1992wycash}.

The external PhotoCraft project, an early-alpha reimplementation of
Photoshop, reports 631 of 631 menu items live, while its dated
assessment estimates about 76\% feature breadth and 45\%
readiness for real work~\cite{photocraft2026}. SWE-Maintain finds that agents solve
fewer later tasks when starting from their own earlier changes than from
the developers' changes, even though each task starts in a fresh
session~\cite{pan2026swemaintain}.

\subsection{Competing explanations for the case}\label{app:rivals}

The case records support several competing explanations
(Table~\ref{tab:rivals}). Other plausible alternatives lack discriminating
evidence (Table~\ref{tab:rivalsadditional}). None is ruled out by missing
records. In particular, the assessed release has not been linked to the
failed deployment (Section~\ref{sec:audittrail}). The named issues,
decisions, selector changes and participant accounts are indexed in
\nolinkurl{anc/case-evidence-matrix.json}.

\begin{table}[!htbp]
\centering\footnotesize
\caption{Rival explanations with direct case records.}\label{tab:rivals}
\begin{tabularx}{\textwidth}{@{}L{2.7cm}YYY@{}}
\toprule
Explanation & Case support & Counter-evidence or limits & Treatment \\
\midrule
Migration decision & ADR 0006 (issue \#196) froze feature work in the MVP and
prioritised a new layout. & A coordinator summary reported a healthy revision;
sign-in remained unresolved. &
Migration and coordination changed together; autonomy alone is not identified. \\
Capstone rubric & Research and documentation are rewarded; full completeness
marks coexist with blocked sign-in. & A working demonstrator is required;
release/scope linkage is pending. & An incentive explanation remains a hypothesis. \\
Owner overrides and priorities & Committed RPI records, full gates and delegated
merging; PR \#445 required an exception. & The participant reports MVP use;
coordinator instructions retained human approval boundaries. & Owner choices are part of the case. \\
Model and tool changes & Recorded model selectors changed during the case. & No matched rerun
under fixed versions. & Actual session configuration remains partly unknown. \\
CI and cloud budgets & Issue \#413 moved validation locally; costs prompted
a higher ceiling and reduced resources. & Merges continued without established
deployed usability. & More budget is not proven to resolve delivery. \\
Identity and sign-in setup & Missing registration/staff bindings; issue \#349
distinguishes redirect from sign-in. & Those steps worked in the MVP\@. &
A configuration obstacle is documented; decay and knowledge loss remain unproven. \\
\bottomrule
\end{tabularx}
\end{table}

\FloatBarrier
\begin{table}[!htbp]
\centering\footnotesize
\caption{Additional plausible rivals and unperformed discriminating checks.}\label{tab:rivalsadditional}
\begin{tabularx}{\textwidth}{@{}L{2.7cm}YYY@{}}
\toprule
Explanation & Reason considered & Counter-evidence or limits & Evidence still needed \\
\midrule
External services and deployment drift & Model/identity configuration differs
from the MVP\@. & No provider outage is established; health checks were narrower. &
Versioned deployment/configuration and dependency-health records for the journey. \\
Scope expansion & RAI and consumption criteria added to issues; further
resident-preview obligations. & Expansion predates autonomy; some controls
are justified. & Original/added scope, sequencing and time-to-outcome comparison. \\
Inadequate acceptance tests & Checks passed without completed real-user sign-in;
scorer read artefacts. & Bounded, honest tests need not be poor tests. &
Capability-to-task-oracle mapping and execution evidence. \\
Rate and context limits & Parallel sessions and a large read path. & No runtime
rate-limit, truncation or attention dataset analysed. & Actual inputs and
blocked/retried request events. \\
Specification ambiguity & Selective review of long, cross-referenced requirements.
& Important domain/consent decisions survived; ambiguity not independently coded.
& Independent interpretation and conflicting acceptance examples. \\
Ordinary prototype debt & Rapid MVP followed by restructuring and rebuilding.
& MVP worked intermittently; no comparable human-led migration. & Failure
analysis distinguishing inherited debt from introduced faults. \\
\bottomrule
\end{tabularx}
\end{table}
\FloatBarrier

\subsection{Evidence checks for the five explanations}\label{app:mechanismaudit}

Each candidate mechanism has a definition, counter-evidence and a proposed
check against competing explanations. Tables~\ref{tab:mechanisms} and
~\ref{tab:audittrail} give the case indicators. The five-arm experiment did
not isolate these mechanisms, and their applicability to other settings
remains untested.

\subsubsection{Specification inflation}

{\small
\noindent
\begin{tabularx}{\textwidth}{@{}L{2.6cm}Y@{}}
\toprule
Element & Audit \\
\midrule
Definition & Requirements and acceptance obligations expand faster than
independently validated behaviour, without evidence that the added scope is
needed for the next usable increment. The relevant comparison is added
obligations versus accepted tasks, such as staff sign-in and case access;
document length alone cannot supply either outcome. \\
Counter-evidence & Growth began before autonomy and sometimes followed the
owner's feedback; regulatory and consent controls can justify detail. \\
Rival explanations & Genuine scope expansion, specification ambiguity,
capstone incentives and an ambitious migration rather than agent-driven
inflation. \\
Experimental implication & Compare smaller requirements horizons at matched
scope and resources. The current fixed-suite experiment cannot identify this
mechanism, even if consolidation helps. \\
\bottomrule
\end{tabularx}
\par
}

\subsubsection{Proxy capture}

{\small
\noindent
\begin{tabularx}{\textwidth}{@{}L{2.6cm}Y@{}}
\toprule
Element & Audit \\
\midrule
Definition & A completion proxy becomes a decision criterion and remains
positive when an independent test of the intended outcome is negative.
Testing this case would require pairing a completion assessment with
sign-in on the assessed release; that link remains unavailable. \\
Counter-evidence & Agents explicitly reported unmet acceptance; neither
fabricated approval nor deliberate deception is established. \\
Rival explanations & An inadequate scorer or acceptance oracle, different
evaluation targets, and a deployment failure invisible to repository review. \\
Experimental implication & Disagreement between visible completion and hidden
behavioural tests would expose proxy limits. It would not by itself establish
intentional test gaming or explain why a gate was chosen. \\
\bottomrule
\end{tabularx}
\par
}

\subsubsection{Learning evaporation}

{\small
\noindent
\begin{tabularx}{\textwidth}{@{}L{2.6cm}Y@{}}
\toprule
Element & Audit \\
\midrule
Definition & A relevant recorded lesson does not reach a later worker in a
usable form, or the worker does not apply it, and the same problem recurs.
The staff-binding lesson would need to
be traced into a later worker's inputs and deployment decisions;
its absence from a citation list cannot establish lost knowledge. \\
Counter-evidence & Important domain decisions survived; the lesson could have
been known but deferred. Actual session attention was not measured. \\
Rival explanations & Deliberate migration sequencing, deployment-specific
configuration, context limits, or conflicting requirements rather than lost
knowledge. \\
Experimental implication & Trace actual inputs and repeated faults. B--C
cannot isolate memory quality because DESIGN and REVIEW notes and acting
permissions differ. A matched-context study would be needed. \\
\bottomrule
\end{tabularx}
\par
}

\subsubsection{Process-gate inflation}

{\small
\noindent
\begin{tabularx}{\textwidth}{@{}L{2.6cm}Y@{}}
\toprule
Element & Audit \\
\midrule
Definition & Additional procedural obligations increase per-change burden
without demonstrated proportional improvement in the targeted outcome.
PR \#445's size-limit correction and interrupted gate identify a place
to measure this burden, but neither elapsed gate time nor avoided defects
was measured for the comparison. \\
Counter-evidence & The size-limit correction addressed a bounded implementation
defect; the interrupted gate did not establish a pass. Gate duration and human review effort were not measured per PR\@. \\
Rival explanations & Legitimate safety and privacy requirements, budget
constraints, new deployment complexity and owner-selected risk tolerance. \\
Experimental implication & Measure gate time and accepted outcomes under
matched risk. The pilot measures resources for its treatment package;
the effect of removing this repository's required gates remains untested. \\
\bottomrule
\end{tabularx}
\par
}

\subsubsection{Feedback starvation}

{\small
\noindent
\begin{tabularx}{\textwidth}{@{}L{2.6cm}Y@{}}
\toprule
Element & Audit \\
\midrule
Definition & Delivery decisions receive too little feedback from attempts
to use the product, despite extensive checks of code and documents.
Here health and unauthenticated-request checks left the authorised
staff journey untested. Measuring the gap requires the intended task and
its release, not a count of tests. \\
Counter-evidence & Human oversight continued; agents surfaced blockers;
engineering checks provided real, though narrower, feedback. \\
Rival explanations & Identity and service configuration, migration priority,
incomplete acceptance coverage, prototype debt or insufficient resources. \\
Experimental implication & Hidden retention tests and CR1 supply behavioural
feedback to the evaluator. The experiment does not test live deployment
recovery, product selection or a human-use checkpoint cadence. \\
\bottomrule
\end{tabularx}
\par
}

\begin{table}[!htbp]
\centering\footnotesize
\caption{Accumulation, transmission and correction in the working model.}\label{tab:theory}
\begin{tabularx}{\textwidth}{@{}L{3.4cm}YL{4.6cm}@{}}
\toprule
Process & Question & In this case \\
\midrule
Accumulation & What obligations and change burden does each feature leave
behind?~\cite{lehman1980laws,lehman1997metrics} &
No scheduled cumulative consolidation before the checkpoint request;
restructuring planned as backlog items \\
Transmission & Which implementations, rules and lessons shape the next
change?~\cite{sinha2025illusion,pan2026swemaintain} &
Growing specifications and read path; previously working identity setup not preserved \\
Selection and correction & Do review and tests expose the cumulative
consequences before more work depends on them?~\cite{cohen2006review,yu2026habituation} &
Completion assessments, the interrupted PR \#445 gate and requests for
a running website; their separate effects remain unmeasured \\
\bottomrule
\end{tabularx}
\end{table}
\FloatBarrier

\subsection{Measures and lexicon}\label{app:measures}

\paragraph{Log entries.} The knowledge-base log
(\texttt{docs/reference/log.md}) was split into entries at each second-level
heading that begins with a date. An entry belongs to the date in its heading.
Word counts use whitespace tokenisation of the entry, heading included.

\paragraph{Evidence-boundary phrases.} Each entry was searched, case-insensitively and
on word boundaries, for these patterns: \emph{unmet}; \emph{not accepted};
\emph{not inferred}; \emph{is inferred}; \emph{remain{(s)} open};
\emph{remain{(s)} unproved}; \emph{remain{(s)} unverified}; \emph{not claimed};
\emph{not established}; and \emph{no} followed, within forty letters, spaces,
commas, slashes or hyphens, by \emph{is} or \emph{was} and then
\emph{claimed} or \emph{inferred}. Density is matches per 1,000 words.

\paragraph{Reference tracing.} An entry cites the simplicity requirements if
it contains an identifier from SA-001 to SA-010 or the word
\emph{simplicity}. It cites the development profile if it contains
\emph{deploymentProfile}, \emph{development profile} or
\emph{allowPublicNetworkAccess}.

\paragraph{Repository counts.} Issue and pull-request counts come from the
GitHub search API on 24 October 2026 (for example,
\texttt{is:pr is:merged merged:>=2026-10-15}). Document sizes are whitespace
word counts at the original case snapshot (\sourceref{R01}).

\paragraph{Proposed measures.} \emph{Time to first human use}: elapsed time
from the start of a delivery cycle to the first time a person uses the running
increment for its purpose\@. \emph{Age of the last usable increment}: time since
a person last verified that the deployed system could be used for its main
journey\@. \emph{Behaviour-changing merge ratio}: the share of merged pull
requests that change what a user can observe in the running increment.
\emph{Cost per accepted outcome}: total model, tool and cloud cost divided by
the number of outcomes a human accepted in the same period. Predeclare the
outcome unit, counting rule, attribution and period. With zero accepted
outcomes the ratio is undefined; report the total cost and zero count.

\subsection{Consolidation step prompt}\label{app:prompt}

The verbatim frozen v4 prompt comes from execution source \sourceref{R31}.
It records that historical version; the amended study uses a later prompt.
The archived bytes are preserved unchanged
(digest \sourceref{R35}).
Display wrapping does not alter its text. Braces mark runner parameters.

\begin{lstlisting}[basicstyle=\footnotesize\ttfamily,
  columns=fullflexible,keepspaces=true,showstringspaces=false,upquote=true,
  breaklines=true,breakatwhitespace=false]
<!-- markdownlint-disable-file -->
# Consolidation step prompt (arm B)

Do not add or change features. Your job is to leave the code better than you found it without changing its behaviour.

1. Run `make check`, then read reports/quality.md (functions above complexity 10, duplicated blocks, lint warnings, file sizes).
2. Review the last feature's diff (`git diff feature-start feature-end`) and the hotspots against this checklist:
   * duplicated logic;
   * long or complex functions;
   * unclear names;
   * dead code;
   * responsibilities in the wrong place, or rules from one feature leaking into another;
   * tests that check implementation details instead of behaviour.
3. Make small, behaviour-preserving changes. Apply every change through `./tcr "<message>"`: it runs all tests and commits if they pass, or reverts if they fail.
4. Never modify or delete files in tests/examples/. You may add tests that pin current behaviour before you refactor.
5. Rewrite DESIGN.md in at most 300 words: the current structure, where each business rule lives, and known debts. Replace its content; do not append.
6. Stop when the quality budget is met (no function above complexity 10, duplicated lines below 3%, zero warnings in the configured lint rules) or when your time budget ({consolidation_budget}) is used.
7. Finish with a summary of at most five lines: what you changed and what debt remains.
\end{lstlisting}

\FloatBarrier
\subsection{An assumed payback model}\label{app:economics}

The earlier cost model assumes that consolidation reduces a growing
implementation premium (Section~\ref{sec:costprojection}). Its parameters
were not fitted to the ten-feature observations. For example, its
$k=5,\delta=0.5$ scenario means spending five credits on consolidation
every cycle while the assumed excess implementation cost without it
increases by half a credit per successive cycle. Those assumptions give
break-even at cycle 21; the experiment does not estimate that growth rate.

Let $N\geq1$ be an integer number of feature cycles, $k$ the assumed average consolidation
investment in AI credits per cycle, and $\delta\cdot(t-1)$ the assumed excess
feature-implementation cost in A relative to B at cycle $t$, excluding
B's consolidation session. Shared baseline implementation costs cancel.
Under the explicit simplifying assumptions of constant $k$, a linearly
growing relative premium $\delta\geq0$, and comparable delivered behaviour,
the cumulative net difference is
\begin{equation}
\Delta C(N)=C_A(N)-C_B(N)
=\sum_{t=1}^{N}\bigl[\delta(t-1)-k\bigr]
=\delta\frac{N(N-1)}{2}-kN.
\label{eq:costprojection}
\end{equation}
The first term sums the assumed extra implementation cost without
consolidation; the second subtracts B's consolidation investment.
A positive value means A costs more within this assumed model.
Experimental savings would require measured, quality-matched costs.
For $k>0$ and $\delta>0$, the first cycle count with $\Delta C(N)\geq0$ is
$N_{\mathrm{BE}}=\lceil1+2k/\delta\rceil$; when $\delta=0$, the investment never pays back in
this credit-only model.
The parameter $\delta$ describes how the cost gap between A and B grows;
it does not measure either arm's total debt.

\begin{figure}[!htbp]
\centering
\includegraphics[width=\linewidth]{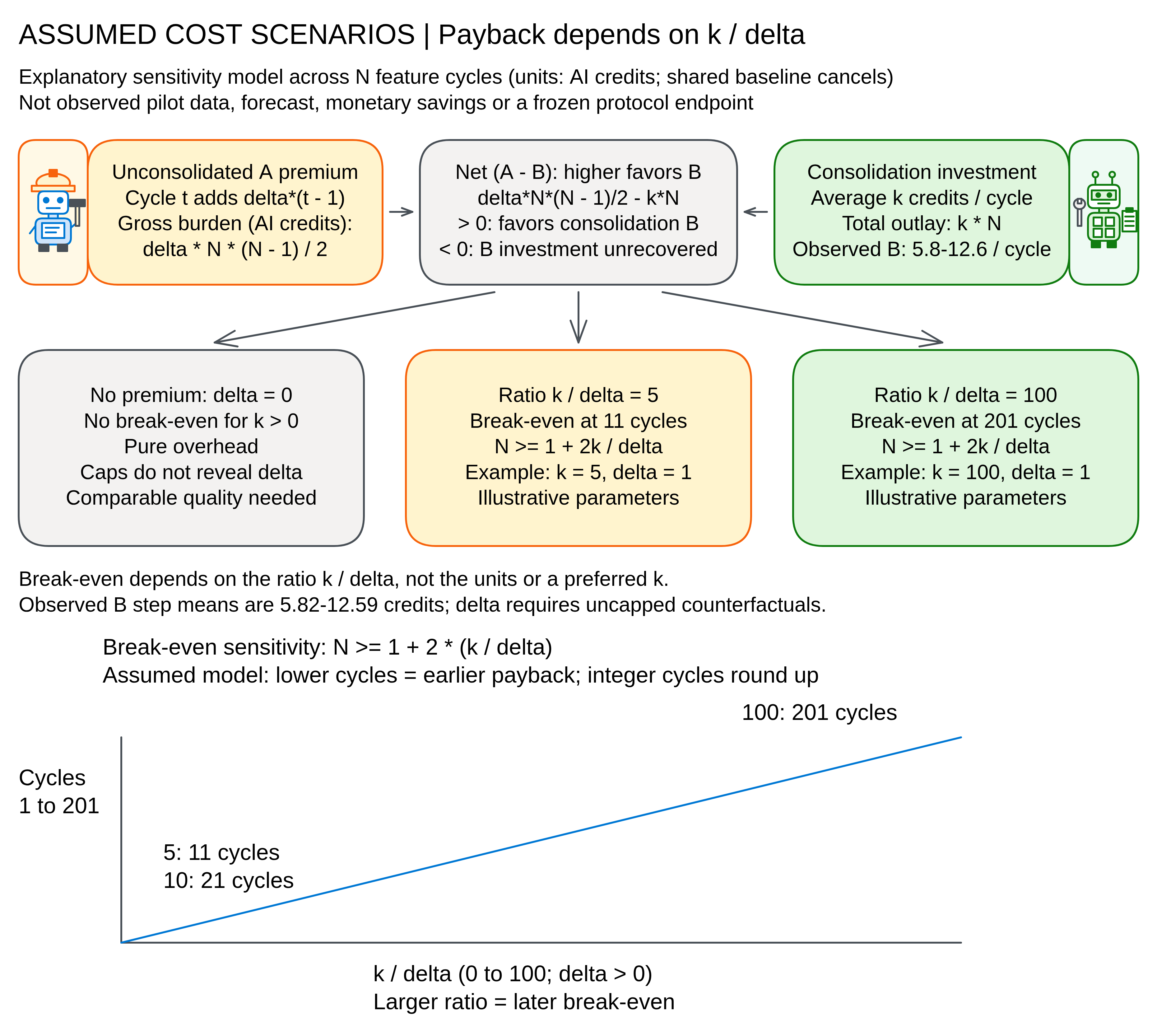}
\caption{Assumed cost-of-inaction mechanism and break-even sensitivity.
The figure uses illustrative parameters throughout. A linearly increasing relative
implementation premium accumulates across feature boundaries; recurring
consolidation has an explicit cost. Payback depends on $k/\delta$: a ratio of 5 breaks even at 11 cycles
and a ratio of 100 at 201; $\delta=0$ gives no payback for $k>0$.
Parameters are invented for transparent sensitivity,
not fitted to study data. No dollar conversion or human-work estimate is
implied. Higher $\Delta C(N)$ favours B in the assumed model; a lower
break-even cycle count means earlier recovery of the investment. Increasing
$k/\delta$ delays that break-even within the assumed model.
Observed cost and quality effects require empirical estimates.}\label{fig:refactoringeconomics}
\end{figure}

\begin{table}[!htbp]
\centering\small
\caption{Illustrative net A--B cost in AI credits from
Equation~\ref{eq:costprojection}, with assumed $k=5$ credits/cycle.
Negative values mean B's investment has not paid back; zero and positive
values mean break-even and a conditional credit advantage.
None of the scenarios is an observed or preferred estimate.}\label{tab:costscenario}

\begin{tabular}{@{}lrrr@{}}
\toprule
Assumed premium increase $\delta$ & 10 cycles & 20 cycles & 40 cycles \\
\midrule
0 credits per successive cycle & -50 & -100 & -200 \\
0.5 credits per successive cycle & -27.5 & -5 & 190 \\
1 credit per successive cycle & -5 & 90 & 580 \\
\bottomrule
\end{tabular}
\end{table}

\FloatBarrier
\subsection{Retained follow-up design candidates}\label{app:futuredesign}

The numerical candidates behind Section~\ref{sec:futureexperiment} are
retained here for a separate study. Their task scales, thresholds, allocations
and analysis choices require design review, feasibility estimates, spending
approval and prospective registration. They do not amend the completed
experiment, and a longer task could still show stable retention.

\paragraph{Longer trajectories and greater coupling.}
In the assumed cost model, break-even depends on investment $k$ and the
relative cost-growth slope $\delta$: $N_{\mathrm{BE}}=\lceil 1+2k/\delta\rceil$
for positive parameters (21 cycles at illustrative $k=5,\delta=0.5$).
Measured consolidation sessions cost $5.82$--$12.59$ credits, but the pilot
does not estimate long-run $\delta$ or when coupling and context limits matter.
Its ten greenfield features produced roughly 440--780 Python source lines.
A follow-up could extend
trajectories to $N=30$--$40$ sequential feature and schema-evolution cycles
on a multi-module brownfield service repository ($3{,}000$--$10{,}000$
source lines) with persistent storage, cross-cutting permission invariants
and scheduled interface migrations, recording retention and cost slopes at
$P_{10},P_{20},P_{30}$ and $P_{40}$.

\paragraph{One final change request and unequal starting functionality.}
Evaluating adaptation on one held-out CR1 after $P_{10}$ produced wide
binomial intervals ($\pm18.8$ to $\pm48.0$ percentage points within models)
and coupled retention obligations to whether F9 ranking had been acquired.
A follow-up protocol could schedule $K\geq4$ orthogonal change requests in
a Latin-square design across checkpoints $P_{10},P_{20},P_{30}$ and $P_{40}$.
Each change request should execute on an isolated branch forked from $P_t$
so adaptation measurement does not alter the main delivery trajectory, and
should be scored both unconditionally and on a core contract set whose
prerequisites are verified before the fork.

\paragraph{Separating the parts of consolidation.}
Arm~B combined behaviour-preserving code refactoring, new agent-authored
tests under an immutable test-file rule and \texttt{DESIGN.md} replacement,
while Arm~C changed both editing rights and checklist content. A $2\times2$
factorial core crossing \emph{behaviour-preserving code refactoring} (with
test files locked) against \emph{explicit invariant and boundary review}
(\texttt{REVIEW.md} targeting authorization, state transitions and
cross-module contracts), paired with a controlled factor comparing
\emph{immutable versus specification-revisable agent tests}, could help
distinguish test-rule effects from refactoring and boundary review.
Component fidelity, interaction effects and other treatment differences
would still need explicit controls.

\paragraph{Decision thresholds, dependence and recovery rules.}
The earlier candidate thresholds were $\geq5\%$ open-regression loss among
acquired contracts by $P_{30}$ or per-cycle implementation-credit growth
of $\delta\geq0.5$ credits per cycle for P1, a task-calibrated minimum
useful B--D coverage gain $\Delta_*$ for P2, and cumulative net credits
$\Delta C(N)>0$ at matched behavioural coverage for P3.
These criteria await protocol review and approval; no estimates have been
collected under them.
A new protocol should justify its thresholds and align randomisation and
variance estimation, considering $(\mathrm{model},\mathrm{replicate})$
blocks, a block bootstrap or a mixed model as appropriate to the actual
allocation and dependence. Bounded transport retries, missing-data rules
and the primary estimand must be settled before collection.

\paragraph{Full RPI versus corrective components.}
I also propose comparing the full research, plan, implement and review
workflow with feature-only delivery.
Testing the package would not by itself identify review's contribution.
A component comparison must specify instructions, editing and test rights,
handover, context, sessions and budgets, and record actual resource use.
No RPI variant, allocation, effect threshold or spending is approved here.

\paragraph{Reporting requirements for the next study.}
The completed study does not supply a graded CR1 panel, a full
attempt-versus-retained loss decomposition, or a complete prior-exposure
record. A new protocol must require the following before collection:
\begin{itemize}[leftmargin=*]
\item Publish each full risk set, observed subset and missing count
$M^p_{r,t}=|\mathcal{R}_{r,t}|-|\mathcal{O}^p_{r,t}|$, with the
corresponding full-risk loss bounds.
\item Publish witnessed workflow risk-set sizes, losses and unavailable
counts alongside final scenario outcomes.
\item Retain onset, recovery, recurrence and uncertainty across missing
checkpoints; distinguish same-cycle post-step losses, rejected attempts
and losses in accepted code.
\item Report CR1 contracts passed out of 15, previously passing contracts
lost and observed/missing denominators, in addition to composite success.
\item Preserve scheduling-block identifiers and align uncertainty analysis
with the allocation and observed execution dependence.
\item Record actual information access and prior exposure for executors,
analysts and judges.
\item Report per-model binary-success counts and stated interval methods,
preserving model strata in aggregate contrasts.
\end{itemize}
These requirements do not retrospectively amend the executed design.

\subsection{Untested hybrid practice candidate}\label{app:practicecandidate}

The report proposes combining the structural cleanup, semantic review and
audited repair represented separately by Arms~B, C and E.
Automated \texttt{./tcr} refactoring with \nolinkurl{tools/quality.py}
eliminated all complexity-$>10$ hotspots across all 63 Arm~B trajectories
and was associated with greater GPT~6.1 Sol F9 acquisition. Read-only review
(\texttt{REVIEW.md} in Arm~C) had higher observed Gemini~4 Argon CR1 success
($76.2\%$ versus $38.1\%$), and Arm~E's audited
\texttt{REPAIR.json} rule achieved $95.2\%$ CR1 success on GPT~6.1 Sol.
The proposed combined step would be triggered when cyclomatic complexity
exceeds 10, duplication reaches $3\%$, lint is nonzero, a feature session
halts uncommitted, or periodically every three to five features. It would
include C's semantic-invariant checklist and an audited test-assertion
escape hatch. The commands and thresholds come from the report's proposal. The combined
policy's efficacy, stopping behaviour and cost have not been measured.

\subsection{Deferred intervention measures}\label{app:deferredmeasures}

Help and administrative records determine trajectory eligibility and cost,
but the study does not report intervention rates. The proposed measures
distinguish cycles receiving corrective help from cycles needing it. HG-DAgger and
ThriftyDAgger provide precedents for measuring intervention in other domains;
their findings do not establish effectiveness for coding agents
~\cite{kelly2019hgdagger,hoque2022thriftydagger}.

For run $r$, $n_r\geq1$ is attempted feature cycles, including assigned
post-steps and failed cycles, and $h_{rj}$ qualifying corrective-help
events. The full-run rates below require complete help annotation for every
attempted cycle. Otherwise report annotation coverage and leave the full-run
rate unknown. Observed human intervention rate would be
\begin{equation}
\mathrm{HIR}_r=\frac{1}{n_r}\sum_{j=1}^{n_r}
\mathbf{1}\{h_{rj}>0\}.
\label{eq:hir}
\end{equation}
HIR answers how often a cycle received corrective help, regardless of how
many events occurred or how long they took. Record event counts and active
human minutes separately. Let $d_{rj}\in\{0,1\}$ indicate an explicit
help request or a block needing corrective human action. The
intervention-demand rate (IDR) is the fraction of cycles requiring that help:
\begin{equation}
\mathrm{IDR}_r=\frac{1}{n_r}\sum_{j=1}^{n_r}d_{rj}.
\label{eq:intervention-demand}
\end{equation}
IDR counts cycles needing help, including requests that received none.
A timeout alone does not establish such a request; zero attempts gives
undefined rates. The experiment's eighteen GPT F9 sessions ending with
a request to extend the immutable \texttt{READ\_METHODS} test are
concrete demand candidates. Their recorded requests do not show that
anyone supplied the requested help; coding them would still require
cycle-level annotation and a complete denominator
(Section~\ref{app:executiondetail}).
A future annotation protocol would need to locate each event, cycle and
phase, check coverage and adjudicate unclear cases. Missing records leave
the rate unknown. Routine launches, planned approvals and safety stops are
not corrective rescue under these proposed definitions.

The original-case history provides qualitative steering examples: requests
for RPI tracking, migration, resource controls and functional checkpoints.
Within the retained coordinator session, 1,198 rows span the selected case
period. Of these, 1,066 user-message strings begin with the exact prefix
\texttt{[Scheduled prompt \#}; the other 132 rows have not been exhaustively
classified as manually authored corrective help. Repeated scheduled strings
occur among these records. The case evidence index retains the session,
date selection, turn locators and versioned launch prescriptions, together
with adverse assistant summaries of unresolved readiness and checkpoint
failures. Those summaries lack the underlying execution receipts.
Attempted-cycle denominators and complete help annotations remain
unavailable, so neither HIR nor IDR can be calculated. This case history is
separate from the completed experiment's owner-attested unassisted
trajectories.

\subsection{Historical freeze provenance}\label{app:freezeprovenance}

The reported case-tool snapshot is HVE Core \texttt{plugins-v3.3.106},
snapshot \sourceref{R29}, generated from source \sourceref{R30}.
These identifiers establish the reported tool source, but not its
installation date or use in individual sessions.

The historical v4 execution source, manifest-only revision, archive and
runtime image are \sourceref{R31}--\sourceref{R34}, respectively.
They identify the inputs to that historical version. The five-arm
amendment has its own source and approval history. The original registration at
\url{https://osf.io/ywsrv/} is owner-reported approved after submission
on 28 October; the inspected v4 prospective amendment flag was false
and its completion date null. New approval and actual prospective
recording cannot be inferred or backdated.
The later registered base at \url{https://osf.io/xcpt4} and current disclosed
unregistered execution are described in Section~\ref{sec:studyexecution}
and Data availability. They do not rewrite these historical freeze records.

\subsection{Measurement fields and provenance}\label{app:measurementprovenance}

The measurement definitions below come from public source \sourceref{R05}
in \url{https://github.com/jmservera/ama-thought-mve}. This is a source
inventory based on inspected code; private experimental data were not
remeasured. The collection path is
\nolinkurl{harness/runner.py} $\rightarrow$
\nolinkurl{harness/measure.py} $\rightarrow$
\nolinkurl{skeleton/tools/metrics_core.py}; pilot outcomes are derived in
\nolinkurl{harness/pilot_analysis.py}, with consumption parsing in
\nolinkurl{harness/analyze.py}.
In the inspected collection code, the numerical evaluator does not invoke an AI
judge. The separate post-run instrument below adds supplementary ratings
without changing that collection path.

\begin{table}[!htbp]
\small
\centering
\caption{Active raw fields, scales and scope. Missing measurements are not
zero. Structural counts and model-generated notes are not independent
efficacy outcomes.}\label{tab:measurementfields}

\begin{tabularx}{\linewidth}{@{}L{0.28\linewidth}Y@{}}
\toprule
Field or record & Definition and reporting limit \\
\midrule
\texttt{hidden}, \texttt{hidden\_passed}, \texttt{hidden\_total} &
Fixed test identities with passed/failed/error/skipped/timeout outcomes,
pass count and discovered denominator. Final coverage requires the exact
feature panel and complete binary endpoint results. \\
\texttt{open\_losses}, \texttt{density}, \texttt{episodes} &
Counts and fraction failing among previously acquired unchanged contracts;
onset, recovery, recurrence and uncertainty. Conservative current summaries
can be unavailable when acquisition or observations are incomplete. \\
\texttt{cr\_success}, \texttt{cr1\_cause} &
Composite success over CR1 and retained passing feature contracts, plus
recorded cause/disposition. An explicitly evidenced terminal agent failure
can resolve an otherwise unavailable success as false; this is not an
AI-generated quality grade. \\
\texttt{mean\_ccn}, \texttt{max\_ccn}, \texttt{share\_ccn\_over\_10} &
Unweighted mean and maximum per-function cyclomatic complexity; fraction
of source functions above 10, not percentage unless multiplied by 100.
Mean and fraction are rounded to three and four decimals respectively. \\
\texttt{dup\_pct} &
The clone detector's duplicated-line percentage\@. jscpd's line denominator
is not automatically the independently counted source NLOC denominator. \\
\texttt{lint\_warnings}, \texttt{lint\_per\_kloc} &
Diagnostic count and $1000\times\mathrm{warnings}/\mathrm{source\ NLOC}$;
the latter is undefined for an empty source denominator. No separate
suppression count is emitted. \\
\texttt{nloc\_src}, \texttt{nloc\_tests}, \texttt{files\_src}, \texttt{functions} &
Python files under source/test directories; lines excluding blank and
full-line comments, source-file and function counts. These are not
instrumented test-coverage measures. \\
\texttt{md\_words}, \texttt{test\_counts}, \texttt{diff}, \texttt{commits} &
Whitespace-delimited Markdown words excluding generated reports/dot
directories; AST test/assertion counts; Git additions/deletions by area
and post-step commits. Semantic quality requires examining those artifacts. \\
Usage and timing &
\texttt{totalNanoAiu}/$10^9$ supplies AI credits; input, output, cache-read
and cache-write token counts remain separate. \texttt{agent.seconds} is
host-observed session wall time; \texttt{totalApiDurationMs} is available
API duration, not necessarily provider-active time. Unknown or incomplete
usage cannot become zero-cost work. \\
Integrity and assistance &
Example-file hashes, protected-test changes, violations and heuristic leak
flags are recorded. Explicit annotation/evidence determines assisted,
unassisted or unknown status; no help is inferred from silence. HIR/IDR
remain excluded/deferred, not automatic zero-valued scores. \\
\bottomrule
\end{tabularx}
\end{table}

\paragraph{Tool settings and repeatability.}
The public host configuration requests Lizard, jscpd and Ruff; per-run
configuration and version records, not package presence alone, establish
the tools actually used. The declared pins are Lizard 1.24.3, Ruff 0.17.3
and jscpd 5.4.1. The code records backend names and supports explicitly
different internal approximations; these are not interchangeable instruments.
Lizard measures each Python source function\@. jscpd runs from an empty
configuration directory in Python mode with minimum 50 tokens and five
lines. Ruff runs with isolated configuration, fixed
\texttt{E,F,W,C90,B,SIM} selection, JSON output, no cache and
\texttt{--exit-zero}. The engine counts diagnostics; the exit code alone
cannot establish a clean-lint result. Sonnet B's 2.26 warnings per 1,000
source lines, for example, remain nonzero despite successful report
generation and the requested zero-warning target.

The invocation omits \texttt{--ignore-noqa}, leaving inline suppressions
effective. The dictionary contains neither a suppression count nor a
maintainability index. A tool's presence or a qualitative review statement
does not establish that either variable was measured.
With no detected functions the engine emits zero mean/maximum/share
summaries; these are empty-input sentinels, not measured zero complexity,
and the mean/share denominator is undefined.
The measurement engine, collector, pilot analyzer, failure-attribution
module and C prompt have identical source blobs in the three disclosed
implementation epochs inspected here. Other parts of the implementation
changed, and execution conditions could differ between versions.

\paragraph{From observations to summaries.}
The collector evaluates at feature and retained post-step checkpoints, and
again after CR1. A/D have no separate post-step; B/C/E retain their assigned
post-step. It writes outcomes, provenance, source HEAD, raw metrics,
integrity and diff records. The current final score is unavailable when
sequence/panel identities or binary endpoint observations are incomplete.
Section~\ref{sec:decaymeasurement} identifies the loss and CR1 summaries
that the inspected source did not yet export. The final-export disposition
in Section~\ref{app:reportingdisposition} distinguishes available retained
F/P summaries and descriptive prerequisite strata from still-unavailable
graded CR1 reporting.
Credits require complete consumption flags; all-attempt delivery/CR1 totals
need a matching attempt journal. Aggregate statistics and seeded resampling
operate on these records, not on AI opinions. Model-specific allocation,
missingness and observed runtime variation remain distinct from whether
the scoring algorithm is programmatic.

\subsection{Supplementary judging instrument and interpretation}\label{app:postrunjudging}

The 9 November proposal in \nolinkurl{judging/README.md}
(source \sourceref{R36}, file \sourceref{R37}) preceded the completed
post-collection assessment. That assessment used the v3 source and runtime
identified below. It is not a replacement for the development experiment. Preparation used a stopped local archive,
retained every allocated model/arm/replicate cell and selected recorded
measurement HEADs.
Missing $P_{10}$ or CR1 observations cannot be replaced by an earlier
successful checkpoint. The reported replay used read-only snapshots and
evaluator assets, with each scenario starting from a fresh in-memory
Workbench. The revision checked supplied records; it did not re-execute archived
candidates or independently qualify the image.

\begin{table}[!htbp]
\centering\small
\caption{Fixed supplementary workflow panel. Feature labels denote
dependencies, not attribution of a failure to a particular feature.}\label{tab:judgingpanel}

\begin{tabularx}{\linewidth}{@{}L{0.15\linewidth}Y@{}}
\toprule
Workflow & Exercised requirements \\
\midrule
F1 & Registration, combined filtering, atomic import and recovery \\
F2 & Lifecycle, invalid transitions, history and stage counts \\
F3 & Handoff, decline and acceptance with ownership preservation \\
F4 & Rework and reopening with ownership and reasons retained \\
F5 & Safety holds, permitted rework and role-controlled release \\
F6 & Consent withdrawal, reinstatement and safety precedence \\
F7 & Offered/confirmed appointments, verified escalation and reopening reset \\
F8 & Live contractor reads, protected fields and read-only enforcement \\
F9 & Stable priority ranking with interacting hold, consent and vulnerability reasons \\
F10 & Audit verification, tamper detection and visibility-preserving export \\
Journey & F1--F10 delivery/reopening with consent, handoff, appointments and audit \\
CR1 & Region-scoped reads, rankings, exports and approved cross-region handoff \\
\bottomrule
\end{tabularx}
\end{table}

\paragraph{Witnessed endpoint loss.}
For trajectory $r$ and endpoint $e\in\{P_{10},\mathrm{CR1}\}$, let
$\mathcal{W}_{r,e}$ contain panel scenarios witnessed passing at an
earlier eligible end-of-cycle checkpoint, and let $S_{r,e}(w)$ be the
observed binary endpoint outcome for scenario $w$. For this measure, the risk set consists of those previously passing
scenarios. When $|\mathcal{W}_{r,e}|>0$ and every member has an available
endpoint, the supplementary retention density is
\begin{equation}
d^{\mathrm{workflow}}_{r,e}=
\frac{\sum_{w\in\mathcal{W}_{r,e}}\mathbf{1}\{S_{r,e}(w)=0\}}
{|\mathcal{W}_{r,e}|}.
\label{eq:workflowloss}
\end{equation}
The numerator counts previously passing scenarios that fail at the
selected endpoint. Dividing by the number of earlier witnessed passes
gives a dimensionless fraction from zero to one. An empty risk set
or any unavailable risk-set endpoint leaves the density undefined.
The next-study reporting requirements in Section~\ref{app:futuredesign}
cover witnessed risk sets and unavailable outcomes.
Treatment can affect acquisition. A repository with few implemented
capabilities can lose little while still failing many final requirements.
This conditional loss density therefore cannot alone rank treatment efficacy
or replace the hidden-contract loss measure. A GPT repository that never
acquired ranking could have zero witnessed ranking loss and still fail
the final ranking workflow.

The instrument distinguishes retained and observed-loss scenarios,
endpoint passes without an earlier witness, complete histories with no
observed pass, incomplete histories with no observed pass, and unavailable
endpoints. F10 first becomes eligible at $P_{10}$, so a failure there
cannot establish earlier-capability loss; its $P_{10}$ pass can establish
a post-CR1 baseline. The full delivery journey likewise first becomes
eligible at $P_{10}$; CR1's region workflow first becomes eligible after
CR1. The maximum earlier-witness risk sets therefore contain nine scenarios
at $P_{10}$ and eleven after CR1. Post-CR1 loss uses any earlier eligible
$P$ witness, so it can include a failure already present at $P_{10}$.
An incremental $P_{10}$-to-CR1 loss would require a passing $P_{10}$
baseline and remains distinct from this endpoint measure.
A witnessed pass followed by an endpoint failure
supports endpoint loss despite missing intervening observations, while
leaving onset time unresolved. A final pass can conceal a recovered
regression. Comparing an endpoint with an earlier witness establishes neither
uninterrupted retention nor a failure's duration, recurrence or cause.

\paragraph{Ratings, blinding and calibration.}
Functional completeness, workflow coherence, error recovery and API
usability are separate ordinal dimensions: 0 absent/unusable; 1 major
blocking failures; 2 material gaps; 3 working with minor evidenced
limitations; 4 coherent fulfilment of exercised requirements.
Unsupported dimensions are null, not zero, and every rating requires a
rationale with citations to supplied final execution observations.
There is no composite software-quality score. A perfect completeness
rating is rejected when execution demonstrates a workflow failure.

Random packet identifiers and shuffled submission order conceal development
arm/model labels. Judges receive requirements, rubric, response shape
and execution evidence, excluding original run IDs, commit hashes,
source trees, treatment notes, transcripts and usage logs.
The inspected packet also includes an earlier passing witness and its
retention label. Historical and final observations reuse scenario-local
IDs; validation checks citation membership in final evidence, without
establishing which occurrence influenced a rationale. A citation to a
scenario-local ID can therefore identify a final observation even though
the judge also saw its earlier passing occurrence. Historical anchoring
and dimension-specific support therefore need human inspection.
Endpoint-only rating packets or checkpoint-qualified IDs would require
a separately versioned instrument; they are not claimed implemented here.
The separate judge adapter disables tool/configuration discovery and
checks its tool-free transport. Residual performance patterns and returned
values can still suggest treatment or model; the preparing researcher is
not automatically blinded. Scenario instrumentation is a software-behaviour
assessment, not proof against deliberate candidate tampering.

The original plan called for a blinded human-reviewed calibration sample
covering successful, incomplete and regressed workflows. The executed
sample and its limits are recorded below: no witnessed-loss packets were
available, the author review remained in progress and guidance changed
after initial AI scoring. The judge family overlaps the GPT development model; no between-model
judge comparison was run. Fresh repeated
sessions provide descriptive means and ranges, not established inter-rater
reliability. The completed assessment used one judge model and two repeats;
the proposal's three-repeat example remains historical. With one judge
model, repeat variation does not measure between-model agreement.

\paragraph{Aggregation and resources.}
A scenario/dimension mean requires all scheduled judge/repeat ratings.
A trajectory mean describes its available scenario means and retains
available/scheduled counts and a completeness flag. The ratings are ordinal. A 4 ranks above a 3, but that difference need not
equal the gap between 2 and 1. Means are descriptive summaries; equal
spacing has not been calibrated.
Arm differences require complete trajectory scores for both compared
arms in every allocated development model; otherwise the contrast is
unavailable rather than estimated from complete survivors. Complete-frame
contrasts equally weight within-model differences, as in
Equation~\ref{eq:primarycontrast}, with B--D the matched-ceiling package
comparison and B--A, B--C and E--B secondary. The tool supplies no
bootstrap intervals, hypothesis tests or formal verdicts. The 14 November
companion extraction adds exploratory intervals without changing that
strict instrument or its whole-frame gate. It resamples whole trajectories
independently within development-model/arm strata (10,000 draws, seed 2027),
equally weighting model differences. ``Complete pair'' refers to two arms with all allocated observations.
The bootstrap samples arms independently; it
does not match trajectories into statistical pairs. The reported intervals are not multiplicity-adjusted.
Repeated scenarios, snapshots and judge sessions remain nested within
trajectory; implementation epochs and shared scheduling/service effects
retain the unresolved interpretation limits in
Section~\ref{sec:protocollimits}. The inspected report does not export
dedicated block/epoch identifiers. For the completed collection, public
run IDs allow the published model/replicate and source-history records to
be joined; controller timing remains in the retained operator archive
(Section~\ref{app:executiondetail}). Those records identify scheduling groups and execution history.
Determining whether trajectories shared provider conditions, and how
their errors depend on one another, requires further evidence.

The original plan required recording judging source/scenario/rubric hashes,
runtime and model versions, repetitions, budget, missingness, calibration
and prior outcome exposure. The supplied completed-assessment record below
retains the available provenance and the later guidance change.
Local timestamps and hashes record when files were retained and identify
their contents. They leave preregistration status to be independently verified.
No preparation-cost measurement is reported; the formula counts judge calls
only. For $N$ complete trajectories with CR1,
$J$ judge configurations and $R$ repetitions, at most $2NJR$ sessions
are scheduled for the two endpoints; wholly unavailable specimens trigger
no calls. Paid judging needs separate explicit operator authorisation.
Recorded judge credits and unknown-charge reservations remain separate
from development/treatment costs. Admission reservations do not establish
a hard provider spending cap. Configuration and request hashes bind
resume; interrupted, invalid, failed or overspent calls block new admission
pending documented operator recovery rather than automatic replacement.
Per-call admission also does not establish full-frame affordability;
budget exhaustion can leave required comparisons unavailable. Calibration
costs require their own allowance and accounting.
Public reports require review and sanitisation; raw packets and mappings
remain private.

\begingroup
\raggedright
\paragraph{Executed post-collection assessment provenance.}
The completed supplementary evaluation (Section~\ref{sec:resultsjudging})
operated on the pre-recovery 313-trajectory study state (\sourceref{R38})
across 630 scheduled endpoint packets (626 evaluable, four unavailable from
two transport-quarantined trajectories). Judge configuration
\nolinkurl{gpt6-luna-judge-v3} (\nolinkurl{gpt-6-luna} at \texttt{medium} reasoning
effort, adapter \nolinkurl{sdk1.0.28-cli1.0.105-judge3}, $J=1$, $R=2$ fresh
repeats) executed under a 3,000-credit combined admission envelope:
Assessment~v1 (source \sourceref{R39}, ledger \sourceref{R40})
stopped on call~1 before prompt dispatch because of a CommonJS/ESM SDK
tool-filter mismatch, with zero settled and 30 reserved credits held.

Assessment~v2 (source \sourceref{R41}, ledger \sourceref{R42})
completed 15 calls (5.282763 settled credits, 30 reserved credits held and
0.3196095 diagnostic credits) before stopping on call~16 for a response
missing \nolinkurl{evidence_ids}.

Assessment~v3
(source \sourceref{R43}, ledger \sourceref{R44})
completed all 1,252 scheduled calls for 418.635894 settled AI credits and
zero held credits within the 2,934-credit remaining ceiling.

Human
calibration (\texttt{S01..S08}, 90 rows, 79 scorable rows across 16 distinct
evidence contexts) used \nolinkurl{human-calibration-guidance/v3} authored after
initial AI judging began, with \nolinkurl{review_status=in_progress} and
disclosed AI-assisted failure/mismatch triage.
\par
\endgroup

\FloatBarrier
\subsection{What the final exports can establish}\label{app:reportingdisposition}

The checks below use only the public
\nolinkurl{runs.csv}, \nolinkurl{retention.csv}, \nolinkurl{episodes.csv}
and \nolinkurl{attempts.csv} under
\nolinkurl{results/pilot-estimation/} at source \sourceref{R17}.
Their SHA-256 digests and derived counts are retained in
\nolinkurl{anc/publication-evidence-2026-11-24.json}. No candidate code, hidden
tests or judge was rerun, and no score, sample size or interval was changed.

\paragraph{PF1: retained feature histories.}
The export contains 6,300 rows: ten F and ten P checkpoints for each of
315 trajectories. All rows declare complete risk-set histories and binary
observations; 630 first-cycle F/P rows have empty prior-pass risk sets and
undefined loss density. A/D rows repeat the same retained state at F/P.
The independent-unit assumption applies to trajectories; checkpoints
remain repeated observations within them.
The conservative analyzer therefore supplies complete retained-state
counts for this dataset. Its absent observed-subset/missing-count output
does not conceal missing feature observations here; it remains a reporting
limitation for future incomplete histories.

\begin{table}[!htbp]
\centering\small
\caption{Retained loss reconciliation. Each arm has 630 F and 630 P rows.
Loss-checkpoints sum failing previously acquired contracts over time,
counting a persisting failure again at each observed checkpoint.}\label{tab:retentionreconciliation}

\begin{tabular*}{\textwidth}{@{\extracolsep{\fill}}lrrrr@{}}
\toprule
Arm & F loss-checkpoints & P loss-checkpoints & Affected trajectories & Episodes \\
\midrule
A & 0 & 0 & 0 & 0 \\
B & 0 & 0 & 0 & 0 \\
C & 0 & 0 & 0 & 0 \\
D & 0 & 0 & 0 & 0 \\
E & 3 & 4 & 1 & 1 \\
\bottomrule
\end{tabular*}
\end{table}

The single episode is the paused-case consent contract in Gemini E-07.
It first fails at $P_7$, remains open through $F_8/P_8$, $F_9/P_9$ and
$F_{10}/P_{10}$, and is right-censored without an observed recovery at
the end of feature delivery. Seven failed contract-checkpoints thus describe
one persisting episode. Rejected edits and failures inside sessions are
outside this retained-checkpoint series. The existing progression figure
plots retained P means. Reconstructing every attempted edit would require
additional within-session evidence.

\paragraph{PF2: CR1 and its prerequisites.}
Table~\ref{tab:crprerequisites} joins the F9/F10 group counts at $P_{10}$
to the existing composite CR1 outcome. ``Complete'' requires every F9
and F10 contract to pass before CR1; the other group can still contain
partially implemented prerequisites. For GPT, the incomplete group lacks
both F9 and F10 completion; Sonnet's groups have different deficits.
Conditioning on these treatment-affected states is descriptive.
It does not repair the unequal preservation burden or estimate a causal
adaptation effect.

\begin{table}[!htbp]
\centering\small
\caption{CR1 successes/allocated trajectories by pre-CR1 F9/F10
completeness. Unknown outcomes remain in the denominator and are listed
separately. A dash denotes an empty subgroup, for which a success rate
is undefined.}\label{tab:crprerequisites}

\begin{tabular*}{\textwidth}{@{\extracolsep{\fill}}lcllr@{}}
\toprule
Model & Arm & Both complete & Not both complete & Unknown (latter group) \\
\midrule
Sonnet 5.5 & A & 12/12 & 9/9 & 0 \\
Sonnet 5.5 & B & 13/13 & 8/8 & 0 \\
Sonnet 5.5 & C & 13/13 & 8/8 & 0 \\
Sonnet 5.5 & D & 11/11 & 9/10 & 0 \\
Sonnet 5.5 & E & 17/17 & 3/4 & 0 \\
\midrule
Gemini 4 Argon & A & 6/21 & \textendash{} & 0 \\
Gemini 4 Argon & B & 8/21 & \textendash{} & 0 \\
Gemini 4 Argon & C & 16/21 & \textendash{} & 0 \\
Gemini 4 Argon & D & 7/21 & \textendash{} & 0 \\
Gemini 4 Argon & E & 11/21 & \textendash{} & 0 \\
\midrule
GPT 6.1 Sol & A & 5/5 & 14/16 & 0 \\
GPT 6.1 Sol & B & 12/12 & 5/9 & 1 \\
GPT 6.1 Sol & C & 8/8 & 9/13 & 1 \\
GPT 6.1 Sol & D & 6/6 & 13/15 & 0 \\
GPT 6.1 Sol & E & 12/12 & 8/9 & 0 \\
\bottomrule
\end{tabular*}
\end{table}

The two same-state-ranking test errors stay unknown; neither final coverage
nor a successful workflow scenario resolves their cause. The exports used
here do not supply a complete graded CR1 panel separating new contracts
passed from previously passing feature contracts lost. That richer endpoint,
complete attempt-level failure classification and independently reviewed
assistance coverage are deferred measurement work. The original owner
attestation and composite success bounds remain unchanged.

\paragraph{PF3: incomplete all-attempt costs.}
All 5,372 attempt rows remain included in the accounting frame.
Table~\ref{tab:heldbillingcells} shows why costs from fully billed runs
cannot represent every allocated run. Several held charges can belong to
the same trajectory;
their counts are not counts of missing repositories. The 19,200 reserved
credits do not cap the eventual invoice, and the 783.16186-credit partial
lower bound does not resolve it. We report the already settled subset
means with their existing denominators. All-allocation costs and
quality-matched efficiency remain unidentified. Final-coverage/cost scatterplots
retain this subset scope.

\begin{table}[!htbp]
\centering\small
\caption{Attempts with held billing by development model and arm,
including feature, post-step and CR1 attempts. Zero means no held attempt,
not zero spending.}\label{tab:heldbillingcells}

\begin{tabular*}{\textwidth}{@{\extracolsep{\fill}}lrrrrrr@{}}
\toprule
Model & A & B & C & D & E & Total \\
\midrule
Sonnet 5.5 & 0 & 0 & 0 & 0 & 1 & 1 \\
Gemini 4 Argon & 1 & 4 & 4 & 4 & 6 & 19 \\
GPT 6.1 Sol & 4 & 7 & 10 & 5 & 9 & 35 \\
\midrule
Total & 5 & 11 & 14 & 9 & 16 & 55 \\
\bottomrule
\end{tabular*}
\end{table}

No further experimental round, judge batch or retrospective registration
is used to fill these gaps. Future work needs a calibrated task with
headroom, explicit acquisition and adaptation denominators, retained and
rejected failure measures, aligned randomisation/inference, fully reconciled
costs and an independently reviewed rating instrument. Those are requirements for a later study, not gaps that the present
results can resolve.

\FloatBarrier
\subsection{Additional endpoint and resource statistics}\label{app:resultsdetail}

These tables retain distributional, size and judging detail from the
completed 315/315 empirical record (\nolinkurl{results/EXPERIMENT_RESULTS.md}).
They complement the separate coverage, CR1, structural and cost tables in
the main results. All fifteen cells have 21 allocated and 21 completed
trajectories (315/315 total).

\begin{table}[!htbp]
\centering\small
\caption{Additional $P_{10}$ coverage and retention statistics across all
21/21 completed trajectories per cell. Losses are mean open contract counts
at $P_{10}$; loss cycles sum those counts over $P_1$--$P_{10}$.}\label{tab:resultsdistribution}
\begin{tabular*}{\textwidth}{@{\extracolsep{\fill}}lcrrrr@{}}
\toprule
Model & Arm & Median (\%) & Minimum (\%) & Mean losses & Loss cycles \\
\midrule
Sonnet 5.5 & A & 100.00 & 96.85 & 0.000 & 0.000 \\
Sonnet 5.5 & B & 100.00 & 98.43 & 0.000 & 0.000 \\
Sonnet 5.5 & C & 100.00 & 97.64 & 0.000 & 0.000 \\
Sonnet 5.5 & D & 100.00 & 96.85 & 0.000 & 0.000 \\
Sonnet 5.5 & E & 100.00 & 97.64 & 0.000 & 0.000 \\
\midrule
Gemini 4 Argon & A & 99.21 & 96.06 & 0.000 & 0.000 \\
Gemini 4 Argon & B & 99.21 & 99.21 & 0.000 & 0.000 \\
Gemini 4 Argon & C & 100.00 & 96.85 & 0.000 & 0.000 \\
Gemini 4 Argon & D & 99.21 & 96.85 & 0.000 & 0.000 \\
Gemini 4 Argon & E & 99.21 & 98.43 & 0.048 & 0.190 \\
\midrule
GPT 6.1 Sol & A & 88.98 & 88.98 & 0.000 & 0.000 \\
GPT 6.1 Sol & B & 100.00 & 88.98 & 0.000 & 0.000 \\
GPT 6.1 Sol & C & 88.98 & 88.98 & 0.000 & 0.000 \\
GPT 6.1 Sol & D & 88.98 & 88.98 & 0.000 & 0.000 \\
GPT 6.1 Sol & E & 100.00 & 88.98 & 0.000 & 0.000 \\
\bottomrule
\end{tabular*}
\end{table}

The descriptive B--D coverage differences are $+0.26$, $+0.15$ and $+3.15$
percentage points for Sonnet, Gemini and GPT, respectively; B--A differences
are $+0.22$, $+0.34$ and $+3.67$; B--C differences are $+0.07$, $-0.19$ and
$+2.10$; E--B differences are $+0.07$, $+0.04$ and $0.00$. Their equal-model
averages are $+1.19$, $+1.41$, $+0.66$ and $+0.04$ points. The pooled
within-cell SD over all 315 completed endpoints is 3.1278 points with 300
degrees of freedom (3.1119 points with 298 degrees of freedom across the
313 pre-recovery endpoints). Within-model cell SD ranges are 0.47--0.94 for
Sonnet, 0.39--0.98 for Gemini and 4.81--5.59 for GPT\@. The pooled SD may inform future planning, but does not establish a common
cell variance or calculate observed power.

\begin{table}[!htbp]
\centering\small
\caption{Final-coverage contrasts (\texttt{H1}), in percentage points,
equally weighting the three models across all 63 versus 63 completed
unassisted trajectories per arm. Point differences and 95\% bootstrap
intervals use 10,000 stratified draws (seed 2027); with zero missing
endpoints, full-allocation bounds coincide with the point differences.}\label{tab:resultsh1contrasts}

\begin{tabular*}{\textwidth}{@{\extracolsep{\fill}}llrll@{}}
\toprule
Contrast & Observed & Gain & 95\% interval & Allocation bounds \\
\midrule
B -- D & 63/63 & +1.1874 & [+0.1000, +2.2247] & [+1.1874, +1.1874] \\
B -- A & 63/63 & +1.4123 & [+0.3250, +2.4497] & [+1.4123, +1.4123] \\
B -- C & 63/63 & +0.6624 & [-0.4499, +1.7498] & [+0.6624, +0.6624] \\
E -- B & 63/63 & +0.0375 & [-1.0749, +1.1374] & [+0.0375, +0.0375] \\
\bottomrule
\end{tabular*}
\end{table}

\begin{table}[!htbp]
\centering\small
\caption{Model-specific coverage contrasts (\texttt{H1}), in percentage
points, across all 21 versus 21 completed unassisted trajectories per arm.}\label{tab:resultsh1modelintervals}

\begin{tabular*}{\textwidth}{@{\extracolsep{\fill}}llrll@{}}
\toprule
Model & Contrast & Gain & 95\% interval & Allocation bounds \\
\midrule
Sonnet 5.5 & B -- D & +0.2625 & [-0.1500, +0.7499] & [+0.2625, +0.2625] \\
Sonnet 5.5 & B -- A & +0.2250 & [-0.1875, +0.7124] & [+0.2250, +0.2250] \\
Sonnet 5.5 & B -- C & +0.0750 & [-0.2250, +0.4124] & [+0.0750, +0.0750] \\
Sonnet 5.5 & E -- B & +0.0750 & [-0.2625, +0.3750] & [+0.0750, +0.0750] \\
\midrule
Gemini 4 Argon & B -- D & +0.1500 & [-0.1500, +0.5249] & [+0.1500, +0.1500] \\
Gemini 4 Argon & B -- A & +0.3375 & [-0.0750, +0.8249] & [+0.3375, +0.3375] \\
Gemini 4 Argon & B -- C & -0.1875 & [-0.4874, +0.1875] & [-0.1875, -0.1875] \\
Gemini 4 Argon & E -- B & +0.0375 & [-0.2250, +0.3000] & [+0.0375, +0.0375] \\
\midrule
GPT 6.1 Sol & B -- D & +3.1496 & [0.0000, +6.2992] & [+3.1496, +3.1496] \\
GPT 6.1 Sol & B -- A & +3.6745 & [+0.5249, +6.8241] & [+3.6745, +3.6745] \\
GPT 6.1 Sol & B -- C & +2.0997 & [-1.0499, +5.2493] & [+2.0997, +2.0997] \\
GPT 6.1 Sol & E -- B & 0.0000 & [-3.1496, +3.1496] & [0.0000, 0.0000] \\
\bottomrule
\end{tabular*}
\end{table}

Table~\ref{tab:resultsh1contrasts} reports all four pooled 63-versus-63
coverage contrasts now unlocked by complete 315/315 unassisted collection.
The pooled B--D and B--A intervals exclude zero, whereas B--C and E--B
span zero. Table~\ref{tab:resultsh1modelintervals} shows the heterogeneity:
the Sonnet and Gemini coverage intervals span zero for every contrast, and
GPT's B--D interval touches zero at its lower endpoint.

\begin{table}[!htbp]
\centering\small
\caption{Exact worst-/best-case identification bounds for within-model
and equal-model-weight CR1 differences (\texttt{H6}), rounded to two
decimals in percentage points. These bounds assign the two unadjudicated
GPT~6.1 Sol CR1 test errors (\texttt{B-08} and \texttt{C-14}) adversarially;
they are not sampling confidence intervals.}\label{tab:resultscrbounds}
\begin{tabular*}{\textwidth}{@{\extracolsep{\fill}}lll@{}}
\toprule
Model & Contrast & Difference bounds (points) \\
\midrule
Sonnet 5.5 & B -- D & [+4.76, +4.76] \\
Sonnet 5.5 & B -- A & [0.00, 0.00] \\
Sonnet 5.5 & B -- C & [0.00, 0.00] \\
Sonnet 5.5 & E -- B & [-4.76, -4.76] \\
\midrule
Gemini 4 Argon & B -- D & [+4.76, +4.76] \\
Gemini 4 Argon & B -- A & [+9.52, +9.52] \\
Gemini 4 Argon & B -- C & [-38.10, -38.10] \\
Gemini 4 Argon & E -- B & [+14.29, +14.29] \\
\midrule
GPT 6.1 Sol & B -- D & [-9.52, -4.76] \\
GPT 6.1 Sol & B -- A & [-9.52, -4.76] \\
GPT 6.1 Sol & B -- C & [-4.76, +4.76] \\
GPT 6.1 Sol & E -- B & [+9.52, +14.29] \\
\midrule
Equal-model & B -- D & [0.00, +1.59] \\
Equal-model & B -- A & [0.00, +1.59] \\
Equal-model & B -- C & [-14.29, -11.11] \\
Equal-model & E -- B & [+6.35, +7.94] \\
\bottomrule
\end{tabular*}
\end{table}

\begin{table}[!htbp]
\centering\small
\caption{CR1 success differences (\texttt{H6}) for cells with 21 observed
Boolean outcomes in both arms, in percentage points. Every row reports
registered bounds using 97.5\% Clopper--Pearson component intervals and
opposite endpoints, a post hoc Newcombe hybrid-score 95\% interval and
a post hoc two-sided Fisher exact $p$ value,
conditional on independent binomial outcomes. All comparisons are exploratory
and unadjusted for multiplicity.}\label{tab:resultscrintervals}

\begin{tabularx}{\textwidth}{@{}L{1.8cm}L{1.4cm}rYrl@{}}
\toprule
Model & Contrast & Gain & Registered bounds (CP 97.5\%, opposite-endpoint) & Fisher $p$ & 95\% interval \\
\midrule
Sonnet 5.5 & B -- D & +4.7619 & [-18.77, +26.77] & 1.00000 & [-11.1902, +22.6694] \\
Sonnet 5.5 & B -- A & 0.0000 & [-18.83, +18.83] & 1.00000 & [-15.4639, +15.4639] \\
Sonnet 5.5 & B -- C & 0.0000 & [-18.83, +18.83] & 1.00000 & [-15.4639, +15.4639] \\
Sonnet 5.5 & E -- B & -4.7619 & [-26.77, +18.77] & 1.00000 & [-22.6694, +11.1902] \\
\midrule
Gemini 4 Argon & B -- D & +4.7619 & [-43.88, +51.67] & 1.00000 & [-22.7012, +31.2676] \\
Gemini 4 Argon & B -- A & +9.5238 & [-39.16, +54.73] & 0.74421 & [-18.0105, +35.2120] \\
Gemini 4 Argon & B -- C & -38.0952 & [-77.05, +14.69] & 0.02778 & [-59.8800, -8.1786] \\
Gemini 4 Argon & E -- B & +14.2857 & [-37.24, +60.67] & 0.53590 & [-14.7411, +40.2154] \\
\bottomrule
\end{tabularx}
\end{table}

The two unadjudicated GPT~6.1 Sol CR1 test errors (\texttt{B-08} and
\texttt{C-14}) prevent complete Boolean-pair intervals for the GPT CR1
contrasts; their exact allocation bounds remain in
Table~\ref{tab:resultscrbounds}. All eight Sonnet and Gemini CR1 contrasts
now have complete 21-versus-21 Boolean intervals in
Table~\ref{tab:resultscrintervals}, whose broad widths caution against
treating single-task binomial differences as precise population estimates.

The handoff's exploratory inspection identified a recurring Sonnet F10
failure to record the expected audit target on
\nolinkurl{test_f10_hidden::test_h10_targets_and_actors}: 8 A, 7 B, 8 C,
9 D and 4 E trajectories. A single miss corresponds to 126/127 passing
tests (99.21\%); some trajectories had further misses, as the minimum
coverage column shows.

All GPT endpoints at 88.98\% passed 113/127 tests and missed the fourteen
F9 ranking tests. The reported earlier test used a fixed
\nolinkurl{READ_METHODS} list and expected \texttt{NotAllowed} outside it;
\nolinkurl{view.rank_cases()} conflicted with that expectation under the
immutable accepted-file rule. Gemini's recurring view failure appeared
in 57 trajectories (15 A, 13 B, 5 C, 14 D and 10 E), on
\nolinkurl{test_f08_hidden::test_h08_every_write_is_refused_on_a_view}
and later \nolinkurl{test_cr1_hidden::test_hcr1_user_views_are_read_only}.
The newly introduced write methods included \nolinkurl{define_region},
\nolinkurl{add_user} and \nolinkurl{approve_handoff}.
These explanations come from exploratory inspection in the source report.
The mechanisms were not separately randomised, and this revision adds no
new annotation audit.

\begin{table}[!htbp]
\centering\small
\caption{Additional size and complexity companions at $P_{10}$ across all
21/21 completed trajectories per cell. Complexity is the mean per function;
source lines exclude blank/full-comment lines, and Markdown words exclude
generated reports and dot directories.}\label{tab:resultssize}

\begin{tabular*}{\textwidth}{@{\extracolsep{\fill}}lcrrr@{}}
\toprule
Model & Arm & Mean complexity & Source lines & Markdown words \\
\midrule
Sonnet 5.5 & A & 2.79 & 439.6 & 106.0 \\
Sonnet 5.5 & B & 2.66 & 463.2 & 351.3 \\
Sonnet 5.5 & C & 2.78 & 449.5 & 378.8 \\
Sonnet 5.5 & D & 2.80 & 435.6 & 106.0 \\
Sonnet 5.5 & E & 2.79 & 438.0 & 106.0 \\
\midrule
Gemini 4 Argon & A & 2.68 & 600.6 & 106.0 \\
Gemini 4 Argon & B & 2.26 & 620.1 & 403.1 \\
Gemini 4 Argon & C & 1.78 & 778.5 & 371.7 \\
Gemini 4 Argon & D & 2.63 & 596.5 & 106.0 \\
Gemini 4 Argon & E & 2.74 & 605.1 & 106.0 \\
\midrule
GPT 6.1 Sol & A & 2.55 & 519.7 & 1747.1 \\
GPT 6.1 Sol & B & 2.49 & 557.0 & 2002.4 \\
GPT 6.1 Sol & C & 2.58 & 519.6 & 2028.9 \\
GPT 6.1 Sol & D & 2.56 & 520.2 & 1732.2 \\
GPT 6.1 Sol & E & 2.60 & 545.6 & 1859.1 \\
\bottomrule
\end{tabular*}
\end{table}

Sonnet and Gemini retained the skeleton's 106 Markdown words in A, D
and E\@; B/C added the design or review handover. GPT produced longer
documentation even without those post-steps. Word counts describe how much text was retained. Assessing its accuracy
or use by later sessions requires separate evidence. In particular, a
\texttt{REVIEW.md} count cannot establish that the next feature session
acted on a warning about write methods in a read-only view.
Completely billed B consolidation sessions averaged 5.82 credits on
Sonnet (210 sessions), 12.59 on Gemini (208) and 12.04 on GPT (208).
These settled step means do not include unknown charges or identify a
stable, quality-matched long-run premium for the economic sensitivity model.

\FloatBarrier
\subsection{Detailed post-collection ratings and calibration}\label{app:judgingresultsdetail}

Tables~\ref{tab:resultsjudgemeans} and~\ref{tab:resultsjudgemeanscr1}
retain available-scenario cell means and all-scenario pass counts.
Functional completeness, workflow coherence, error recovery and API
usability remain separate 0--4 ordinal dimensions. Averaging is descriptive;
some coherence/recovery dimensions are null after a crashed scenario.
The reported means therefore do not establish complete-frame contrasts
for those dimensions. Every B--C and E--B complete-frame contrast is
unavailable because of the incomplete C/E trajectories; coherence and
recovery complete-scenario contrasts are unavailable even for B--A and B--D.

\paragraph{Companion estimates and their scope.}
The separate \nolinkurl{extracted-complete-pair-and-judging-contrasts/v1}
export supplements, rather than replaces, those strict contrasts.
Table~\ref{tab:judgeobservedcontrasts} compares each trajectory's mean over
its available scenarios. It reports trajectories with at least one
unrated scenario dimension alongside the difference and interval.
Every compared arm has 63 observed trajectories, but its rated scenario
mix can differ. The available-scenario comparison answers a different question from a
comparison covering every scheduled scenario. Its scenario mix can differ
between treatments, and counting abstentions does not recover missing ratings.

\begin{table}[!htbp]
\centering\small
\caption{Observed-scenario coherence and recovery differences (0--4)
for complete trajectory pairs. All arms have 63 allocated and observed
trajectories. NA counts are trajectories with any unrated scenario
dimension, focal/comparator, not wholly missing trajectories. The 95\%
bootstrap intervals resample trajectories, not individual ratings.}\label{tab:judgeobservedcontrasts}

\begin{tabular*}{\textwidth}{@{\extracolsep{\fill}}lllrlr@{}}
\toprule
Endpoint & Contrast & Dimension & Difference & 95\% interval & NA F/C \\
\midrule
$P_{10}$ & B -- D & Coherence & +0.0180 & [-0.0327, +0.0663] & 2/8 \\
$P_{10}$ & B -- D & Recovery & +0.0039 & [-0.0234, +0.0307] & 9/16 \\
$P_{10}$ & B -- A & Coherence & +0.0559 & [+0.0019, +0.1060] & 2/4 \\
$P_{10}$ & B -- A & Recovery & +0.0068 & [-0.0213, +0.0355] & 9/17 \\
\midrule
CR1 & B -- D & Coherence & +0.0376 & [+0.0016, +0.0728] & 2/1 \\
CR1 & B -- D & Recovery & -0.0112 & [-0.0265, +0.0022] & 8/16 \\
CR1 & B -- A & Coherence & +0.0425 & [+0.0073, +0.0767] & 2/0 \\
CR1 & B -- A & Recovery & -0.0069 & [-0.0227, +0.0079] & 8/18 \\
\bottomrule
\end{tabular*}
\end{table}

Tables~\ref{tab:judgepartialp10} and~\ref{tab:judgepartialcr1} instead
retain the C/E missing trajectories. They calculate the lowest and highest possible contrasts by allowing
each wholly missing trajectory score to range from 0 to 4, with 21
allocations per model/arm and equal model weights. For coherence and recovery, this bounds only
missing trajectories' available-scenario summaries, not the unrated
dimensions within observed trajectories. Positive completeness/API B--C
bounds describe this finite allocation; they are not sampling confidence
intervals or evidence that the whole scheduled scenario panel was rated.
Conditional bootstrap intervals over the 62 observed C/E trajectories
are not substituted for the full-allocation bounds here.

\begin{table}[!htbp]
\centering\small
\caption{Partial-frame $P_{10}$ judge contrasts (0--4). B--C observes
63/62 trajectories and E--B 62/63, from 63 allocated to each arm.
Observed differences use available-scenario means. Bounds vary the wholly
missing trajectory; NA F/C counts partly unrated observed trajectories.}\label{tab:judgepartialp10}

\begin{tabular*}{\textwidth}{@{\extracolsep{\fill}}llrlr@{}}
\toprule
Contrast & Dimension & Observed difference & Allocation bounds & NA F/C \\
\midrule
B -- C & Completeness & +0.0442 & [+0.0382, +0.1017] & 0/0 \\
B -- C & Coherence & +0.0288 & [+0.0240, +0.0875] & 2/5 \\
B -- C & Recovery & +0.0009 & [-0.0009, +0.0626] & 9/13 \\
B -- C & API & +0.0437 & [+0.0375, +0.1010] & 0/0 \\
\midrule
E -- B & Completeness & -0.0036 & [-0.0671, -0.0036] & 0/0 \\
E -- B & Coherence & +0.0024 & [-0.0611, +0.0024] & 2/2 \\
E -- B & Recovery & +0.0116 & [-0.0514, +0.0121] & 11/9 \\
E -- B & API & +0.0051 & [-0.0584, +0.0051] & 0/0 \\
\bottomrule
\end{tabular*}
\end{table}

\begin{table}[!htbp]
\centering\small
\caption{Partial-frame CR1 judge contrasts, with the same allocation,
scale, bounds and NA convention as Table~\ref{tab:judgepartialp10}.
Neither the bounds nor the observed-scenario differences establish
complete-scenario or population effects.}\label{tab:judgepartialcr1}

\begin{tabular*}{\textwidth}{@{\extracolsep{\fill}}llrlr@{}}
\toprule
Contrast & Dimension & Observed difference & Allocation bounds & NA F/C \\
\midrule
B -- C & Completeness & +0.0258 & [+0.0218, +0.0853] & 0/0 \\
B -- C & Coherence & +0.0278 & [+0.0242, +0.0877] & 2/1 \\
B -- C & Recovery & -0.0035 & [-0.0043, +0.0592] & 8/13 \\
B -- C & API & +0.0271 & [+0.0231, +0.0866] & 0/0 \\
\midrule
E -- B & Completeness & +0.0040 & [-0.0595, +0.0040] & 0/0 \\
E -- B & Coherence & +0.0001 & [-0.0634, +0.0001] & 1/2 \\
E -- B & Recovery & +0.0067 & [-0.0563, +0.0072] & 9/8 \\
E -- B & API & +0.0046 & [-0.0589, +0.0046] & 0/0 \\
\bottomrule
\end{tabular*}
\end{table}

\FloatBarrier

\begin{table}[!htbp]
\centering\footnotesize
\setlength{\tabcolsep}{4pt}
\caption{Descriptive $P_{10}$ judge means and workflow passes over evaluable
endpoints. Scores are functional completeness, coherence, recovery and
API usability. ``Pass'' counts endpoints passing all eleven workflows.}\label{tab:resultsjudgemeans}

\begin{tabular*}{\textwidth}{@{\extracolsep{\fill}}lcrrrrrr@{}}
\toprule
Model & Arm & Evaluable & Pass & Completeness & Coherence & Recovery & API \\
\midrule
Sonnet 5.5 & A & 21 & 20 & 3.985 & 3.991 & 3.972 & 3.987 \\
Sonnet 5.5 & B & 21 & 21 & 3.998 & 4.000 & 3.974 & 4.000 \\
Sonnet 5.5 & C & 21 & 21 & 4.000 & 4.000 & 3.974 & 4.000 \\
Sonnet 5.5 & D & 21 & 20 & 3.989 & 3.991 & 3.965 & 3.987 \\
Sonnet 5.5 & E & 21 & 21 & 4.000 & 4.000 & 3.961 & 4.000 \\
\midrule
Gemini 4 Argon & A & 21 & 20 & 3.991 & 3.994 & 3.948 & 3.991 \\
Gemini 4 Argon & B & 21 & 21 & 4.000 & 4.000 & 3.968 & 4.000 \\
Gemini 4 Argon & C & 21 & 21 & 4.000 & 4.000 & 3.965 & 4.000 \\
Gemini 4 Argon & D & 21 & 21 & 4.000 & 4.000 & 3.972 & 4.000 \\
Gemini 4 Argon & E & 20 & 20 & 4.000 & 4.000 & 3.968 & 4.000 \\
\midrule
GPT 6.1 Sol & A & 21 & 5 & 3.563 & 3.633 & 3.893 & 3.548 \\
GPT 6.1 Sol & B & 21 & 12 & 3.760 & 3.785 & 3.892 & 3.742 \\
GPT 6.1 Sol & C & 20 & 7 & 3.625 & 3.699 & 3.891 & 3.611 \\
GPT 6.1 Sol & D & 21 & 6 & 3.595 & 3.740 & 3.885 & 3.576 \\
GPT 6.1 Sol & E & 21 & 12 & 3.747 & 3.793 & 3.939 & 3.758 \\
\bottomrule
\end{tabular*}
\end{table}

\begin{table}[!htbp]
\centering\footnotesize
\setlength{\tabcolsep}{4pt}
\caption{Descriptive post-CR1 judge means and workflow passes. ``Pass''
counts endpoints passing all twelve workflows, a different instrument
from the hidden-contract CR1 composite in the main results.}\label{tab:resultsjudgemeanscr1}

\begin{tabular*}{\textwidth}{@{\extracolsep{\fill}}lcrrrrrr@{}}
\toprule
Model & Arm & Evaluable & Pass & Completeness & Coherence & Recovery & API \\
\midrule
Sonnet 5.5 & A & 21 & 20 & 3.990 & 3.990 & 3.976 & 3.990 \\
Sonnet 5.5 & B & 21 & 21 & 4.000 & 4.000 & 3.966 & 4.000 \\
Sonnet 5.5 & C & 21 & 21 & 4.000 & 4.000 & 3.970 & 4.000 \\
Sonnet 5.5 & D & 21 & 20 & 3.984 & 3.986 & 3.962 & 3.984 \\
Sonnet 5.5 & E & 21 & 20 & 3.992 & 3.992 & 3.964 & 3.992 \\
\midrule
Gemini 4 Argon & A & 21 & 20 & 3.992 & 3.992 & 3.958 & 3.992 \\
Gemini 4 Argon & B & 21 & 21 & 4.000 & 4.000 & 3.968 & 4.000 \\
Gemini 4 Argon & C & 21 & 21 & 4.000 & 4.000 & 3.976 & 4.000 \\
Gemini 4 Argon & D & 21 & 21 & 4.000 & 4.000 & 3.974 & 4.000 \\
Gemini 4 Argon & E & 20 & 20 & 4.000 & 4.000 & 3.969 & 4.000 \\
\midrule
GPT 6.1 Sol & A & 21 & 5 & 3.740 & 3.746 & 3.976 & 3.738 \\
GPT 6.1 Sol & B & 21 & 12 & 3.829 & 3.856 & 3.955 & 3.831 \\
GPT 6.1 Sol & C & 20 & 7 & 3.752 & 3.772 & 3.954 & 3.750 \\
GPT 6.1 Sol & D & 21 & 6 & 3.740 & 3.757 & 3.987 & 3.740 \\
GPT 6.1 Sol & E & 21 & 12 & 3.849 & 3.864 & 3.977 & 3.853 \\
\bottomrule
\end{tabular*}
\end{table}

\FloatBarrier
Gemini's high workflow ratings do not resolve its hidden permission-test
failures. The workflow's \nolinkurl{privacy:3} rejected
\nolinkurl{view.open_case}; it did not test the newly added
\nolinkurl{view.define_region} or \nolinkurl{view.approve_handoff} paths.
Sonnet E had one endpoint fail the newly eligible CR1 regions workflow
(20/21 all-pass). Such a failure has no earlier eligible passing witness
and is distinct from loss of an acquired scenario.

\paragraph{Human spot-check and repeated contexts.}
Eight anonymised packets (\texttt{S01..S08}) were selected before AI scoring:
six $P_{10}$ packets with eleven scenarios and two CR1 packets with twelve,
giving 90 rows. Four packets came from the incomplete/failing stratum and
four from the passing stratum. No witnessed-loss packets were available.
S01's eleven rows were unavailable, leaving 79 scorable rows and 158
human-versus-AI repeat comparisons per dimension. Sixty-nine rows passed
every check cleanly. The rows represented sixteen distinct full scoring
contexts; S02/S03 shared final evidence, as did S05/S06/S08.

Servera Bondroit's \nolinkurl{human-calibration-guidance/v3} record remains
\nolinkurl{review_status=in_progress} and
\nolinkurl{review_completed_yes_no=no}. It discloses AI-assisted selection
of failure/mismatch cases rather than individual inspection of every
passing row. Guidance v3 followed an AI usability dry run and the start
of initial AI scoring. The agreement figures describe an incomplete, AI-assisted spot-check,
not independent inter-rater reliability.

\begin{table}[!htbp]
\centering\footnotesize
\setlength{\tabcolsep}{4pt}
\caption{Human-versus-AI agreement (\%). Each dimension has 158 scorable
repeat comparisons over 79 rows; context weighting uses sixteen distinct
contexts. Both sides abstained on all eleven unavailable rows. The last
column counts null-versus-numeric disagreements on scorable rows.}\label{tab:resultscalibration}

\begin{tabular*}{\textwidth}{@{\extracolsep{\fill}}lrrrrr@{}}
\toprule
Dimension & \shortstack{Raw\\exact} & \shortstack{Raw\\within 1} & \shortstack{Context\\exact} & \shortstack{Context\\within 1} & \shortstack{Null vs.\\numeric} \\
\midrule
Completeness & 98.7 & 100.0 & 96.9 & 100.0 & 0 \\
Coherence & 97.5 & 98.7 & 93.8 & 96.9 & 2 \\
Recovery & 94.3 & 98.7 & 87.5 & 95.3 & 2 \\
API usability & 98.1 & 100.0 & 92.2 & 100.0 & 0 \\
\bottomrule
\end{tabular*}
\end{table}

Across 632 dimension-repeat comparisons there were fourteen one-point
numeric differences, four null-versus-numeric differences and no numeric
differences of two or more points. One boundary concerned whether a final
\texttt{ValidationError} rejection in \nolinkurl{ranking:3}, without a later
recovery step, justified recovery rating 3 or 4. Other differences concerned
abstention after an uncaught \texttt{NotAllowed} or \texttt{AttributeError}.

\FloatBarrier
\subsection{Task, collection and accounting record}\label{app:executiondetail}

The S1 task was the standard-library Python package \texttt{workbench}
in \nolinkurl{suites/s1-retrofit}. Each trajectory began with the same
skeleton: \texttt{Makefile}, \nolinkurl{pyproject.toml} requesting maximum
complexity 10, \texttt{./tcr} and \nolinkurl{tools/quality.py}.
Features covered registration/atomic CSV import (F1), the six-stage
lifecycle (F2), single-owner handoff (F3), rework/reopening (F4), hazard
holds (F5), consent with hold precedence (F6), appointments/no-access
escalation (F7), role-redacted read-only views (F8), priority ranking (F9)
and SHA-256-chained audit/export (F10). CR1 added postcode-prefix regions,
region-scoped users and manager-approved cross-region handoffs.

The handoff records SDK 1.0.28, CLI 1.0.105 (protocol 3), medium reasoning
effort, disabled configuration/skill/MCP discovery, excluded built-in/MCP
tools and the single \nolinkurl{isolated_exec} tool. The reported runtime
image is identified by \sourceref{R45} in the end register.
The external evaluator supplied 127 feature tests and fifteen new CR1
tests, alongside Lizard 1.24.3, jscpd 5.4.1 and Ruff 0.17.3
(\texttt{--isolated --select E,F,W,C90,B,SIM}). These are supplied source
and execution records, not newly reproduced runtime qualification.

The 126 A/D trajectories scheduled eleven steps each; the 189 B/C/E
trajectories scheduled twenty-one, totalling 5,355. At the 13--14 November
snapshot, the \nolinkurl{study-resilience-policy/v1} supervisor quarantined
two trajectories after unfinished unknown-charge transport failures rather
than replaying uncertain state:
\nolinkurl{pilot-estimation-gemini-4-argon-E-15} had measured 10/21 steps
(eleven unattempted after the feature-six HTTP stream interruption) and
\nolinkurl{pilot-estimation-gpt-6.1-sol-C-20} had measured 18/21 steps
(three unattempted after the feature-ten WebSocket fallback interruption).
Both trajectories were subsequently completed via audited source-pinned
recovery transitions (\sourceref{R22}, \sourceref{R23};
\nolinkurl{E-15} experienced one further HTTP stream timeout on
\nolinkurl{f09-feature-1} at 16/21 steps before completing), yielding
all 315/315 completed trajectories and 5,355/5,355 measured steps while
retaining all historical attempt holds in the study ledger.

\paragraph{Adjudication and unknown outcomes.}
Read-only re-evaluation under \nolinkurl{analysis-adjudications/v1}, at
source \sourceref{R50}, reclassified
proven candidate exceptions from error to failure in 87 step records
(98 individual test outcomes). Original metrics were preserved and
development agents were not rerun. Two CR1 outcomes remain unknown:
\nolinkurl{gpt-6.1-sol-B-08} and \nolinkurl{gpt-6.1-sol-C-14} retained an
unadjudicated error on
\nolinkurl{test_f09_hidden::test_h09_same_state_same_ranking}.
Their unknown outcomes widen the GPT CR1 bounds by 4.76 percentage points
($1/21$).

\paragraph{Outstanding credits and in-band interaction.}
Across 5,372 total study attempts, 5,317 settled with complete billing
totaling 62,299.75904 AI credits (62,299.76; compared with 5,303 attempts
and 62,052.6069 credits at the 5,357-attempt 313/315 snapshot). The
55 unresolved attempts comprise 38 native-retry holds and 17 historical
recovery-interrupted holds, with zero remaining quarantine holds (versus
15 recovery holds and one quarantine hold among 54 holds at the 313/315
snapshot). They retain 19,200.0 reserved credits (18,900.0 prior to the
\nolinkurl{E-15-f09-feature-1} timeout hold) and a 783.16186-credit
certified partial lower bound. The final invoice remains incomplete, and no currency price is available.

The audit in
\nolinkurl{results/pilot-estimation-judging/intervention-log-audit-summary.json}
covers all 5,372 retained \nolinkurl{transport/events.jsonl} study-attempt
journals (5,387 session-start and final-message events in the directory
scan; 5,357 journals at the pre-recovery snapshot).
Disabled user-input/elicitation interfaces produced zero interactive RPC
events. Of 42,952 tool calls in the 5,372-journal audit (42,664 in the
report summary and 42,594 at the 5,357-journal snapshot), 42,878 (99.83\%;
42,590 and 42,520 in those earlier summaries) requested
\nolinkurl{isolated_exec}. All 74 other requests (53 \texttt{create},
21 \texttt{edit}, all Sonnet, with 50 entries retained in the summary array)
were rejected as nonexistent tools; none requested an interactive question tool.

Thirty-six sessions (0.67\%) across 29/315 trajectories (9.21\%) ended
with a permission question, leaving source unchanged in 35 cases.
GPT supplied eighteen such sessions out of 1,795 study attempts (1,805 in
the earlier directory count), all at F9, with source unchanged in all
eighteen, asking to extend an immutable earlier \nolinkurl{READ_METHODS}
test. Sonnet supplied eighteen out of 1,786 attempts (1,785 earlier;
thirteen consolidation, four review, one feature), seventeen leaving source
unchanged, concerning shell edits or \texttt{sh ./tcr} without built-in
editing tools or \nolinkurl{/usr/bin/env bash}. Gemini supplied none out of
1,791 attempts (1,767 before \nolinkurl{E-15} completion). Separately, 513
non-review measured steps left HEAD unchanged (509 pre-recovery). These
journal observations do not independently establish out-of-band intervention
coverage.

\paragraph{Owner attestation and updated report binding.}
Isabel Cabezas Martin supplied the unassisted-run attestation. She executed
and collected the experiment but did not conduct its analysis; as a
coauthor, she was not an independent auditor. This attribution was
confirmed by the corresponding author during the final editorial pass.
The final \nolinkurl{results/pilot-estimation/provenance.json} records
\nolinkurl{intervention-coverage/v1} as \texttt{owner-attested} at
\nolinkurl{2026-11-15T09:08:12.811511+00:00}: all 315 covered trajectories,
zero assisted, with manifest \sourceref{R27} scoped to final state \sourceref{R26}
(superseding the 14 November 313-trajectory attestation at
\nolinkurl{2026-11-14T02:48:20.351742+00:00}, manifest \sourceref{R46},
pre-recovery state \sourceref{R38}).
The published provenance links the attestations to study states.
Out-of-band intervention coverage and the private manifest's contents
have not been independently verified.

The 14 November report refresh (\sourceref{R14}, \sourceref{R47}) and
15 November completed report (\sourceref{R17}--\sourceref{R21}) retain
separate identities in the register. Final development results and the
earlier judging snapshot must be read on their respective observation frames.
Scheduling/epoch dependence, scenario abstention and multiplicity remain
interpretation limits.

\paragraph{Retained timing and version metadata.}
Read-only inspection of the operator archive on 14 November found
numeric start times, controller durations, attempt IDs and source commits
for all 5,357 attempts in the pre-recovery
\nolinkurl{.study/pilot-estimation/state.json} (\sourceref{R38}, matching
Assessment~v3), and the committed
\nolinkurl{results/pilot-estimation/attempts.csv} and
\nolinkurl{provenance.json} at \sourceref{R17} record all 5,372 attempts
and all fifteen source transitions through final state \sourceref{R26}.
These records support reconstruction of execution order and source epochs
without rerunning candidates; they contain no separate finish or
boot/launch IDs in the inspected attempt objects and do not establish
provider-service independence.

The retained v3 configuration records \nolinkurl{gpt-6-luna}, two repeats
and \nolinkurl{sdk1.0.28-cli1.0.105-judge3}; its instrument freeze and
replacement lineage agree on the previously reported study and judging
source commits. The human-calibration completion flag remains false.
Recorded SDK/CLI selectors and configurations are distinguished from
unexposed backend build identifiers. Starting a new session would produce
new observations, rather than recover an unlogged historical version.

\begin{table}[!htbp]
\centering\footnotesize
\caption{Execution sources across the completed 315/315 study. The 5,372
attempts include 5,355 returned steps and 17 recovery interruptions (zero
remaining infrastructure-failed or quarantine-held attempts; the 13 November
handoff through E14 had 5,341 returned steps and 5,357 attempts across
313 completed trajectories). At the historical 8 November checkpoint, E1--E3
had 651, 227 and 369 reported steps (1,247 total). Source identifiers do
not establish collection times or independent replicate counts. Record
links resolve to full identifiers in the end register.}\label{tab:executionepochs}

\begin{tabularx}{\textwidth}{@{}L{0.9cm}YL{1.6cm}L{1.6cm}@{}}
\toprule
Epoch & Execution source and operational correction & Returned steps & Total attempts \\
\midrule
E1 & \sourceref{R48}; pinned CLI runtime fix & 651 & 652 \\
E2 & \sourceref{R49}; source-bound estimation recovery & 227 & 228 \\
E3 & \sourceref{R05}; failed-request turn-charge binding (369 at 8 Nov) & 604 & 605 \\
E4 & \sourceref{R50}; sandbox-rejection recovery and audited evidence & 600 & 601 \\
E5 & \sourceref{R51}; correlated HTTP 503 same-turn retry handling & 207 & 208 \\
E6 & \sourceref{R52}; correlated WebSocket reset and recovery holds & 113 & 114 \\
E7 & \sourceref{R53}; usage-unknown WebSocket closure accounting & 124 & 125 \\
E8 & \sourceref{R54}; mixed-boundary retry history and provenance export & 165 & 166 \\
E9 & \sourceref{R55}; unaccounted native-retry source recovery & 31 & 33 \\
E10 & \sourceref{R56}; unbilled provider-reset continuation and paired halts & 143 & 145 \\
E11 & \sourceref{R57}; correlated Gemini 503 retries and halted charges & 608 & 609 \\
E12 & \sourceref{R58}; Gemini HTTP stream-timeout reporting and holds & 54 & 55 \\
E13 & \sourceref{R59}; audited bootstrap for retained study state & 1,388 & 1,390 \\
E14 & \sourceref{R07}; WebSocket fallback validation and halt quarantine & 426 & 426 \\
E15 & \sourceref{R22}; quarantine-completion recovery for E-15 and C-20 & 9 & 10 \\
E16 & \sourceref{R23}; 5,350-step recovery completing E-15 after F9 HTTP timeout & 5 & 5 \\
\midrule
Total & Sixteen execution sources; 315 allocated trajectories, 315 complete & 5,355 & 5,372 \\
\bottomrule
\end{tabularx}
\end{table}

\FloatBarrier
\subsection{Execution order and what remains unknown}\label{app:scheduling}

The 15 November read-only inspection used the public source and exports at
\sourceref{R17}. In \nolinkurl{harness/study.py}, lines 286--323 shuffle
model order within each replicate and arm order within each model using
the configured seed. Lines 2141--2180 dispatch eligible trajectories through
a worker pool, admitting new work as workers finish and deferring routes
subject to backoff. Thus the intended allocation interleaves arms, while
actual overlaps can cross scheduling groups.

The published \nolinkurl{runs.csv} contains all 315 trajectories in 63
model/replicate groups, each containing A--E. The 5,372-row
\nolinkurl{attempts.csv} retains source assignments across sixteen epochs,
but has no per-attempt start or finish timestamp columns.
The separate \nolinkurl{provenance.json} records parallelism changing from
ten to eight and controller-start metadata. CSV export order does not establish execution order. The earlier operator
inspection found per-attempt starts and durations in retained state;
reconciling those records with amendments and outcome access remains
necessary. Neither the scheduler nor the export grouping establishes
independent provider-service conditions or resolves the bootstrap choice.
For this paper the complete amendment/outcome-access chronology remains
unavailable; the old requirement is dispositioned as a limit on inference,
not reported as satisfied. Reconstructing it would require approved retained
operator records. A later study must record
decision times, outcome access and execution overlaps prospectively.

\subsection{Historical resource planning scenarios}\label{app:resources}

These scenarios retain the 2--5 November planning record. Their full
ten-feature calibration differs from the short calibration disclosed for
the unregistered execution. They are not measured costs, a current
affordability approval or evidence that historical launch gates were met.

Each study trajectory contains ten feature cycles and one CR1 session,
the latter capped at 45 minutes and 300 nominal credits in every arm.
A/D therefore have eleven sessions per trajectory; B/C/E have twenty-one.
A allows at most 5.75 agent-hours and 3,300 nominal credits per study
trajectory; B/C/D/E each allow 8.25 hours and 6,300 credits.
Across three models and five arms, the study has $15n$ trajectories,
$3n[2(11)+3(21)]=255n$ sessions,
$3n[5.75+4(8.25)]=116.25n$ maximum agent-hours and
$3n[3300+4(6300)]=85500n$ nominal session credits.

The historical excluded cost-only calibration scenario retains one trajectory per model per arm:
15 trajectories, ten features each and no CR1. A/D have ten sessions and
B/C/E twenty; A allows five hours and 3,000 credits, and B/C/D/E each
allow 7.5 hours and 6,000 credits. Thus calibration adds 240 sessions,
105 maximum agent-hours and 81,000 nominal credits, independent of $n$.

\begin{table}[!htbp]
\centering\small
\caption{Historical five-arm resource scenarios for the offered choices of $n$ per
model per arm; the historical default column is bold. These are summed nominal
session allowances and time ceilings, not expenditure or enlarged budgets.
Every choice exceeds the aggregate envelope at nominal session caps.}\label{tab:pilotresources}

\begin{tabularx}{\textwidth}{@{}Y>{\bfseries}rrr@{}}
\toprule
Quantity & $n=12$ (default) & $n=21$ & $n=33$ \\
\midrule
Study trajectories per arm, all models & 36 & 63 & 99 \\
Study trajectories, all arms & 180 & 315 & 495 \\
Study sessions, including CR1 & 3,060 & 5,355 & 8,415 \\
Study maximum agent-hours & 1,395.00 & 2,441.25 & 3,836.25 \\
Study nominal session credits & 1,026,000 & 1,795,500 & 2,821,500 \\
Excluded calibration trajectories, no CR1 & 15 & 15 & 15 \\
Excluded calibration sessions & 240 & 240 & 240 \\
Calibration maximum agent-hours & 105.00 & 105.00 & 105.00 \\
Calibration nominal session credits & 81,000 & 81,000 & 81,000 \\
Combined trajectories & 195 & 330 & 510 \\
Combined sessions & 3,300 & 5,595 & 8,655 \\
Combined maximum agent-hours & 1,500.00 & 2,546.25 & 3,941.25 \\
Combined nominal session credits & 1,107,000 & 1,876,500 & 2,902,500 \\
\bottomrule
\end{tabularx}
\end{table}

Combined nominal allowances are $85500n+81000$ credits and maximum
agent-hours are $116.25n+105$. At default $n=12$, 1,107,000 nominal credits
and 3,300 sessions including calibration do not authorize spending beyond
the unchanged 300,000-credit envelope (US\$3,000) or 270,000-credit admission ceiling.
Excluded calibration informs cost and time feasibility only, not outcome-
or variance-driven sample resizing. Its cost forecast must show that the
entire selected allocation, including CR1 and all-attempt liabilities, fits
remaining admission and the deadline after prior spend and unknown-usage
holds. Otherwise halt before study collection. A different allocation
requires a separately authorized prospective amendment and a new saved
study, never a changed allocation on resume or a silent reduction to fit
the budget. An interrupted or unaffordable study retains its selected
allocation and is reported incomplete.

\FloatBarrier
\subsection{Active AI review and consolidation prompt examples}\label{app:activeprompts}

These are the exact B/C templates checked in at the inspected active
source commit. They are not a new judge prompt or a protocol amendment.
The provider request also contains session context outside these templates. The runner removes HTML-comment lines and the first
Markdown heading, substitutes the configured post-step budget in minutes,
and saves the actual supplied text as the step's \texttt{prompt.md}.
The public configuration has a 900-second post-step budget, rendering
\textquotedblleft{}15 minutes\textquotedblright{}; this inspection does not verify every private run's rendered
prompt or provider/system context. Each step is a fresh session under its
assigned study model/effort. The model can reason qualitatively over the
visible report and diff, but hidden outcomes are measured outside it.

Arm B requests behaviour-preserving changes and structural stopping
targets. Arm C requests a 300-word issue-oriented \texttt{REVIEW.md}
without code/test changes. The host enforces the permitted-file and
retained-test rules; it does not parse either note into a numerical quality
score. Differences between review notes are variation in the treatment.
The notes were not used to measure reviewer reliability or calibrate the
outcome judge.
The template's ``optional placebo'' heading is retained verbatim.
Arm C supplies information that can affect later development, so it
cannot be assumed to be an inert placebo.
The historical v4 prompt in Section~\ref{app:prompt} remains a separate
unchanged source record.

\paragraph{Arm B\@: consolidation.}
Source \nolinkurl{harness/prompts/consolidation-step.prompt.md}
(\sourceref{R60}).
Display wrapping does not change the retained template bytes.

\begin{lstlisting}[basicstyle=\footnotesize\ttfamily,
  columns=fullflexible,keepspaces=true,showstringspaces=false,upquote=true,
  breaklines=true,breakatwhitespace=false]
<!-- markdownlint-disable-file -->
# Consolidation step prompt (arm B)

Do not add or change features. Your job is to leave the code better than you found it without changing its behaviour.

1. Run `make check`, then read reports/quality.md (functions above complexity 10, duplicated blocks, lint warnings, file sizes).
2. Review the last feature's diff (`git diff feature-start feature-end`) and the hotspots against this checklist:
   * duplicated logic;
   * long or complex functions;
   * unclear names;
   * dead code;
   * responsibilities in the wrong place, or rules from one feature leaking into another;
   * tests that check implementation details instead of behaviour.
3. Make small, behaviour-preserving changes. Apply every change through `./tcr "<message>"`: it runs all tests and commits if they pass, or reverts if they fail.
4. Never modify or delete files in tests/examples/ or any accepted Python test file (including imports, fixtures and constants). Add tests that pin current behaviour only in new files.
5. Rewrite DESIGN.md in at most 300 words: the current structure, where each business rule lives, and known debts. Replace its content; do not append.
6. Stop when the quality budget is met (no function above complexity 10, duplicated lines below 3%, zero warnings in the configured lint rules) or when your time budget ({consolidation_budget}) is used.
7. Finish with a summary of at most five lines: what you changed and what debt remains.
\end{lstlisting}

\noindent\begin{minipage}{\linewidth}
\paragraph{Arm C\@: read-only review.}
Source \nolinkurl{harness/prompts/review-only-step.prompt.md}
(\sourceref{R61}).
This is a qualitative treatment instruction, not an outcome-grading rubric.

\begin{lstlisting}[basicstyle=\footnotesize\ttfamily,
  columns=fullflexible,keepspaces=true,showstringspaces=false,upquote=true,
  breaklines=true,breakatwhitespace=false]
<!-- markdownlint-disable-file -->
# Review-only step prompt (optional placebo arm C)

Do not change any code or tests. Your job is to review what was just built.

1. Run `make check`, then read reports/quality.md.
2. Review the last feature's diff (`git diff feature-start feature-end`) and the hotspots against this checklist: duplicated logic; long or complex functions; unclear names; dead code; responsibilities in the wrong place; tests that check implementation details instead of behaviour.
3. Rewrite REVIEW.md in at most 300 words: the main issues you see and where they are. Replace its content; do not append. Do not modify any other file.
4. Time budget: {consolidation_budget}. Commit REVIEW.md and stop.
\end{lstlisting}
\end{minipage}\par

\subsection{Moved protocol and inspection detail}\label{app:editorialaudit}

The following passages preserve the earlier execution and inspection account
moved out of the main article during the final editorial pass. Dates,
version-specific statements and historical requirements describe their
original scope. The completed results and consolidated threats govern the
paper's present interpretation. Unfulfilled reporting instructions are
collected as requirements for the next study in
Section~\ref{app:futuredesign}.

\subsubsection{Registration and execution-version chronology}

Isabel Cabezas Martin conducted the experiment and collected its data
independently across the 315 allocated trajectories under the predefined
exploratory protocol and disclosed amendments. Juan Manuel Servera Bondroit
conceived and designed the study, defined the evaluation criteria, analysed
the results, conducted the human-calibration spot-check and prepared the
manuscript. This division of work was intended to reduce experimenter bias.
Both authors are Microsoft employees; execution was separate from design
and analysis, but not institutionally independent.

An earlier three-arm protocol was registered at \url{https://osf.io/ywsrv/overview},
and a 7 November 2026 OSF-registered base at
\url{https://osf.io/xcpt4}. The five-arm design and allocation options were
registered before collection; later model-stratified scheduling and mid-study resilience
recoveries (short calibration, parallel execution, WebSocket/HTTP retry
handling, VM restarts and quarantine-completion transitions) were publicly
disclosed in version control, culminating in initial handoff commit
\sourceref{R07} and final quarantine-recovery sources
\sourceref{R22} and \sourceref{R23} in the end register.
The completed study estimates exploratory differences under an evolving
protocol. The registered base covers only part of that execution history.
The public protocol leaves its formal hypothesis list empty and marks the
current amendment's registration fields false or null. Those fields concern
the amendment.

The active profile is \nolinkurl{pilot-estimation.json},
\texttt{model-stratified-v2-five-arm}, suite \texttt{s1-retrofit}.
GPT~6.1 Sol, Claude Sonnet~5.5 and Gemini~4 Argon use medium effort across
arms A--E, F1--F10 and held-out CR1, with seed 2027. The runtime selection
I supplied is $n=21$ trajectories per model per arm despite the
file default of 12: $3\times5\times21=315$ planned trajectories.
A trajectory is the sequence of work on one repository. Its feature steps,
sessions and completion counters describe work within that sequence and
cannot be counted as additional independent replicates.

Collection used sixteen successive execution-source versions.
The first three were inspected on 8 November;
the 13--14 November snapshot had 313 completed trajectories, and audited
recovery completed all 315. Table~\ref{tab:executionepochs} links each version
to its operational correction and step counts; the end register supplies
the full identifiers. Failures and unknowns remain in the record.
Implementation changes and scheduling can be confounded with time, arm
and model; source-commit times alone do not establish collection times.
The retained amendment records document decisions and source transitions,
but do not provide a complete decision-by-decision record of outcomes
already inspected when each change was made. That chronology is unavailable
in the public exports used here. We retain the resulting descriptive
comparisons without claiming prospective outcome-blind adaptation.

The 1 November five-arm amendment, 2 November allocation revision and
5 November v6 source findings below remain dated planning and inspection
records. Historical v4/v6/v7 freezes and diagnostics retain their own
scope. Earlier power/resource scenarios, including full ten-feature
calibration, are preserved unchanged; they do not establish power or
expenditure for the active F10/$n=21$ amendment.
The results below retain all 315 allocated trajectories, historical
quarantines, credit holds and execution-version boundaries.

\subsubsection{Historical implementation inspection}

The source-published v6 kit implemented all five arms and an immutable
replicate selector, superseding an earlier A/B/C-only capability account.
The source inspection, historical gate status and unchanged v4 freeze are
retained in the dated revision record and source register.

\subsubsection{Historical design diagram}

\begin{figure}[!htbp]
\centering
\includegraphics[width=\linewidth]{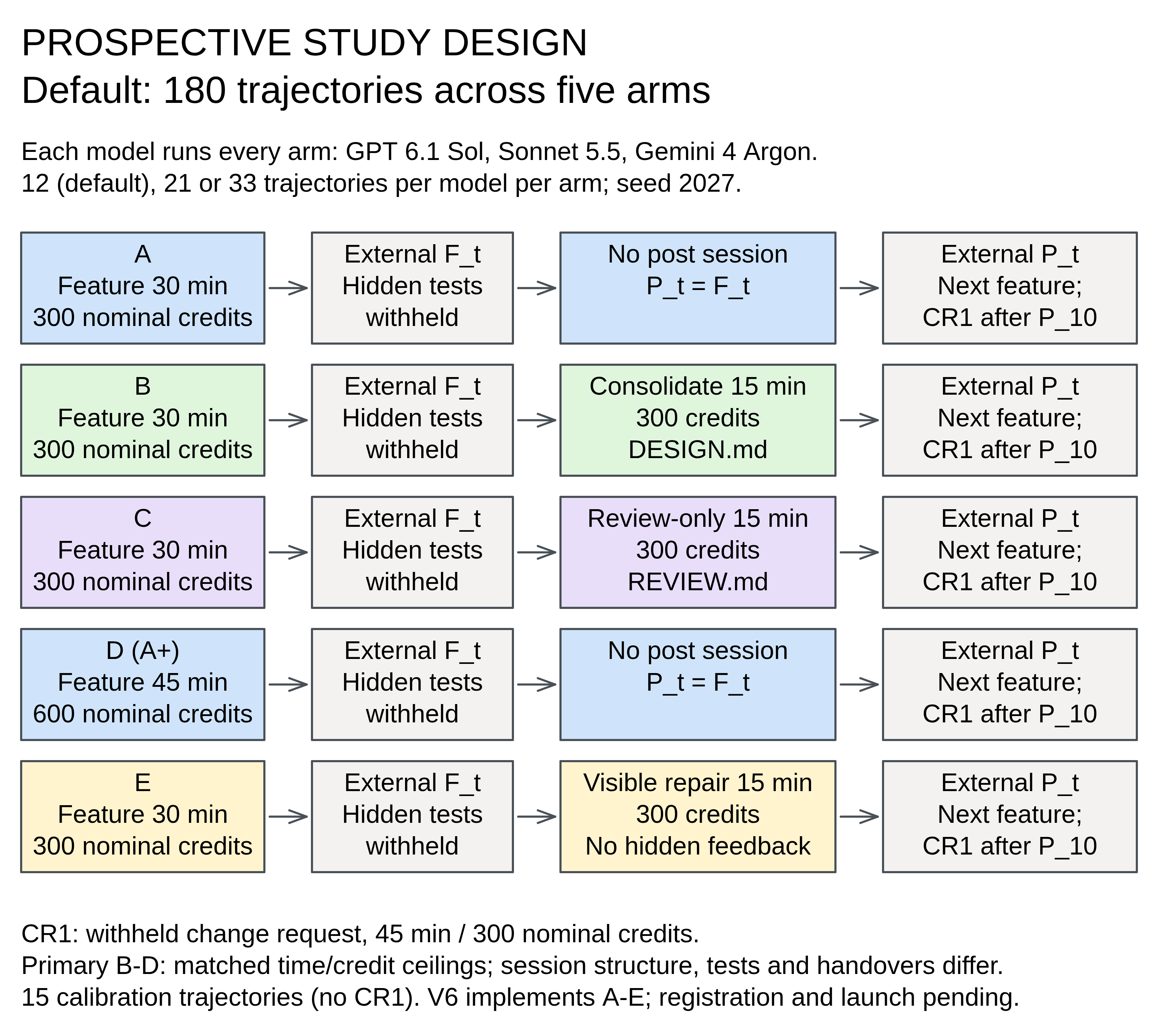}
\caption{Historical v6 design diagram. Its default allocation, full
calibration and pending-registration/launch labels describe that source
record; current $n=21$ execution and amendments are in
Section~\ref{sec:studyexecution}. Every model runs all five arms.
The external evaluator is separate from candidate tools. CR1 content
remains withheld until its session; privacy and isolation reduce exposure
without proving that a model has no prior knowledge.}\label{fig:pilotdesign}
\end{figure}

\subsubsection{Container capabilities and collection dispositions}

The broker's candidate-execution containers disable networking, use a
read-only root filesystem and set \texttt{no-new-privileges}, with CPU,
memory, process and time limits. Docker uses \texttt{--cap-drop=ALL}
but re-adds four for its protected supervisor; the candidate child runs
as UID/GID 1001 with cleared effective, permitted and inheritable
capabilities. Those four supervisor capabilities remain at the container level even
though they are cleared for the candidate child.
Development workspace mounts remain writable; replay snapshot/evaluator
mounts are read-only. Copilot authentication stays with the trusted host transport.
An isolation or collection failure leaves the checkpoint unavailable.
If collection succeeds and the candidate fails, that failure remains in
the evidence. It does not automatically make the checkpoint unavailable
or assign a zero judge rating. These boundaries describe the inspected
source. Their enforcement was not adversarially proven or verified across
every archived execution version.

\subsubsection{Retention analyser inspections and reporting status}

The v6 analyzer inspected on 5 November uses a stricter summary rule: incomplete prior
acquisition or any unobserved eligible result makes its loss count and
density unavailable, with completeness flags. Its standard export does
not emit the observed-subset counts, missing-count fields and bounds
specified above. Those are prospective reporting requirements retained
from the earlier design. The 8 November read-only source inspection
(\sourceref{R05}) confirms that the conservative rule remains in
\texttt{pilot\_analysis.retention}; it does not establish the planned
observed-subset export or amend the reporting rule.
The final public export permits a bounded reporting disposition
(Supplement Section~\ref{app:reportingdisposition}): all 6,300 F/P rows
declare complete risk sets and observations, so no partial-observation
bounds are needed for these retained feature checkpoints. This does not
implement or validate the richer pipeline for future incomplete histories.

\subsubsection{Graded adaptation reporting and prerequisite strata}

The earlier reporting plan also requested results stratified by whether
F9 and F10 were acquired before CR1. Less complete repositories have fewer
passing contracts to preserve. Treatment can affect prerequisite acquisition,
so conditioning on these strata introduces post-treatment selection.
One fixed CR1 is task-specific. A prewritten, randomly selected request
pool belongs to a separately registered subsequent study.
The summary inspected on 5 November computes composite CR1 success. The
active-source inspection on 8 November likewise finds composite
\texttt{cr\_success}, not a completed graded or F9/F10-stratified CR1
reporting pipeline. Supplement Section~\ref{app:reportingdisposition}
adds descriptive F9/F10 strata from existing exports. A complete graded
15-contract CR1 panel and prior-pass loss decomposition are unavailable in
those summaries and deferred to the deeper study; composite outcomes and
the two unknowns remain the reported endpoints. Retained raw vectors alone
do not establish a completed reporting pipeline.

\subsubsection{Accounting instruments, lint settings and telemetry}

AI-credit accounting must include every delivery and CR1 attempt, including
failures, assistance and infrastructure retries; phases are reported
separately. Lower spending is useful only when read alongside what the
agent delivered. Credits, tokens, elapsed
time and currency are distinct. Lizard measures complexity, Ruff lint
and jscpd duplication. The inspected Ruff invocation uses
\texttt{--isolated} with the fixed \texttt{E,F,W,C90,B,SIM} rule selection,
ignoring repository configuration files, but does not include
\texttt{--ignore-noqa}. Inline \texttt{\# noqa} can therefore suppress
diagnostics, and the active metric dictionary exports no separate suppression
count. The omitted flag limits the lint measurement's protection against
suppression, although the inspection did not establish that agents used it.
The earlier description overstated the safeguard. The clarification
changes no running instrumentation; a fix would require its own source,
epoch and amendment disclosure.
SDK request wall time and tool time are observable. Provider-active
time and rate-limit waits require explicit telemetry; absent telemetry
is not zero.

\subsubsection{Allocation choices and minimum useful effect}

The 2 November planning record offers $n\in\{12,21,33\}$ independent
repository trajectories per model per arm, with \textbf{12 as the default};
current execution selects 21. That record calls for the choice, rationale,
planning SD, power/precision trade-off, date and source-bound configuration
to be recorded before collection. It does not establish that the current
amendments were prospectively registered. Selecting a sample size does not
establish that its associated SD assumption is true.
The saved allocation is immutable on resume. Outcome-driven sample-size
changes, additions and replacements are prohibited; interim coverage,
loss, CR1 success or other outcomes cannot determine $n$.

Servera Bondroit selected a minimum useful absolute coverage gain of ten
percentage points, approximately thirteen test-weighted contracts.

\subsubsection{Conditional planning variance and sample-size calculation}

For balanced cells of $n_{\mathrm{cell}}$ independent trajectories and a
common within-cell variance $\sigma^2$, its planning variance is
$2\sigma^2/(M n_{\mathrm{cell}})$, with $n_{\mathrm{cell}}=n$.
For one two-sided comparison with $\alpha=0.05$, target power
$1-\beta=0.80$ and target difference $\Delta_*=10$ points, the normal
planning minimum, conditional on the assumed SD, is
\begin{equation}
n_{\mathrm{plan}}(\sigma)=
\left\lceil\frac{2{(z_{1-\alpha/2}+z_{1-\beta})}^2\sigma^2}
{M\Delta_*^2}\right\rceil.
\label{eq:samplesize}
\end{equation}

\begin{table}[!htbp]
\centering\small
\caption{Historical conditional normal planning for the ten-point B--D contrast,
three equally weighted models, $\alpha=0.05$ and target power 0.80.
The SDs are planning assumptions; power remains conditional on their
validity and the stated model.}\label{tab:replicationchoices}

\begin{tabular}{@{}rr@{}}
\toprule
Assumed within-cell SD (points) & Planning minimum per model per arm \\
\midrule
15 & \textbf{12 (default allocation)} \\
20 & 21 \\
25 & 33 \\
\bottomrule
\end{tabular}
\end{table}

Actual within-cell variability was unknown at planning. Target power
depends on independent trajectories,
the common variance, normal approximation and equal-model estimand; it is
not guaranteed by any offered choice. In particular, default $n=12$ is
\emph{not} an 80\%-power design under the existing SD of 25 points.
The bounded score's worst-case SD is 50, requiring 131 per cell under
the same approximation; that is outside the offered allocation choices.
These planning requirements and assumptions are retained from the earlier
record. Neither the registered base nor selection of $n=21$ establishes
their validity for evolving-source exploratory data.

\subsubsection{Analysis availability, dependence and historical reporting requirements}

The inspected v6 code uses independent arm/model draws; it
does not preserve joint scheduling-block draws.
The inspected primary analysis requires metadata for every allocated
trajectory across the five arms and all models, known unassisted status
and observed final coverage for the contributing B/D trajectories.
Otherwise it reports the primary estimate unavailable. All allocations and
costs remain in the record; a secondary summary of only observed models or
complete cases cannot replace the primary estimate. An earlier missing H2
checkpoint can coexist with an observed final coverage or CR1 endpoint.
This availability rule also needs explicit adoption or prospective
revision in the approved study record.
Model-specific estimates and other contrasts are secondary, with no
family-wide coverage claim.
 The earlier source-bound review
identified an unresolved choice between this independent-trajectory
variance/bootstrap scheme and model/replicate blocking.
Trajectories with the same scheduling label may share execution conditions,
but the label alone cannot establish their dependence or independence. Scheduling-only blocks with independent
inference require justification; block-aware analysis would require a
separately approved, disclosed choice and aligned planning variance.
Shared block/service effects remain unmeasured; the v6 source freeze
does not resolve this choice. We preserve the analysis actually performed
as a conditional exploratory summary. Its nominal 95\% coverage depends
on within-cell trajectory independence and does not account for
execution-epoch confounding or shared provider conditions. An interval
excluding zero is therefore not a confirmed causal or generalisable
treatment effect.
Public run identifiers allow the model/replicate scheduling groups and
source epochs to be reconstructed. Retained operator metadata also contain
start times and controller durations for every attempt (Supplement
Section~\ref{app:executiondetail}). Those intervals describe controller activity. They do not show whether
requests shared provider conditions or whether their errors were independent. The reported companion intervals retain the analysis actually
performed; a dependence-sensitive comparison would be a separately
disclosed post-collection analysis.
No dependence-sensitive reanalysis or additional experimental round is
included in this paper. The incomplete outcome-access chronology and
dependence justification are recorded limitations; a deeper study needs
prospective timing, source and exposure records and inference aligned
with its randomisation. The present estimand covers only these three
selected models, this task and this evolving execution.
Chen, Zaharia and Zou's dated ChatGPT comparison motivates monitoring
service drift despite exposed names~\cite{chen2023drift}; it does not
measure drift in this roster or make a version pin an immutability guarantee.
The completed pooled SD is reported in Supplement Section~\ref{app:resultsdetail}.

\subsubsection{Resource admission, operational readiness and information access}

The experiment distinguishes resources an agent could use from resources
it actually consumed. A larger allowance may leave spending unchanged
when a session exits early; a transport interruption may leave a charge
unknown even after the development work is complete. The results report
observations and unresolved liabilities separately. Historical planning
scenarios remain in Supplement Section~\ref{app:resources}, including
Table~\ref{tab:pilotresources}; they retain the earlier full-calibration
assumption and do not establish that historical launch gates were met.

The actual aggregate envelope remains 300,000 AI credits, with a
270,000-credit local admission ceiling: the controller must stop admitting
new work when spending and reserved liabilities reach that amount.
Prior spending and provisions for
unknown historical usage remain charged or held against remaining
admission; unknown usage is never zero or free headroom.
Original 300-credit sessions use a 270-credit soft limit; D's
600-credit feature uses 540. A retry shares the original allowance.
In-flight responses can overshoot; these are not provider billing caps.
At GitHub's published rate, 1 GitHub AI credit = US\$0.01~\cite{githubcopilotpricing}.
AI credits remain the primary measured unit. USD figures are list-price
equivalents at the published overage rate; actual out-of-pocket cost depends
on plan allowances.

The historical plan required the selected allocation, calibration and
all-attempt liabilities to fit the remaining envelope and deadline before
collection. It prohibited outcome-driven resizing or silent reductions to
fit the budget. Its detailed arithmetic remains in the Supplement.
The launcher inspected on 5 November enforces attempt and aggregate admission but does
not compute or approve that full-allocation forecast. A dated researcher
forecast and approval are required separately; a nonempty calibration
note is insufficient.
Proposed progression criteria are infrastructure failures at
most 5\%, no missing primary observations, complete billing/model/help
records, zero oracle contradictions in triplicate reference checks,
and no information-boundary failure. Hidden-test mutation assessment
must report its actual mutant denominator and unresolved equivalence.
These operational feasibility criteria cover infrastructure and measurement
readiness before full collection.

Only current and previous visible specifications and examples reach
candidate tools. Hidden tests, scores, identifiers, messages, reference
code and future CR1 remain external. Candidate commands use no-network
containers; the trusted host SDK's provider connection is different.
The earlier plan called for a researcher-managed protected workspace
without the original case repositories, manuscript or session-context
caches. Cabezas Martin's current execution role is separate from design
and manuscript preparation; coauthorship supplies no evidence of blindness
or actual access restrictions.
The earlier repository-privacy and later approved-release plans remain
historical intentions, not inspected current visibility.
OSF status and GitHub privacy are separate; neither removes prior public
or model exposure. The historical registered-freeze path and current
disclosed unregistered path have different registration status.
This manuscript update neither verifies operational approvals nor changes
information boundaries, budgets, configuration or launch permissions.

\subsubsection{Recovery completion, step counts and intervention coverage}

Isabel Cabezas Martin completed two previously transport-quarantined
trajectories through audited recovery (\sourceref{R22}, \sourceref{R23}),
preserving prior unknown-charge holds. They were
{\def\UrlBreaks{\do\-}\def\UrlBigBreaks{}\nolinkurl{pilot-estimation-gpt-6.1-sol-C-20}} and
{\def\UrlBreaks{\do\-}\def\UrlBigBreaks{}\nolinkurl{pilot-estimation-gemini-4-argon-E-15}}. Both finished with 100.0\%
(127/127) $P_{10}$ and 100.0\% (142/142) CR1 hidden pass rates.
The interruption history, task roster and runtime record are in Supplement
Section~\ref{app:executiondetail}.

Collection produced all 5,355 of 5,355 (100.0\%) scheduled steps across
5,372 attempts. The 17 extra attempts are historical recovery-interrupted
holds, so attempts, measured steps and independent repository trajectories
have different denominators. Execution spans sixteen source commits and
disclosed unregistered amendments to the registered base
(Section~\ref{sec:studyexecution}).

The updated provenance records owner-attested intervention coverage
(\nolinkurl{intervention-coverage/v1}) for all 315 completed trajectories,
with none recorded as assisted. This is an operator account of intervention,
corroborated by the 5,372-journal in-band SDK audit rather than an
independent audit of out-of-band edits or hints.

\subsubsection{Final billing-hold accounting}

Billing remains partly held for interrupted provider calls. Across 5,372
attempts, 5,317 settled with complete billing totaling 62,299.76 AI credits.
The remaining 55 attempts retain 19,200.0 reserved credits against unknown
final charges, with 783.16 credits
certified as a partial lower bound. Reservations are not measured spending,
and missing charges are not zero. The Supplement retains the exact amounts,
hold categories and failure-adjudication record; no currency conversion is made.

\subsubsection{Incomplete judging frames and scenario abstentions}

\FloatBarrier
In the frozen 626-packet Assessment~v3 archive, complete-frame judge
contrasts for B--C and E--B remain bounded over $[0,4]$ because C and E
each lacked one trajectory at the time of judging, and workflow-coherence
and error-recovery complete-scenario contrasts remain unavailable where
crashed ranking workflows left dimensions unrated. The companion bounds
those four unjudged packets and reports means over each trajectory's
available scenarios alongside counts of trajectories containing an
abstention (Arm~B cut error-recovery crash abstentions from 16/63 in D and
17/63 in A to 9/63 at $P_{10}$, and from 16/63 and 18/63 to 8/63 at CR1).
Bounds for unjudged packets do not bound unrated scenarios within an
observed trajectory; the Supplement keeps these summaries separate from
complete-frame comparisons.

\subsubsection{SDK interaction and permission-halt audit}

The SDK audit found no interactive user-input RPCs across all 5,372 retained
study-attempt journals, with those interfaces disabled. Of 42,952 tool
requests, 42,878 (99.83\%) used \texttt{isolated\_exec}; all 74 other
requests were rejected Sonnet \texttt{create}/\texttt{edit} attempts.
Thirty-six sessions (0.67\%) across 29 trajectories (9.21\%) ended with a
permission or environment question, leaving \texttt{src/} unchanged in
35/36 cases. The Supplement retains the per-model breakdown and distinguishes
the journal audit's invocation counts from the report summary and earlier
snapshot. These observations describe recorded requests; owner attestation,
including out-of-band intervention, remains a separate evidence source.

\subsubsection{Detailed instrument table}

Table~\ref{tab:measurementprovenance} lists the measurements in the active kit
and how each is produced.

\begin{table}[!htbp]
\small
\centering
\caption{Measurements in the active kit (\sourceref{R05}).
Programmatic scoring is distinct from model-generated treatment text.
Deterministic means repeatable for fixed source, inputs, tools and environment,
not identical across different generated repositories or runtime sessions.}\label{tab:measurementprovenance}

\begin{tabularx}{\linewidth}{@{}L{0.22\linewidth}Y L{0.23\linewidth}@{}}
\toprule
Measure and role & Values and instrument & Provenance \\
\midrule
Final coverage; primary & Hidden feature contracts passed out of 127 at
$P_{10}$, expressed on the 0--100 scale & Programmatic test oracles \\
Retained-contract loss; secondary & Previously acquired unchanged
contracts later failing; counts, density, episodes and accumulated test-cycles
& Checkpoint calculations; incomplete history can be unavailable \\
CR1 adaptation; secondary & All 15 CR1 contracts pass and previously
passing feature contracts remain satisfied & Programmatic tests and
documented disposition rules \\
Complexity; structural proxy & Lizard function-level cyclomatic complexity:
mean, maximum and share above 10 & Deterministic code analysis \\
Duplication; structural proxy & jscpd duplicated-line percentage,
Python mode, minimum 50 tokens and five lines & Deterministic clone detection \\
Lint; structural proxy & Ruff diagnostic counts and warnings per
1,000 source-code lines, frozen selected rules & Deterministic diagnostics;
inline suppression caveat below \\
Size and change; companions & Source/test code lines, files/functions,
Markdown words, test/assertion counts and Git additions/deletions & Deterministic
counts; not proof of code quality \\
Resources; cost companions & Recorded credits, input/output/cache tokens,
session seconds and available API-duration telemetry & Variable observations;
aggregation is programmatic, not AI judgement \\
Assistance and integrity; companions & Recorded help/dispositions,
protected-file hashes, violations and leak indicators & Explicit records and
checks; absent annotation is not zero \\
\bottomrule
\end{tabularx}
\end{table}

\begin{figure}[!htbp]
\centering
\includegraphics[width=\linewidth]{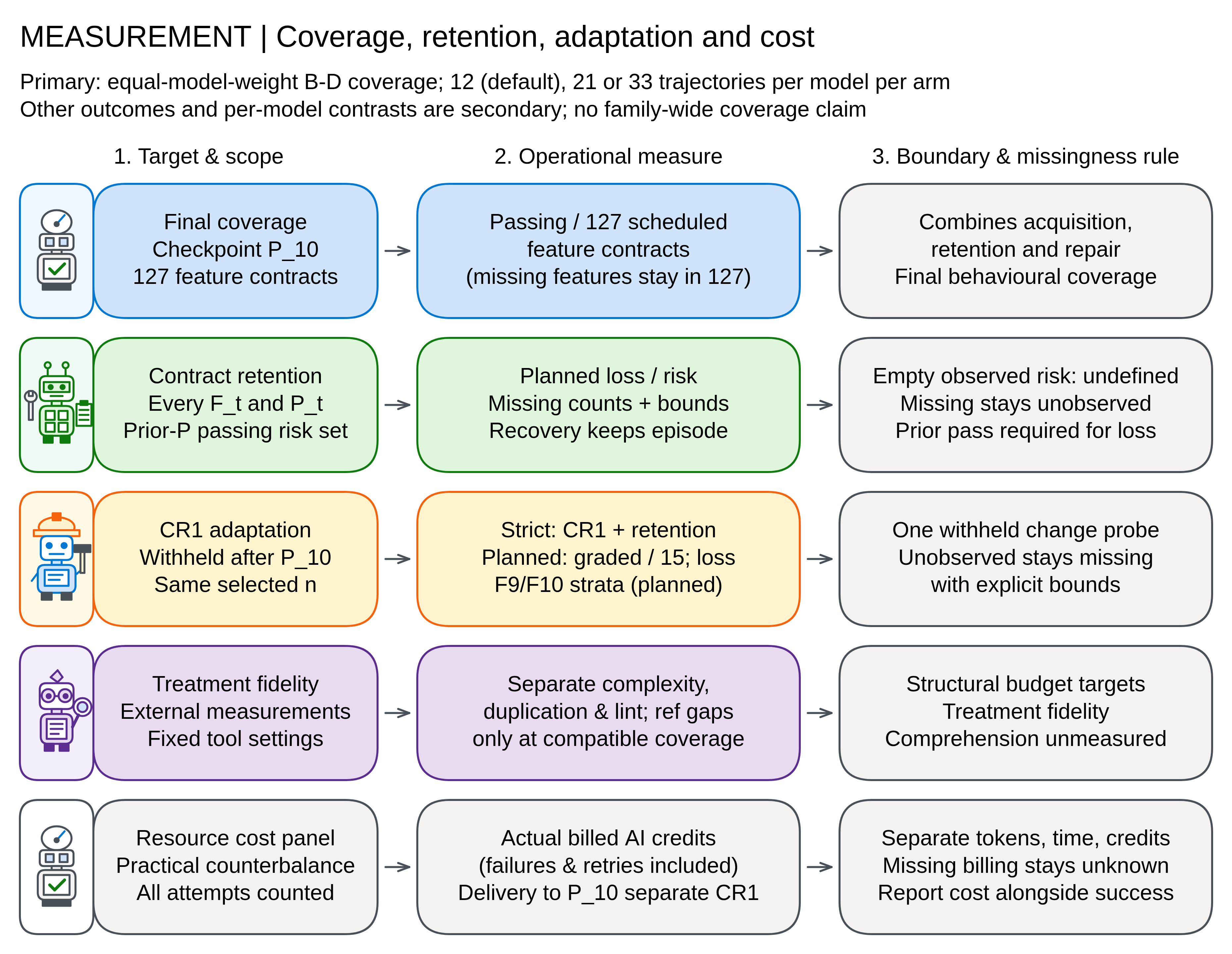}
\caption{Historical measurement design: coverage, retained behaviour, adaptation and all-attempt cost
are separate measures. Structural metrics targeted by B's stopping
budget are treatment-fidelity checks, not independent efficacy outcomes.
Unknown results and billing remain unknown. The observed-subset retention
and graded CR1 panels describe planned reporting requirements; their
implementation was unresolved in the 5 November v6 inspection.
The active-source inspection confirms conservative loss and composite CR1
summaries, not those planned richer exports.}\label{fig:experimentmeasures}
\end{figure}

\subsubsection{Historical deposit and access detail}

This account preceded confirmation that both GitHub repositories are public;
the current Data availability statement governs public and embargo status.

The repository analysed was publicly accessible during the case at
\url{https://github.com/jmservera/ama}. The original case and later follow-up
use separate snapshots (\sourceref{R01}, \sourceref{R02}). The source package
retains the case evidence index, repository observations and metrics, missing
execution links, analysis scripts and supplementary participant-observer context.
These records separate retained source from participant recollection,
and live issue states can change independently. Full source identifiers and
file digests for the core source records appear in
Table~\ref{tab:source-register} at the end of the Supplement. The case
evidence index supplies the recovered session/turn and instruction-version
locators. A digest identifies retained content. Claim support requires examining
that content.

The experiment source and sanitized outputs are at
\url{https://github.com/jmservera/ama-thought-mve}. Historical
\href{https://github.com/jmservera/ama-thought-mve/releases/tag/protocol-frozen-v4}
{v4},
\href{https://github.com/jmservera/ama-thought-mve/releases/tag/protocol-frozen-v6}
{v6} and
\href{https://github.com/jmservera/ama-thought-mve/releases/tag/protocol-frozen-v7}
{v7 releases retain separate protocol archives} and checksum files. Dated
source inspections (\sourceref{R03}--\sourceref{R06}) and the sixteen execution
versions (Table~\ref{tab:executionepochs}) are distinct from those releases.
The source's saved replicate selector permits 12, 21 or 33 trajectories per
model/arm, default 12, and the executed selection was 21. The earlier inspections
checked implementation identities. They did not check runtime-image qualification or
source-declared test totals. Their registration and collection
declarations, as current on the inspection dates, stay date-scoped in
\nolinkurl{anc/revision-report.md}.

The earlier three-arm protocol was registered at
\url{https://osf.io/ywsrv/overview}; I confirm the registered base at
\url{https://osf.io/xcpt4} on 7 November 2026. Both registrations are now
publicly accessible. The five-arm design and resilience recoveries were
publicly committed but unregistered. Source records \sourceref{R07}--\sourceref{R09}
identify the 12 November execution disclosure, and \sourceref{R22}--\sourceref{R27}
identify the final recovery and its state/attestation bindings. A disclosure in Git
cannot retroactively amend an OSF registration.

The completed report is \nolinkurl{results/EXPERIMENT_RESULTS.md}
(\sourceref{R17}--\sourceref{R20}), supported by
\nolinkurl{results/pilot-estimation/}. The judging outputs and companion
contrasts are in \nolinkurl{results/pilot-estimation-judging/}
(\sourceref{R21}). The earlier 13 and 14 November report versions
(\sourceref{R10}--\sourceref{R16}) are retained, because completing the development
dataset did not complete the earlier judging snapshot. The Supplement collects
the outcome, missingness, billing and execution details in one place.

The authors select the experiment GitHub repository as the deposit.
The \href{https://github.com/jmservera/ama-thought-mve/releases/tag/results-exploratory-s1-v1}
{results-exploratory-s1-v1 release} archives the 216 sanitized
files from source \sourceref{R17} with checksum and manifest sidecars
(excluding seven administrative tracking files and an endpoint DLP sidecar),
and Supplement Section~\ref{app:reportingdisposition} binds the public
analysis exports to their SHA-256 digests. No post-collection OSF deposit
was created: the existing OSF registrations link to the repository and
remain pre-collection records, while later amendments are disclosed in Git.
Raw packets, private specimen-to-trajectory mappings and runtime workspaces
stay off-Git in the evidence vault, with the intention of preserving future
reviewer blinding.
The public attempt and provenance exports retain actual arm/model/source
assignments, failures and unknowns. Changing execution versions and service
conditions limit homogeneous pooling. Current access and prior exposure
need their own records, and no independent replication is claimed.

\subsubsection{Detailed judge contrast intervals}

\begin{table}[!htbp]
\centering\small
\caption{Supplementary judge differences and companion 95\% bootstrap
intervals on the 0--4 scale (10,000 whole-trajectory draws, seed 2027),
equally weighting the three development models. B--D compares matched
allowances; B--A adds the consolidation allowance. Both compared arms must
have complete ratings for every allocated trajectory and scheduled scenario:
63 versus 63 at each endpoint. Intervals are exploratory and unadjusted.}\label{tab:resultsjudgecontrasts}

\begin{tabular*}{\textwidth}{@{\extracolsep{\fill}}llrr@{}}
\toprule
Endpoint & Contrast & Completeness (95\% CI) & API (95\% CI) \\
\midrule
$P_{10}$ & B -- D & $+0.0577\;[+0.0007,+0.1097]$ & $+0.0599\;[0.0000,+0.1154]$ \\
$P_{10}$ & B -- A & $+0.0729\;[+0.0173,+0.1248]$ & $+0.0722\;[+0.0137,+0.1263]$ \\
CR1 & B -- D & $+0.0351\;[-0.0046,+0.0741]$ & $+0.0357\;[-0.0040,+0.0747]$ \\
CR1 & B -- A & $+0.0357\;[-0.0013,+0.0728]$ & $+0.0370\;[-0.0007,+0.0747]$ \\
\bottomrule
\end{tabular*}
\end{table}

\subsubsection{Collection and estimability detail}

The final export records \texttt{primary\_estimates\_available=true};
the machine-readable field name is \nolinkurl{primary_estimates_available=true}.

The study completed all 315 of 315 (100.0\%) allocated repository trajectories, each starting from the same Python skeleton and receiving ten
features, followed by regional access control as a withheld extension.
Audited recovery completed the two interrupted trajectories without
discarding their charge holds. All 5,355 scheduled steps completed across 5,372 attempts, with the additional attempts representing recovery work rather than independent replicates.

With all 315 allocated trajectories complete and owner-attested unassisted,
the primary and secondary coverage estimates are available
and pooled comparisons use
63 versus 63 trajectories and within-model comparisons use 21 versus 21, with regression counts also complete. Two completed GPT trajectories retain an unadjudicated CR1 test error, whereas exact binomial intervals describe sampling uncertainty for fully observed
Sonnet and Gemini outcomes. GPT's identification bounds answer which success rates are possible if each unknown outcome is assigned first to failure and then to success, with both uncertainties identified in the detailed tables.

\FloatBarrier
\section{Source and digest register}\label{app:source-register}

Record links such as \sourceref{R17} identify the version used for a claim.
Git commit IDs identify revisions, tree IDs directory contents and blob IDs
individual Git objects. SHA-256 digests identify retained file bytes or an
image. Matching an identifier verifies content identity. Claim support and
independent validation require their own evidence. Unless another repository is named, records refer to
\nolinkurl{jmservera/ama-thought-mve}. Historical and final records remain
separate even when they concern the same file.

\begingroup
\footnotesize
\begin{longtable}{@{}L{0.7cm}L{6.2cm}L{8.1cm}@{}}
\caption{Exact source identifiers and digests. Each distinct identifier is
listed once; descriptions specify its object and revision scope.}\label{tab:source-register}
\\
\toprule
ID & Retained object and use & Identifier type and full value \\
\midrule
\endfirsthead
\multicolumn{3}{@{}l}{\tablename~\thetable\ continued: source and digest register}\\
\toprule
ID & Retained object and use & Identifier type and full value \\
\midrule
\endhead
\midrule
\multicolumn{3}{r@{}}{Continued on the next page}\\
\endfoot
\bottomrule
\endlastfoot
\phantomsection\label{prov:R01}R01 & Original case snapshot; \nolinkurl{jmservera/ama}. & Git commit: \nolinkurl{29c6b512f5e68733bb2cc1882343db39f5eb2cad} \\
\phantomsection\label{prov:R02}R02 & Separate 29 October case follow-up; \nolinkurl{jmservera/ama}. & Git commit: \nolinkurl{9b74ac8daaec2a880172e43161b37ac7f2c3f801} \\
\phantomsection\label{prov:R03}R03 & 5 November inspection: manifest-only child of v6; historical declarations, not current registration status. & Git commit: \nolinkurl{5ab24be4556876500a40e56201f6efc3930cc8b8} \\
\phantomsection\label{prov:R04}R04 & Published v6 implementation; five arms and immutable replicate selector. & Git commit: \nolinkurl{7de2f0d42955c62b52906747de480df2b3f1e0ca} \\
\phantomsection\label{prov:R05}R05 & 8 November active-source inspection; execution version E3 and active B/C prompt source. & Git commit: \nolinkurl{e71037d961cb80fc5c659437796de7736820f6c9} \\
\phantomsection\label{prov:R06}R06 & Directory tree of the 8 November inspected source. & Git tree: \nolinkurl{e7ec1561d4ecee921e7c9eba3711a1ab7e9f4058} \\
\phantomsection\label{prov:R07}R07 & Execution version E14; resilience disclosure on 12 November. & Git commit: \nolinkurl{a9d76bb6dbb118c43315fb2a13b7bcdb732d946d} \\
\phantomsection\label{prov:R08}R08 & E14 protocol/execution manifest. & SHA-256: \nolinkurl{0a7bc8fd073e6029ea6c573c1096a0e8dee5dbf7f602809eb3caeca1ef4f35f9} \\
\phantomsection\label{prov:R09}R09 & E14 execution inputs. & SHA-256: \nolinkurl{bc5d9ce04e5b255bf589d0068a902145bd25cfc384fe0110e2f9ad5ce4319305} \\
\phantomsection\label{prov:R10}R10 & Initial 13 November results integration; published main revision. & Git commit: \nolinkurl{7943e40347862e76e3bc04490040ad5904487c86} \\
\phantomsection\label{prov:R11}R11 & Branch revision retained with that initial integration. & Git commit: \nolinkurl{8591630c556f9f87c952f5d44ca7ce58b2bf1a83} \\
\phantomsection\label{prov:R12}R12 & Directory tree retained with the initial results integration. & Git tree: \nolinkurl{cb542312d8bde5d26bc716f2c8f7a5e8eaef7037} \\
\phantomsection\label{prov:R13}R13 & Initial \nolinkurl{results/WHITEPAPER_REPORT.md}. & Git blob: \nolinkurl{3e44ae10547b6376864089e605e946f74a378d56} \\
\phantomsection\label{prov:R14}R14 & 14 November attested report refresh; local/online identity checked. & Git commit: \nolinkurl{bd75c1c2104d001d62f9fb2bb4feab438b86f763} \\
\phantomsection\label{prov:R15}R15 & Refreshed \nolinkurl{results/WHITEPAPER_REPORT.md}. & Git blob: \nolinkurl{f416316985c8dca20ad8955fe3c596db995c6276} \\
\phantomsection\label{prov:R16}R16 & 14 November \nolinkurl{results/pilot-estimation-judging/extracted-contrasts.json}. & Git blob: \nolinkurl{f1aef11e7094ca284c25594459aecb180212c9cb} \\
\phantomsection\label{prov:R17}R17 & Completed 15 November report and exports; current empirical source. & Git commit: \nolinkurl{d097a10b090baefadf995ecd910eb96c324f0234} \\
\phantomsection\label{prov:R18}R18 & Directory tree of the completed report revision. & Git tree: \nolinkurl{2b07b649758f9f3f42d3adb86e1415b0c6ffb11d} \\
\phantomsection\label{prov:R19}R19 & Final \nolinkurl{results/EXPERIMENT_RESULTS.md}, Git identity. & Git blob: \nolinkurl{fdfb57da92f195d580f2ee3de4840fc0c13cd607} \\
\phantomsection\label{prov:R20}R20 & The same final experiment report, retained-byte digest. & SHA-256: \nolinkurl{e4680e66ba6791c7a23bd484954767bc7d580f881db939cf4fb113d28af10f15} \\
\phantomsection\label{prov:R21}R21 & Final \nolinkurl{results/pilot-estimation-judging/extracted-contrasts.json}; judging still uses its earlier snapshot. & Git blob: \nolinkurl{0ab114c5f02596baa5a6a25939f652dac6f30569} \\
\phantomsection\label{prov:R22}R22 & Execution version E15: audited quarantine-completion recovery for Gemini E-15 and GPT C-20. & Git commit: \nolinkurl{075102667af7b2e8aac926163cd0bcd726b40190} \\
\phantomsection\label{prov:R23}R23 & Execution version E16: final recovery completing all 315 trajectories. & Git commit: \nolinkurl{5481fd8a47aa0bad660e7c0ab68f7fcfeea4137f} \\
\phantomsection\label{prov:R24}R24 & Final recovery manifest; disclosed 14 November. & SHA-256: \nolinkurl{0fb342ff7efe16678e84f140220fd71eb8f01e25fee5b776c22aeb9ee3357577} \\
\phantomsection\label{prov:R25}R25 & Final execution inputs, after recovery. & SHA-256: \nolinkurl{c41f02de9f84843260b53ec59cc0f874b6afaf1555dc38209e66db63949f2f24} \\
\phantomsection\label{prov:R26}R26 & Final study state; 315 completed trajectories. & SHA-256: \nolinkurl{7992b0e21cbacde39b48bb6223d5b663b7013630b265ae64a37223414ab05e4a} \\
\phantomsection\label{prov:R27}R27 & Final owner-attestation manifest; scoped to state R26. & SHA-256: \nolinkurl{d26ebe733e50256184c2d5e4d3d66dda55233c14d54a434a037cde6f4cd64879} \\
\phantomsection\label{prov:R28}R28 & PhotoCraft source snapshot; \nolinkurl{storytold/photocraft}; application not executed here. & Git commit: \nolinkurl{5bb853654196eca807212ff8955ba2eb293071ea} \\
\phantomsection\label{prov:R29}R29 & Reported HVE Core \nolinkurl{plugins-v3.3.106} case-tool snapshot; \nolinkurl{microsoft/hve-core}. & Git commit: \nolinkurl{2df5013e29ebfd2ce61bcd0cf021d8c8186482eb} \\
\phantomsection\label{prov:R30}R30 & HVE Core source used to generate that snapshot; \nolinkurl{microsoft/hve-core}. & Git commit: \nolinkurl{e9af018fe5ffbfbc72321a8bca21fb987359af08} \\
\phantomsection\label{prov:R31}R31 & Historical v4 execution source and frozen consolidation prompt source. & Git commit: \nolinkurl{e8707ecc8965d3f81f5142a2603042bfca1aab15} \\
\phantomsection\label{prov:R32}R32 & Historical v4 manifest-only revision. & Git commit: \nolinkurl{03c50856838b539780d57216af2a4cf6744187d7} \\
\phantomsection\label{prov:R33}R33 & Historical \nolinkurl{osf-v4-2026-10-31.tar.gz} archive. & SHA-256: \nolinkurl{1a42fac40a4f1aa82df59c8c3f7f8d6dce37d22502c5952c398362ac63d6bd65} \\
\phantomsection\label{prov:R34}R34 & Historical v4 runtime image, not the amended study's image. & Image SHA-256: \nolinkurl{090220b0b172ec205f1ea473d7c3a223eec5d7d619eb21acdcf15c9289287fa3} \\
\phantomsection\label{prov:R35}R35 & Retained byte-identical \nolinkurl{anc/consolidation-frozen-v4.prompt.txt}. & SHA-256: \nolinkurl{75f646edbe04e1e3403b092b13edd6b66615a3362e67ad49744d229cdcb94f29} \\
\phantomsection\label{prov:R36}R36 & 9 November judging proposal and source-described runtime figure. & Git commit: \nolinkurl{b7973e200a6edbb1d19355badae3b7e03a066b09} \\
\phantomsection\label{prov:R37}R37 & \nolinkurl{judging/README.md} at the historical proposal revision. & Git blob: \nolinkurl{6cc0c236e4ca093198d38909dceb80f850196faa} \\
\phantomsection\label{prov:R38}R38 & Pre-recovery study state; 313 completed trajectories; Assessment v3 and earlier attestation baseline. & SHA-256: \nolinkurl{fb39a46fc36f33f62f3286fd517b63bb09787e9c7bcd7d276c4a362f256f2de1} \\
\phantomsection\label{prov:R39}R39 & Halted Assessment v1 judging source. & Git commit: \nolinkurl{3637817e7821dbfdeffb577725b4428f71d511f8} \\
\phantomsection\label{prov:R40}R40 & Assessment v1 ledger; first call halted before dispatch. & SHA-256: \nolinkurl{37db622597ca3c02ca74abbebb681c629d3a0a1c7c1110f3e447b15f827caa28} \\
\phantomsection\label{prov:R41}R41 & Halted Assessment v2 judging source. & Git commit: \nolinkurl{e44eaed7ad7e9bea254e7e77c1bb63957b39e68a} \\
\phantomsection\label{prov:R42}R42 & Assessment v2 ledger; 15 calls completed before an invalid response. & SHA-256: \nolinkurl{6fe5de37ea428cbe51ce5e4e6a95b96e483913ace24a50483d44afc842f6f068} \\
\phantomsection\label{prov:R43}R43 & Completed Assessment v3 judging source; one judge model, two repeats. & Git commit: \nolinkurl{c10e6d6f62785a5028406a9963acdd499f6bd6b6} \\
\phantomsection\label{prov:R44}R44 & Assessment v3 ledger; 1,252 completed calls. & SHA-256: \nolinkurl{cf8b71037a68b5cc08863cbabc65f3645416ba026cca45e0b007ac6a3d3d999b} \\
\phantomsection\label{prov:R45}R45 & Reported study runtime image; source/execution record, not newly qualified here. & Image SHA-256: \nolinkurl{61a8adc8e9764d1746e0277c4fb39b27512f972ece2a9c5696246c027e9fe4fb} \\
\phantomsection\label{prov:R46}R46 & Historical 14 November owner-attestation manifest; scoped to state R38, 313 trajectories. & SHA-256: \nolinkurl{66b098851800522d9d5117177aac34525cb1e9a17c0c4bd7d924d512c1ea6b06} \\
\phantomsection\label{prov:R47}R47 & 14 November \nolinkurl{results/WHITEPAPER_REPORT.md}, retained-byte digest. & SHA-256: \nolinkurl{ffde7266c3379446b5cdcc8fa695ee387c99a8a6c31fbd227b62291fd33c200c} \\
\phantomsection\label{prov:R48}R48 & Execution version E1; pinned CLI runtime fix. & Git commit: \nolinkurl{99a1ed170d7c71acb428b52e0e24be8856cc3ea3} \\
\phantomsection\label{prov:R49}R49 & Execution version E2; source-bound estimation recovery. & Git commit: \nolinkurl{3c2ce89b00fe583c1f0715e8ccd3510419d5e734} \\
\phantomsection\label{prov:R50}R50 & Execution version E4; sandbox-rejection recovery and analysis adjudications. & Git commit: \nolinkurl{fb1a90a0347609b1b1c33f4cd1147433a25cab96} \\
\phantomsection\label{prov:R51}R51 & Execution version E5; correlated HTTP 503 same-turn retries. & Git commit: \nolinkurl{355d33389441b325da72363a9c3028dbdefa2b4b} \\
\phantomsection\label{prov:R52}R52 & Execution version E6; WebSocket resets and recovery holds. & Git commit: \nolinkurl{1d663f08af117332773feb22efc8e182734780a3} \\
\phantomsection\label{prov:R53}R53 & Execution version E7; usage-unknown WebSocket closure accounting. & Git commit: \nolinkurl{958d75bc386964edfe6d211117ce53cf5fd53a82} \\
\phantomsection\label{prov:R54}R54 & Execution version E8; mixed-boundary retry history and export. & Git commit: \nolinkurl{03ac696b382e539cad5f8153b91d144bb4c1990b} \\
\phantomsection\label{prov:R55}R55 & Execution version E9; unaccounted native-retry recovery. & Git commit: \nolinkurl{732c4896da7c0a5b0db0f60c02348743a5b2c54e} \\
\phantomsection\label{prov:R56}R56 & Execution version E10; unbilled provider resets and paired halts. & Git commit: \nolinkurl{fc8dd5995cde046b4c2de1f2eb25a0df32ff39eb} \\
\phantomsection\label{prov:R57}R57 & Execution version E11; correlated Gemini 503 retries and halted charges. & Git commit: \nolinkurl{29e5f0834717b3c8a58a248858782174e775b6b8} \\
\phantomsection\label{prov:R58}R58 & Execution version E12; Gemini HTTP stream timeouts and holds. & Git commit: \nolinkurl{73712f7834ca520df543b6a58315bb299ffd38a8} \\
\phantomsection\label{prov:R59}R59 & Execution version E13; audited bootstrap for retained study state. & Git commit: \nolinkurl{0c35eff0e345d04c41722a43420d19b892c7fb8c} \\
\phantomsection\label{prov:R60}R60 & Active \nolinkurl{harness/prompts/consolidation-step.prompt.md} at R05; retained B template. & SHA-256: \nolinkurl{1ee3da712fcb787a7843d1270aabf61b55abb7c8b7df2534d4adc078b0241494} \\
\phantomsection\label{prov:R61}R61 & Active \nolinkurl{harness/prompts/review-only-step.prompt.md} at R05; retained C template. & SHA-256: \nolinkurl{06a8dfcaee30ab52c243c771844619cf08f23854ef866f1bfde5277362391725} \\
\end{longtable}
\endgroup

\end{document}